\documentclass[11pt]{article}
\usepackage[]{authblk}
\usepackage{amsmath}
\usepackage{graphicx}
\usepackage{latexsym}
\usepackage{amssymb}
\usepackage{amsbsy}

\DeclareSymbolFont{msbm}{U}{msb}{m}{n}
\DeclareMathSymbol{\C}{\mathalpha}{msbm}{'103}
\DeclareMathSymbol{\R}{\mathalpha}{msbm}{'122}
\DeclareMathSymbol{\Z}{\mathalpha}{msbm}{'132}
\DeclareMathSymbol{\N}{\mathalpha}{msbm}{'116}

\newtheorem{remark}{Remark}

\def\RR{\mathbb R}

\def\be{\begin{equation}}
\def\ee{\end{equation}}
\def\bea{\begin{eqnarray}}
\def\ba{\begin{array}{l}\displaystyle}
\def\eea{\end{eqnarray}}
\def\ea{\end{array}}

\def\IR{\mathop{\mbox{\rm Iround}}\nolimits}
\begin{document}

\title{The Moment Guided Monte Carlo Method
\thanks{Acknowledgements: This work was supported by the Marie Curie Actions of the European
Commission in the frame of the DEASE project (MEST-CT-2005-021122)
and by the French Commisariat \`{a} l'\'{E}nergie Atomique (CEA) in
the frame of the contract ASTRE (SAV 34160) }}

\author[1,2]{Pierre Degond}
\author[1,2,3]{Giacomo Dimarco\footnote
{Corresponding author address: Institut de Math\'{e}matiques de
Toulouse, UMR 5219 Universit\'{e} Paul Sabatier, 118, route de
Narbonne 31062 TOULOUSE Cedex, FRANCE. \\ \emph{E-mail
addresses}:pierre.degond@math.univ-toulouse.fr,
giacomo.dimarco@unife.it,lorenzo.pareschi@unife.it}}
\author[4]{Lorenzo Pareschi}

\affil[1]{Universit\'{e} de Toulouse; UPS, INSA, UT1, UTM ;

Institut de Math\'{e}matiques de Toulouse ; F-31062 Toulouse,
France.} \affil[2]{CNRS; Institut de Math\'{e}matiques de Toulouse
UMR 5219;

F-31062 Toulouse, France.} \affil[3]{Commissariat  \`a l'Energie
Atomique CEA-Saclay DM2S-SFME;

91191 Gif-sur-Yvette, France.} \affil[4]{University of Ferrara;
Department of Mathematics;

44100 Ferrara, Italy.}  \maketitle

\begin{abstract}
In this work we propose a new approach for the numerical simulation
of kinetic equations through Monte Carlo schemes. We introduce a new
technique which permits to reduce the variance of particle methods
through a matching with a set of suitable macroscopic moment
equations. In order to guarantee that the moment equations provide
the correct solutions, they are coupled to the kinetic equation
through a non equilibrium term. The basic idea, on which the method
relies, consists in guiding the particle positions and velocities
through moment equations so that the concurrent solution of the
moment and kinetic models furnishes the same macroscopic quantities.
\end{abstract}


\tableofcontents

\section{Introduction}
The Boltzmann equation provides a kinetic description of gases and
more generally of particle systems. In many applications, the
correct physical solution for a system far from thermodynamical
equilibrium, such as, for instance, rarefied gases or plasmas
requires the resolution of the Boltzmann equation \cite{cercignani}.
The numerical simulation of the Boltzmann equation with
deterministic techniques presents several drawbacks due to
difficulties in treating the collision terms and to the large
dimension of the problem. The distribution function depends on seven
independent variables: three coordinates in physical space, three
coordinates in velocity space and the time. As a consequence,
probabilistic techniques such as Direct Simulation Monte Carlo
(DSMC) methods are extensively used in real situations due to their
large flexibility and low computational cost compared to finite
volume, finite difference or spectral methods for kinetic equations
\cite{bird, Cf, CPima, Nanbu80}. On the other hand DSMC solutions
are affected by large fluctuations. Moreover, in non stationary
situations it is impossible to use time averages to reduce
fluctuations and this leads to, either poorly accurate solutions, or
computationally expensive simulations.

More generally Monte Carlo methods are frequently used in many real
applications to simulate physical, chemical and mathematical systems
\cite{sliu}. We quote \cite{Cf} for an overview on efficient and low
variance Monte Carlo methods. For applications of variance reduction
techniques to kinetic equation we mention the works of Homolle and
Hadjiconstantinou \cite{Hadji} and \cite{Hadji1}. We mention also
the work of Boyd and Burt \cite{Boyd} and of Pullin \cite{Pullin78}
which developed a low diffusion particle method for simulating
compressible inviscid flows. We finally quote the works of Dimarco
and Pareschi \cite{dimarco2} which worked on the construction of
efficient and low variance methods for kinetic equations in
transitional regimes.

The basic idea described in this work consists in reducing the
variance of Monte Carlo methods by forcing particles to match
prescribed sets of moments given by the solution of deterministic
equations. In order to provide the correct solution, the moment
equations are coupled to the DSMC simulation of the Boltzmann
equation through a kinetic correction term, which takes into account
departures from thermodynamical equilibrium.

We remark that the general methodology described here is independent
from the choice of the collisional kernel (Boltzmann, Fokker-Planck,
BGK etc..). However we point out that additional improvements can be
obtained with hypotheses on the structure of the distribution
function, on the type of considered kernel and on the type of
resolution methods used for the kinetic and fluid equations.

In the present paper we will focus on the basic matching technique,
which consists in matching the kinetic solution to that obtained by
the deterministic solution of the first three moment equations. The
idea is that the deterministic solution of the moment equation
(through finite volume or finite difference techniques) leads to a
more accurate solution, in term of statistical fluctuations, than
the DSMC method. Therefore, we constrain the DSMC method to match
the moments obtained through the deterministic resolution of the
moment equations in such a way that the higher accuracy of the
moment resolution improves the accuracy of the DSMC method. We
experimentally show that this is indeed the case.

We leave an in depth discussion of possible higher order matching
extensions to future work. For simplicity, in the numerical tests,
we will make use of a BGK collision term. However the formulation of
the method is general and extensions to the full Boltzmann
interaction term are possible without changing the structure of the
algorithm as explained in details in the paper. Results in the case
of the Boltzmann operator and improvements of the basic technique
described here will be presented in \cite{dimarco1}.

The remainder of the paper is organized as follows. In the next
section we recall some basic notions on the Boltzmann equations and
its fluid limit. The details of the numerical method are described
in Section 3. In Section 4 numerical examples which demonstrate the
capability of the method are presented. Finally some future
developments and remarks are detailed in the last Section.

\section{The Boltzmann equation and its fluid limit}

We consider equations of the following form \be
\partial_t f + v\cdot\nabla_{x}f
=Q(f,f)\label{eq:1}\ee with initial data \be f|_{t=0}=f_{init}\ee
where $f(x,v,t)$ is a non negative function describing the time
evolution of the distribution of particles with velocity $v \in
\R^{d_v}$ and position $x \in \Omega \subset \R^{d_x}$ at time $ t
> 0$. The operator $Q(f,f)$ describes particles interactions and
is assumed to satisfy the local conservation properties \be\langle
mQ(f,f)\rangle=0\ee where we define integrals over the velocity
space as follows\be\int_{\R^{d_v}} \psi dv=:\langle \psi\rangle\ee
and $m(v)=(1,v,\frac{|v|^2}{2})$ are the collision invariants.
Integrating (\ref{eq:1}) against its invariants in velocity space
leads to the following set of conservations laws \be
\partial_t \langle mf\rangle+\nabla_x
\langle vmf\rangle=0.\label{eq:macr}\ee Equilibrium functions for
the operator $Q(f,f)$ (i.e. solutions of $Q(f,f)=0$) are local
Maxwellian of the form \be M_{f}(\rho,u,T)=\frac{\rho}{(2\pi
T)^{d_v/2}}\exp\left(\frac{-|u-v|^{2}}{2T}\right), \label{eq:M}\ee
where $\rho$, $u$, $T$ are the density, mean velocity and
temperature of the gas at position $x$ and at time $t$\be
 \rho=\int_{\RR^d} fdv, \ u=\frac{1}{\varrho}\int_{\RR^d} vfdv, \ T=\frac{1}{d\varrho}
 \int_{\RR^d}|v-u|^{2}fdv.
\label{eq:Mo} \ee In the sequel we will denote by \be
U=(\rho,u,T),\qquad E[U]=M_f. \ee Clearly we have \be U=\langle
mE[U]\rangle.\ee

Now, when the mean free path between the particles is very small
compared to the typical length scale of the experiment, the operator
$Q(f,f)$ is large and we can rescale the space and time variables in
(\ref{eq:1}) as \be x'=\varepsilon x, \ \ t'=\varepsilon t,\ee to
obtain \be
\partial_t f + v\cdot\nabla_{x}f
=\frac1{\varepsilon}Q(f,f)\label{eq:1b},\ee where $\varepsilon$ is a
small parameter proportional to the mean free path and the primes
have been omitted to keep notations simple.

Passing to the limit for $\varepsilon\rightarrow 0$ leads to
$f\rightarrow E[U]$ and thus we have a closed hyperbolic system of
equations for the macroscopic variables $U$\be
\partial_t U+\nabla_xF(U)=0,\ee
with $F(U)=\langle vmE[U]\rangle$.

\section{The Moment Guided Monte Carlo Methods}

For the sake of simplicity, in this work, we consider the problem in
one dimension both in physical and velocity spaces. Extensions to
multidimensional problems are straightforward and will be considered
in \cite{dimarco1}. The starting point of the method is the
following micro-macro decomposition \be f=E[U]+g.\ee The function
$g$ represents the non-equilibrium part of the distribution
function. From the definition above, it follows that $g$ is in
general non positive. Moreover since $f$ and $E[U]$ have the same
moments we have \be \langle mg\rangle=0.\ee Now $U$ and $g$ satisfy
the coupled system of equations \begin{eqnarray}
\partial_t U+\partial_x
F(U)+\partial_x\langle vmg\rangle&=&0,\label{eq:2}\\
\partial_t f+v\partial_x
f&=&Q(f,f). \label{eq:3} \end{eqnarray}  We skip the elementary
proof of the above statement and refer to \cite{degond2} for details
on the decomposition of the distribution function and the coupled
systems which it is possible to derive.

Our goal is to solve the kinetic equation with a Monte Carlo method,
and concurrently the fluid equation with any type of finite
difference or finite volume scheme, where the correction term
$\partial_x\langle vmg\rangle$ is evaluated using particle moments.
The two equations (\ref{eq:2}-\ref{eq:3}), except for numerical
errors, give the same results in terms of macroscopic quantities. It
is natural to assume that the set of moments obtained from the fluid
system represents a better statistical estimate of the true moments
of the solution, since the resolution of the moment equations does
not involve any stochastic process.

Thus we can summarize the method in the following way. At each
time step $t^n$
\begin{enumerate}
 \item Solve the kinetic equation (\ref{eq:3}) with a Monte Carlo scheme and obtain a first
 set of moments $U^*=\langle mf^*\rangle$.
 \item Solve the fluid equation (\ref{eq:2}) with a finite volume/difference scheme using particles to evaluate $\partial_x\langle vmg\rangle$ and obtain a second set of moments $U^{n+1}$.
 \item Match the moments of the kinetic solution with the fluid solution through a transformation of the
 samples values $f^{n+1}=T(f^*)$ so that $\langle mf^{n+1}\rangle=U^{n+1}$.
 \item Restart the computation to the next time step.
 \end{enumerate}
For Step 1, one can use any Monte Carlo method (or more generally
any low accurate but fast solver). Step 2 and 3 of the above
procedure require great care since they involve the evaluation of
$\partial_x\langle vmg\rangle$ and the moment matching procedure.

Finally let us note that, in principle, it is possible to improve
the method, adding to system (\ref{eq:2}) additional equations for
the time evolution of higher order moments and get \be
\partial_t \langle m_n f\rangle+\partial_x \langle v
m_nf\rangle=\langle m_nQ(f,f)\rangle\label{eq:5}\ee with $m_n=v^n$
and $n\geq 3$. The solution of (\ref{eq:5}) with a finite
volume/difference scheme, which in the general case is not
straightforward, will provide a better estimate of the moments which
are used in the moment matching \cite{Grad}, \cite{Levermore1},
\cite{Ruggeri}.

We will call this general class of methods Moment Guided Monte Carlo
schemes. In the sequel, we briefly focus on steps $2$ and $3$ of the
above procedure.

\subsection{Solution of the Moment Equations}

In this section we discuss the discretization of the moment
equations. We will, at the end of the section, suggest some
approaches that can possibly be used to improve the method in the
nearby future. Our scope, in the construction of the numerical
scheme, is to take advantage from the knowledge of the Euler part of
the moment equations \be \underbrace{\partial_t U+\partial_x
F(U)}_{\hbox{Euler equations}}+\partial_x \langle vmg \rangle=0.
\label{eq:2b}\ee Thus, the method is based on solving first the set
of compressible Euler equations, for which many efficient numerical
methods have been developed in the literature, and then considering
the discretization of the kinetic flux $\partial_x\langle
vmg\rangle$. To that aim, for the space discretization of the
compressible Euler equations we use both a first order central
scheme of Lax-Friedrichs type or a second order MUSCL central
scheme, while a backward discretization is used for the time
derivative in all cases\be \frac{U^{*}_i-U^{n}_i}{\Delta t}
+\frac{\psi_{i+1/2}(U^n)-\psi_{i-1/2}(U^n)}{\Delta
x}=0.\label{eq:discmom}\ee In the case of the second order scheme
the discrete flux reads \be
   \psi_{i+1/2}(U^n)=\frac{1}{2}(F(U^n_{i})+F(U^n_{i+1}))-\frac{1}{2}\alpha(U^n_{i+1}-U^n_{i})+\frac{1}{4}(\sigma^{n,+}_i-\sigma^{n,-}_{i+1})
\ee where \be \sigma^{n,\pm}_i=\left(F(U^n_{i+1})\pm \alpha
U^n_{i+1}-F(U^n_{i})\mp \alpha
U^n_{i}\right)\varphi(\chi^{n,\pm}_i)\ee with $\varphi$ a given
slope limiter, $\alpha$ equal to the larger eigenvalue of the Euler
system and \be \chi^{n,\pm}_i=\frac{F(U^n_{i})\pm \alpha
U^n_{i}-F(U^n_{i-1})\mp \alpha U^n_{i-1}}{F(U^n_{i+1})\pm \alpha
U^n_{i+1}-F(U^n_{i})\mp \alpha U^n_{i}}\ee where the above ratio of
vectors is defined componentwise. In the numerical test section we
used both first and second order discretization techniques to test
their different behaviors when coupled with the DSMC solver. As
explained in that section, second order schemes increase
fluctuations while first order does not. Moreover, since the
numerical diffusion introduced by first order schemes can be
excessive, we also test a switching method which passes from first
to second order accuracy as a function of the ratio between the non
equilibrium term and the equilibrium one. Thus, we define the
following quantities\be
\beta_{i}^{n}=\frac{\lambda_{i}^{n}}{|F_3(U^{n}_i)|}, \
\lambda_{i}=\int_{\R^3}v \frac{|v|^2}{2}
g^{n}_{i}dv\label{eq:crit2}\ee where $F_3(U)$ is the energy flux.
Note that in the one dimensional case the non equilibrium mass and
momentum fluxes are identically zero. The MUSCL second order scheme
is then used when $\beta^n_i$ is small while the first order scheme
is used otherwise. This method, as showed in the tests, does not
increase fluctuations and, at the same time, guarantees a lower
level of numerical dissipation in the results.

We now discuss how to discretize the non equilibrium term
$\partial_x<vmg>$. To this aim, the same space first order discrete
derivative is used as for the hydrodynamic flux $F(U)$. The non
equilibrium term $\langle vmg\rangle=\langle vm(f-E[U])\rangle$ is
computed by taking the difference between the moments of the
particle solution and those of the Maxwellian equilibrium. Thus the
final scheme, for the moment equations, reads\be
\frac{U^{n+1}_i-U^{n}_i}{\Delta t}
+\frac{\psi_{i+1/2}(U^n)-\psi_{i-1/2}(U^n)}{\Delta
x}+\frac{\Psi_{i+1/2}(<vmg^n>)-\Psi_{i-1/2}(<vmg^n>)}{\Delta
x}=0.\label{eq:discmom1}\ee where $\psi_{i+1/2}(U^n)$ can be either
first or second order while $\Psi_{i+1/2}(<vmg^n>)$ is always first
order. In our method, in addition to the mass momentum and energy
equations, we consider the third order moment evolution equation. In
this case, like in (\ref{eq:5}), we have the additional problem of
evaluating the source term that now appears at the right hand side.
At the particle level this can be done by simply measuring the
variations of higher order moments in each cell during particle
collisions. Thus the discretized third order moment equation is
performed in two steps, where the second one reads\be \frac{<m_3
f^{n+1}_i>-<m_3 f^{*}_i>}{\Delta t}
+\frac{\Psi_{i+1/2}(<vm_3f^n>)-\Psi_{i-1/2}(<vm_3f^n>)}{\Delta
x}=0,\label{eq:discmom3}\ee and where $f^*$ is the solution of the
first step, the collision step, which depends on the type of
collisional operator. We will describe this step in the numerical
test section below in the case of a BGK type kernel. The main
advantage of considering additional moment equations is that this
reduces the fluctuations in the evaluations of the macroscopic
quantities $U$. Indeed, in the extended moment system, particles
play a role only in the evaluation of higher order terms $\langle
v^pf \rangle$, $p > 3$ and not directly on the evolution of the
hydrodynamics quantities.

As a conclusion for this section, we discuss some possible
improvements which will be developed in future works
\cite{dimarco1}. To this aim, we observe that the decomposition of
the flux term into an equilibrium and a non equilibrium part can be
further exploited. Indeed, as an effect of the guided Monte Carlo
technique, the only remaining source of fluctuations in the moment
equations is due to the non equilibrium term $\partial_x\langle v m
g\rangle$. Thus, instead of using the same numerical scheme as for
the flux $\partial_x F(U)$, we can develop a specific discretization
method which further reduces the variance of these fluctuations. We
can consider cell averages of the form \be \frac{1}{\Delta
x}\int^{x_{i+1/2}}_{x_{i-1/2}}\partial_x \langle v m
g\rangle\,dx=\frac{\langle v m g\rangle|_{x=x_{i+1/2}}-\langle v m
g\rangle|_{x=x_{i-1/2}}}{\Delta x}, \ m=(1,v,|v|^2),\ee with
$\langle vmg \rangle = \langle vmf \rangle -F(U)$.
The integral over the velocity space can be evaluated by summing
over the particles\be \langle vmf \rangle|_{x=x_{i+1/2}} \approx
\frac{1}{N}\sum_{j\in I_{i+1/2}}B(p_j-x_{i+1/2})m_j\ee where $p_j$
and $\nu_j$ represent the position and velocity of the j-th
particle, $m_j=(1,\nu_j,|\nu_j|^2)$, $I_{i+1/2}$ a given space
interval of size $h$ (typically $h\geq \Delta x$) containing
$x_{i+1/2}$ and $B\geq 0$ is a suitable weight function s.t.
\[
\int_{\RR} B(x)\,dx=1.
\]
For example $B(x)=1/h$ if $|x|\leq h/2$ and $B(x)=0$ elsewhere,
gives rise to a simple sum of the particles moments in the interval
$I_{i+1/2}$ known as the 'Nearest Grid Point' procedure in plasma
physics \cite{birsdall}. Smoother reconstructions can be recovered
by convolving the samples with a bell-shaped weight like a B-spline
\cite{pa-inria}. Note that the value $h$ has a strong influence on
the fluctuations in the reconstructed function, and in general
should be selected as a good compromise between fluctuations and
resolution.



\subsection{The Moment Matching}
\label{sec:mm} In the present work we restrict ourselves to the
following linear transformation: let a set of velocities
$\nu_1,\ldots,\nu_{J}$ with first two moments $\mu_1$ and $\mu_2$ be
given. Suppose better estimates $\sigma_1$ and $\sigma_2$ of the
same moments are available (using the moment equation). We can apply
the transformation described in \cite{Cf} \be
\nu_j^*=(\nu_j-\mu_1)/c+\sigma_1\quad
c=\sqrt{\frac{\mu_2-\mu_1^2}{\sigma_2-\sigma_1^2}},\quad
i=1,\ldots,J \label{eq:4}\ee to get
\[
\frac1{J}\sum_{j=1}^J \nu_j^*=\sigma_1,\qquad
\frac1{J}\sum_{j=1}^J (\nu_j^*)^2=\sigma_2.
\]
Of course this renormalization is not possible for the moment of
order zero (the mass density). Let us denote by $\mu_0$ an estimate
of the zero order moment and by $\sigma_0$ its better evaluation by
the moment equations.

Among the possible techniques that can be used to restore a
prescribed density we choose to replicate or discard particles
inside the cells. Other possibilities are to deal with weighted
particles, move particles among cells according to some
interpolation procedure or reconstruct the probability distribution
starting from samples and resample particles. We leave a deeper
analysis of possible alternate choices to future works.

In order to recover the moment $\sigma_0$, in the case
$\mu_0>\sigma_0$, we can use a discarding procedure. Note that we
would like to eliminate exactly the following number of
particles,\be\widetilde{N_p}=\frac{\mu_0-\sigma_0}{M_p}\ee where
$M_p$ is the mass of a single particle. In general, the precise
match is impossible, since the particles mass is kept fixed in time,
and $\widetilde{N_p}$ can never be an integer. A fixed mass $M_p$
implies that \be \mu_0=N_1M_p, \ \sigma_0=N_2M_p\ee with $N_1$ and
$N_2$ integers such that $N_1>N_2$. $N_1$ and $N_2$ are the number
of particles in the cell before and after the matching. Moreover,
since the estimate $\sigma_0$ is not in general an integer multiple
of $M_p$, a mismatch $e$ such that $e<\pm M_p$ is unavoidable. Thus
we can simply eliminate from the cell a suitable stochastic integer
approximation of $\widetilde{N_p}$ \be
N_p=\IR\left(\frac{\mu_0-\sigma_0}{M_p}\right)\ee where $\IR(x)$ is
a stochastic rounding defined as
\begin{equation}\IR=\left\lbrace
\begin{array}{ll}
\lfloor x\rfloor +1, & \text{with probability}\ \ x-\lfloor x\rfloor\\
\lfloor x\rfloor, & \text{with probability}\ \  1-x+\lfloor
x\rfloor
\end{array}
\right.
\end{equation}
with $\lfloor x\rfloor$ the integer part of $x$.

In the opposite case, in which the mass of the particles inside a
cell is lower than the mass prescribed by the fluid equations
$\mu_0<\sigma_0$, the situation is less simple. In this situation,
since the distribution function is not known analytically, it is not
possible to sample new particles without introducing correlations
between samples. In this case we need to replicate \be
N_p=\IR\left(\frac{\sigma_0-\mu_0}{M_p}\right)\ee particles. Note
that this is done allowing repetitions. After the generation step,
samples are relocated uniformly inside each spatial cell.

Now we briefly discuss the possibility of forcing samples to follow
higher order prescribed moments. To this aim, observe that, the
moment matching procedure has infinite possible solutions, since the
number of particles inside a cell is larger than the number of the
constraints. However, we aim at finding a transformation which
possibly preserves the Gaussian ditribution. The only operations
which obey this constraint are linear transformations like
(\ref{eq:4}), i.e.  shifts and homotheties of the particle
velocities.

However, if we slightly relax the constraint of preservation of the
Gaussian distribution, we can reformulate the problem in the
following terms: find a suitable transformation which leads to the
required moments with the minimal changes in the distribution
function. In the general case, this request has a non trivial answer
which can be recovered by solving an appropriate non linear system
of equations with several constraints at each time step for every
cell. For this reason an efficient implementation of this procedure
is still an open question.


\section{Numerical results}

In the present section we report on some numerical results of the
moment guided method on different test cases obtained using a
simplified BGK model for the kinetic equation. First, we perform an
accuracy test using a smooth periodic solution and then we consider
two classical shock problems. In all the tests, we compare the
moment guided (MG) solution with the standard Monte Carlo (MC)
solution and with the direct deterministic solution to the BGK
equations based on a discrete velocity model (DVM) \cite{Mieussens}.

\subsection*{The Moment Guided DSMC method applied to the BGK model}

In this paragraph we detail a possible algorithm, which merges the
techniques described in the previous sections, in the case of the
simplified BGK collision operator.

As usual the starting point of Monte Carlo methods is given by a
time splitting \cite{pa-inria} between free transport \be
\partial_t f + v\cdot\nabla_{x}f
=0,\ee and collision, which in the case of the BGK operator is
substituted by a relaxation towards the equilibrium \be
\partial_t f=\frac{1}{\varepsilon}(f-E[U]).\label{eq:9}\ee
In Monte Carlo simulations the distribution function $f$ is
discretized by a finite set of particles
\begin{eqnarray}
& & f = \sum_{i=1}^N M_P \, \delta(x-x_i(t)) \delta(v-v_i(t)),
\label{particle}
\end{eqnarray}
where $x_i(t)$ represents the particle position and $v_i(t)$ the
particle velocity. During the transport step then, the particles
move to their next positions according to \be x_i(t+\Delta
t)=x_i(t)+v_i(t)\Delta t\label{transport}\ee where $\Delta t$ is
such that an appropriate CFL condition holds.

The collision step changes the velocity distribution and, in this
simplified case, the space homogeneous problem admits the exact
solution at time $t+\Delta t$ \be f(t+\Delta t)=e^{-\Delta
t/\varepsilon}f(t)+(1-e^{-\Delta
t/\varepsilon})E[U](t).\label{eq:8}\ee The relaxation step of a
Monte Carlo method for the BGK equation consists in replacing
randomly selected particles with Maxwellian particles with
probability $(1-e^{-\Delta t/\varepsilon})$. Thus\be v_i(t+\Delta
t)=\left\lbrace
\begin{array}{ll}
\displaystyle v_i(t), &  \text{with probability} \ e^{-\Delta
t/\varepsilon}\\
\displaystyle E[U](v), &  \text{with probability} \ 1-e^{-\Delta
t/\varepsilon}\\
\end{array}
\right.\label{relax}\ee where $E[U](v)$ represents a particle
sampled from the Maxwellian distribution with moments $U$.

Thus, finally, at each time step the moment guided Monte Carlo
method reads as follows:
\begin{description}
\item[(i)] transport and collide particles
(\ref{transport}-\ref{relax});
\item[(ii)] solve the first three moment equations (\ref{eq:discmom1})
and the additional equation for the third order moment
(\ref{eq:discmom3});
\item[(iii)]  match the computed mass, momentum and energy of the particle solution (section \ref{sec:mm})
 to those computed with the moment
equations.
\end{description}
Moments are reconstructed by simple summation formulas in each cell;
fluxes are then obtained by interpolation on the grid points and
then discretized with Lax-Friedrichs type central schemes of first
or second order as described in the previous sections.

\begin{remark}~
\begin{itemize}

\item Second order methods have been used for the Sod tests
while first order methods have been used for all others tests. We
point out that second order scheme may produce larger fluctuations,
especially when slope limiters are used and that the switching
technique between first and second order schemes as described in
(\ref{eq:crit2}) prevents the onset of these oscillations in the
considered test cases.

\item After the relaxation step (\ref{eq:8}), the perturbation term can be rewritten as
\begin{eqnarray}
   g(t+\Delta t)&=&f(t+\Delta t)-E[U(t+\Delta t)]=e^{-\Delta
t/\varepsilon}f(t)+(1-e^{-\Delta
t/\varepsilon})E[U(t)]+\nonumber  \\
   &-&E[U(t+\Delta t)]=e^{-\Delta
t/\varepsilon}(f(t)-E[U(t)])=e^{-\Delta t/\varepsilon}g(t),
\end{eqnarray}
since $U(t+\Delta t)=U(t)$ in the space homogeneous case. Thus the
discretized moment equations (\ref{eq:discmom1}) can be rewritten as
\begin{eqnarray}
  U^{n+1}_i &=& U^{n}_i-\frac{\Delta
t}{\Delta
x}(\psi_{i+1/2}(U^n)-\psi_{i-1/2}(U^n))+\nonumber\\&-&\frac{\Delta
t}{\Delta
x}e^{-\Delta t/\varepsilon} \Psi_{i+1/2}(<vm(f^n-E[U^n])>)+  \nonumber\\
  &+&\frac{\Delta t}{\Delta
x}e^{-\Delta t/\varepsilon}
\Psi_{i-1/2}(<vm(f^n-E[U^n])>).\label{eq:discmom2BGK}
\end{eqnarray}
As $\Delta t /\varepsilon$ grows, which means that the system
approaches the equilibrium, the contribution of the kinetic term
vanishes even though it is evaluated through particles. This does
not happen if we just compute the kinetic term $\partial_x \langle
vmg\rangle$ from the particles without considering the structure of
the distribution function $f$. This dramatically decreases
fluctuations when the Knudsen number is small.

Since this property is related to the BGK structure and we aim at a
method that can be applied to the full Boltzmann equation we do not
take advantage of it in the numerical results. We leave the
possibility to extend this idea to the Boltzmann equation using time
relaxed Monte Carlo (TRMC) methods\cite{PR} to future
investigations.
\end{itemize}
\end{remark}

\subsection*{Accuracy test} 
\begin{figure}
\begin{center}
\includegraphics[scale=0.39]{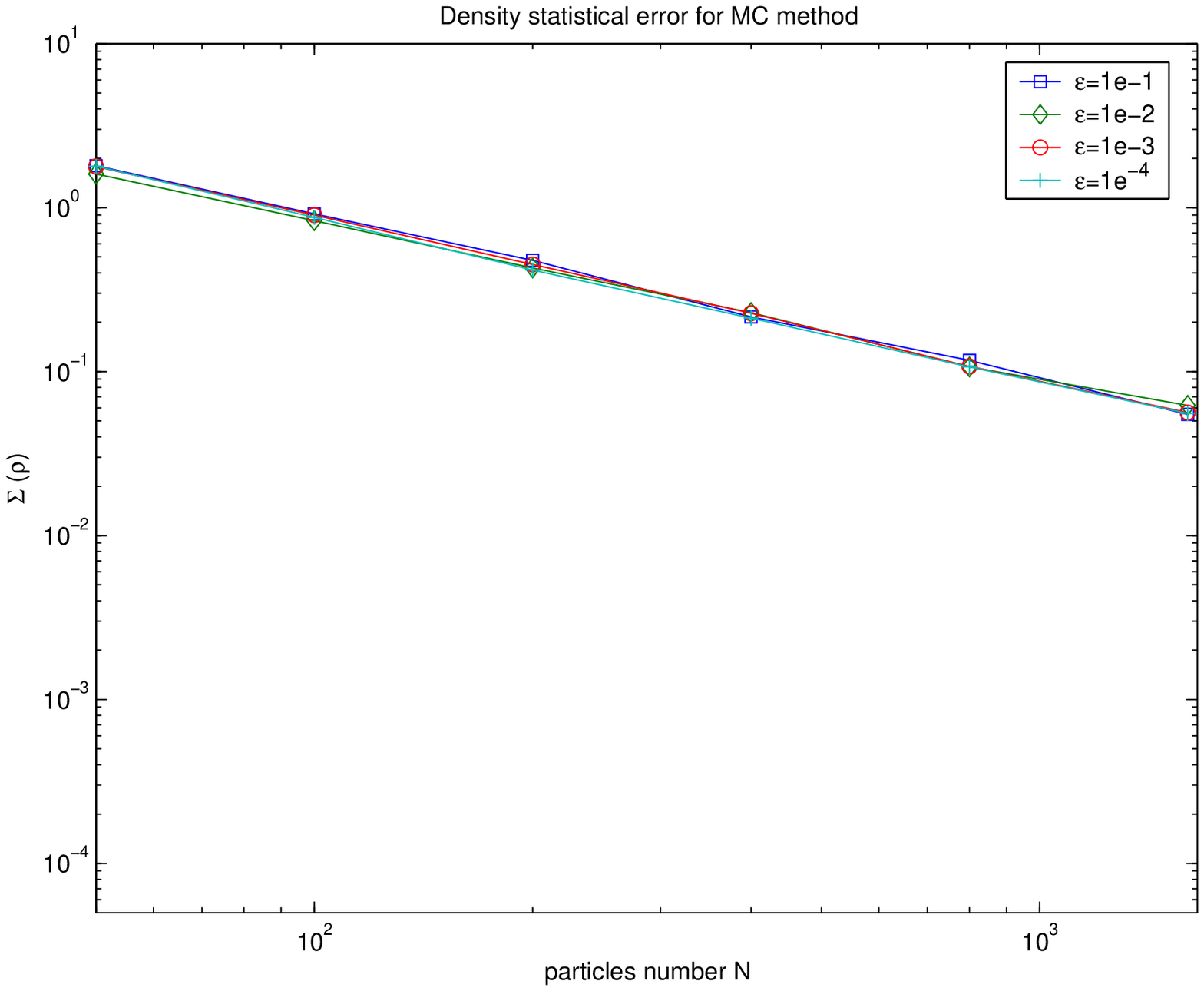}
\includegraphics[scale=0.39]{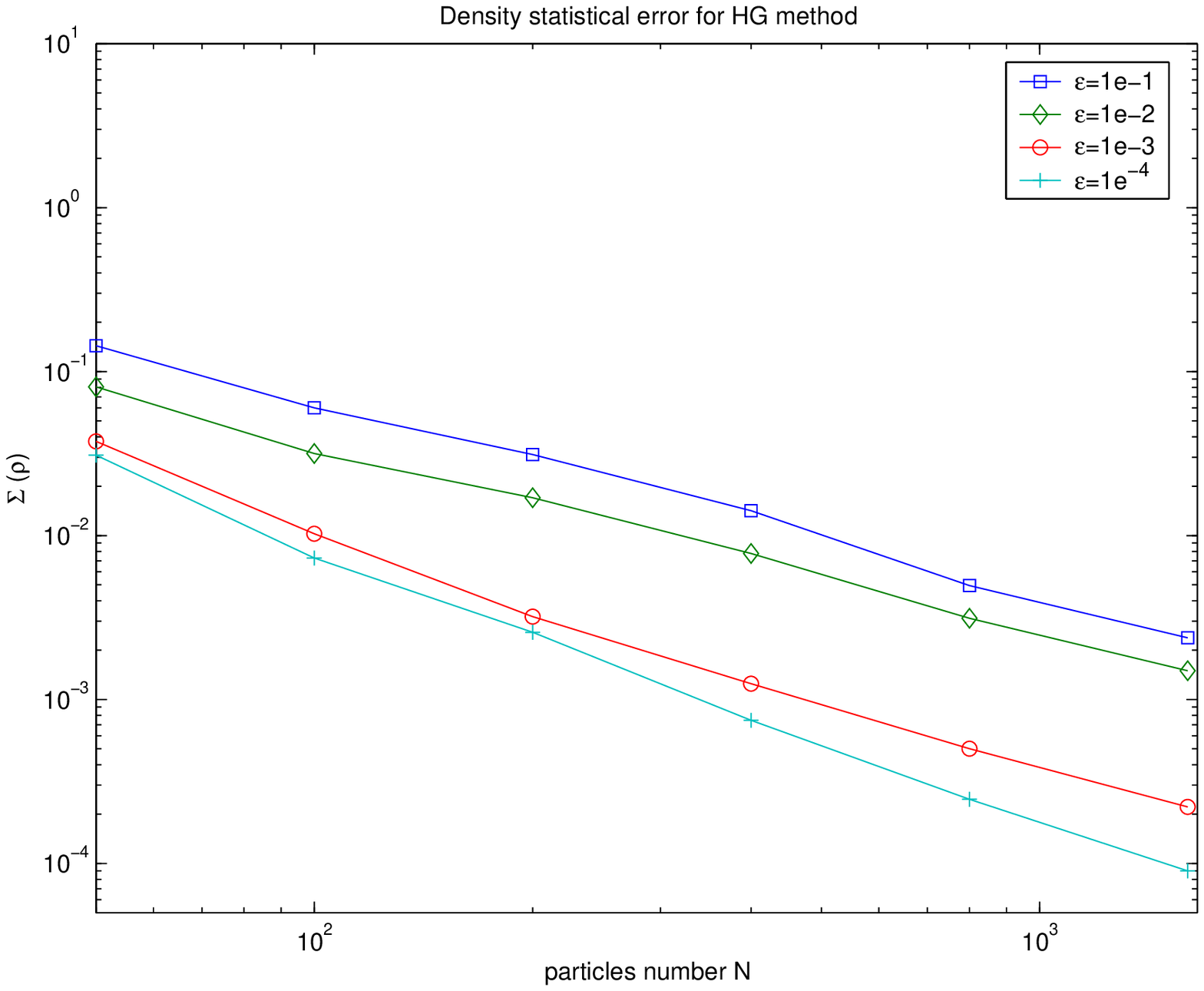}
\includegraphics[scale=0.39]{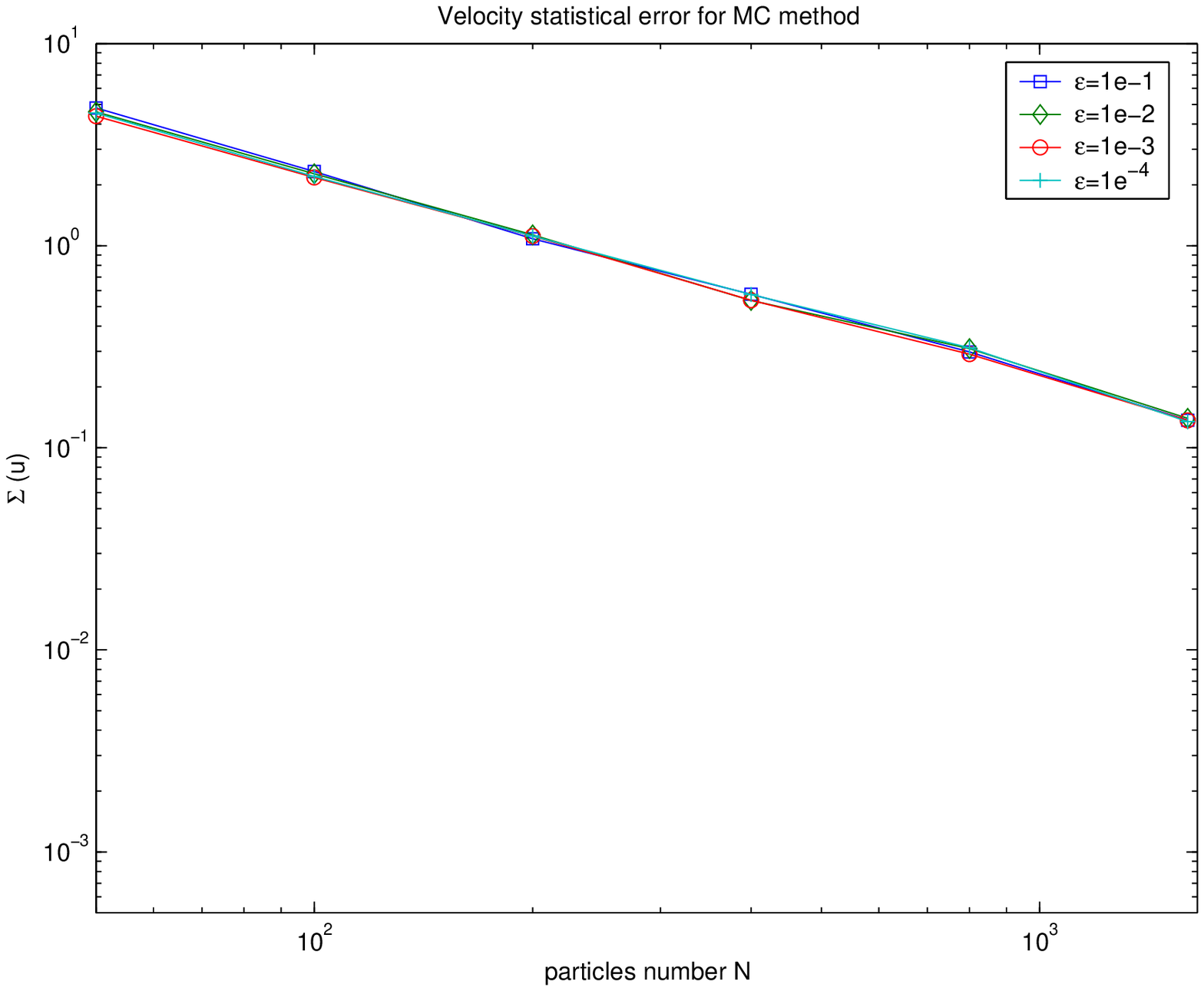}
\includegraphics[scale=0.39]{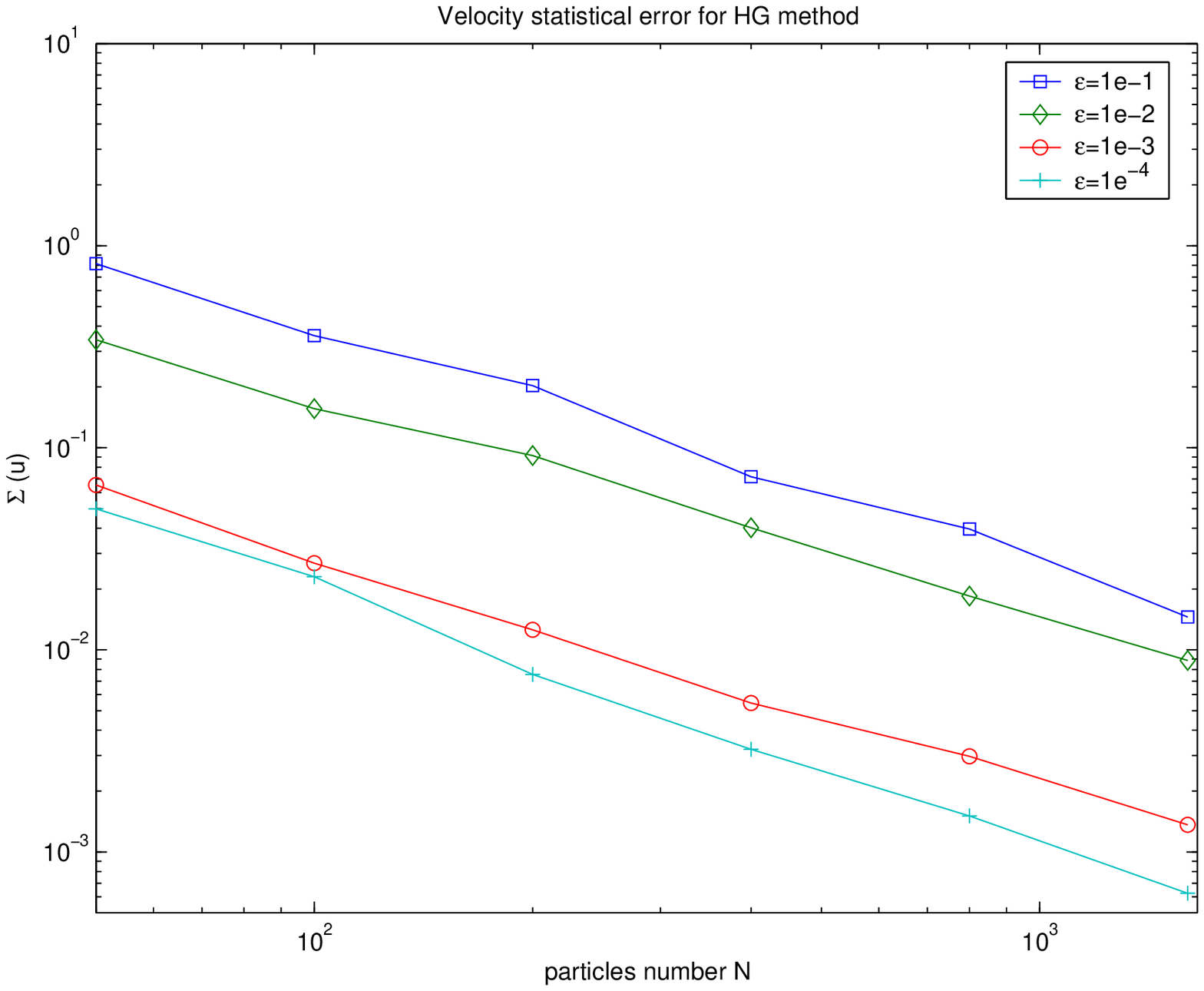}
\includegraphics[scale=0.39]{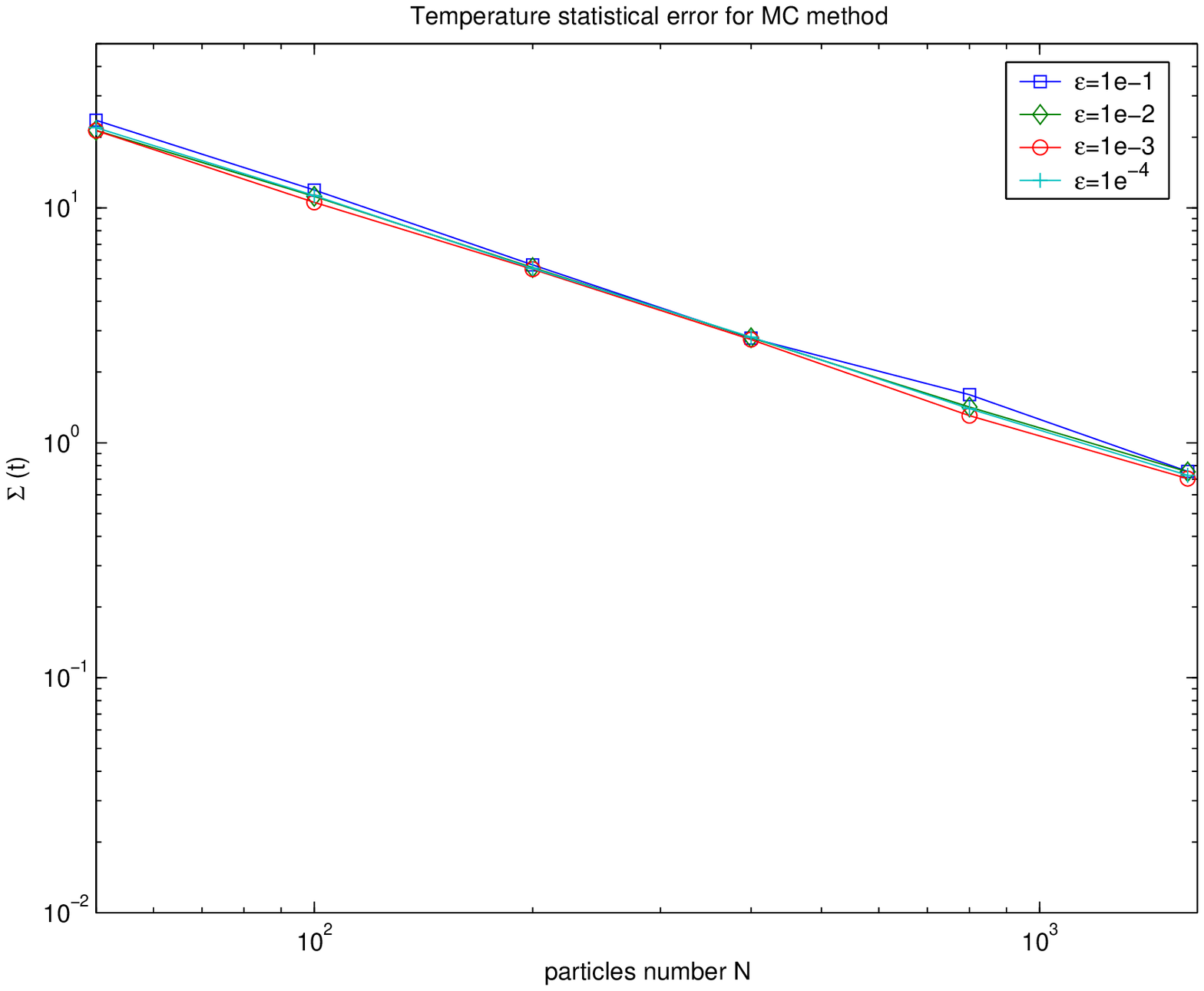}
\includegraphics[scale=0.39]{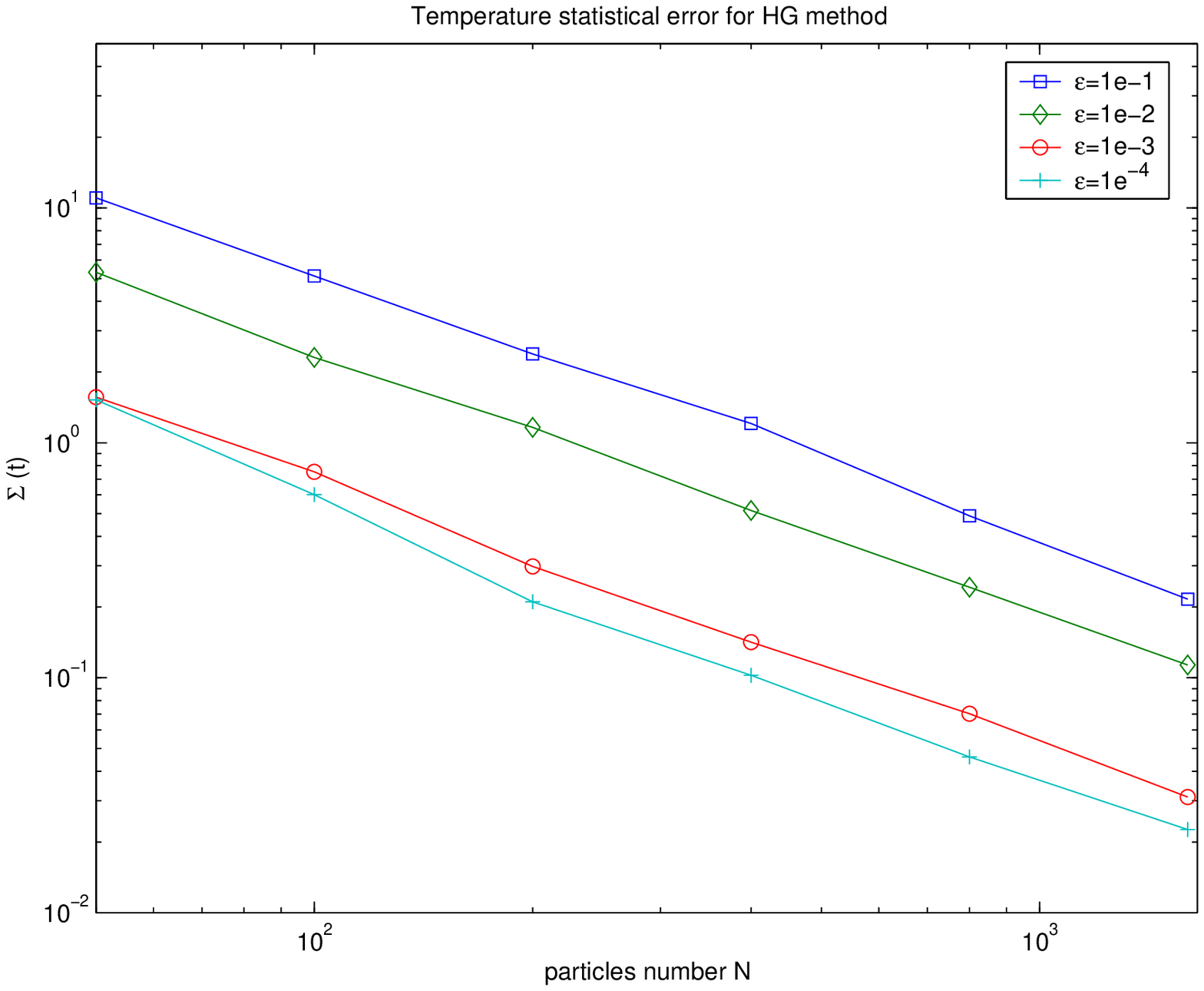}
\caption{Statistical error test: Solution at $t=0.05$ for density
(top), velocity (middle) and temperature (bottom). MC method (left),
Hydro Guided MC method (right). Knudsen number vary from
$\varepsilon=10^{-1}$ to $\varepsilon=10^{-4}$. Squares indicate
errors for $\varepsilon=10^{-1}$, diamonds for
$\varepsilon=10^{-2}$, circles for $\varepsilon=10^{-3}$ while
crosses indicate errors for $\varepsilon=10^{-4}$.} \label{SE}
\end{center}
\end{figure}

First we report on the results of a
stochastic error analysis with respect to the number of particles.
As reference solution we considered the average of $M$ independent
realizations\be
\overline{U}_{MC}=\frac{1}{M}\sum_{i=1}^{M}U_{i,MC}\ee and \be
\overline{U}_{MG}=\frac{1}{M}\sum_{i=1}^{M}U_{i,MG}\ee where the two
subscripts $MC$ and $MG$ indicate respectively the reference
solution for the Monte Carlo method and for the Moment Guided
method. We use two different reference solutions since the two
schemes present different discretization errors and thus they
converge, when the number of particles goes to infinity, to
different discretized solutions. The two reference solutions are
obtained by fixing the time step and mesh size and letting the
number of particles go to infinity. In this way, both reference
solutions contain negligible stochastic error. At the same time,
both solutions involve space and time discretization errors.
However, the amount of such errors does not change when the number
of particles varies. Therefore by comparing solutions obtained with
a given $\Delta t$, $\Delta x$, but with a finite number of
particles to reference solutions obtained with the same $\Delta t$,
$\Delta x$, but with a very large number of particle, we obtain a
true measure of the error originating from the stochastic nature of
the method. Then, we measure the quantity \be
\Sigma^2(N)=\frac{1}{M}\sum_{i=1}^{M}\sum_{j=1}^{j_{max}}(U_{i,j}-\overline{U_{j}})^2\ee
where $\overline{U_j}$ represents the reference solution and
$j_{max}$ represents the number of mesh point. The test consists of
the following initial data \be \nonumber \varrho(x,0)=1+a_\varrho
\sin\frac{2\pi x}{L} \ee \be u(x,0)=1.5+a_u \sin\frac{2\pi x}{L} \ee
\be \nonumber \frac{1}{2}\int f |v|^2dv=W(x,0)=2.5+a_W
\sin\frac{2\pi x}{L} \ee with \be \nonumber a_\varrho=0.3 \ \
a_u=0.1 \ \ a_W=1. \ee This test problem gives rise to a periodic
smooth solution in the interval $t \in [0,5\times 10^{-2}]$. The
results of this test in log-log scale are shown in Figure \ref{SE}.
On the left, we reported the stochastic error for the pure Monte
Carlo and on the right for the Moment Guided method. From top to
bottom, the errors for the three macroscopic quantities are depicted
for different values of the Knudsen number. For the Monte Carlo
case, the stochastic error does not substantially change with
respect to the Knudsen number and shows a convergence rate
approximatively equal to $1/2$. At variance, for the Moment Guided
method, errors decrease as the Knudsen number diminish and the
convergence rate of the method increases achieving values close to
one. This behavior is due to the fact that, for large Knudsen
numbers, the kinetic part of the solution, $g$, is not negligible
and evaluated through the DSMC method. By contrast, close to
thermodynamical equilibrium, $g\rightarrow 0$, which means that the
Monte Carlo component of the solution carries only fluctuations but
no information. It is remarkable that, in all analyzed regimes, the
stochastic error of the Moment Guided method is smaller than that of
the pure particle solver.

\subsection*{Unsteady shock test}

Next we consider an unsteady shock test case. This choice reflects
the fact that the method is specifically aimed at situations in
which the classical variance reduction technique using time
averaging cannot be used or turns out to be useless, since
time-averaging or using more particles leads to the same
computational effort.

\begin{figure}
\begin{center}
\includegraphics[scale=0.39]{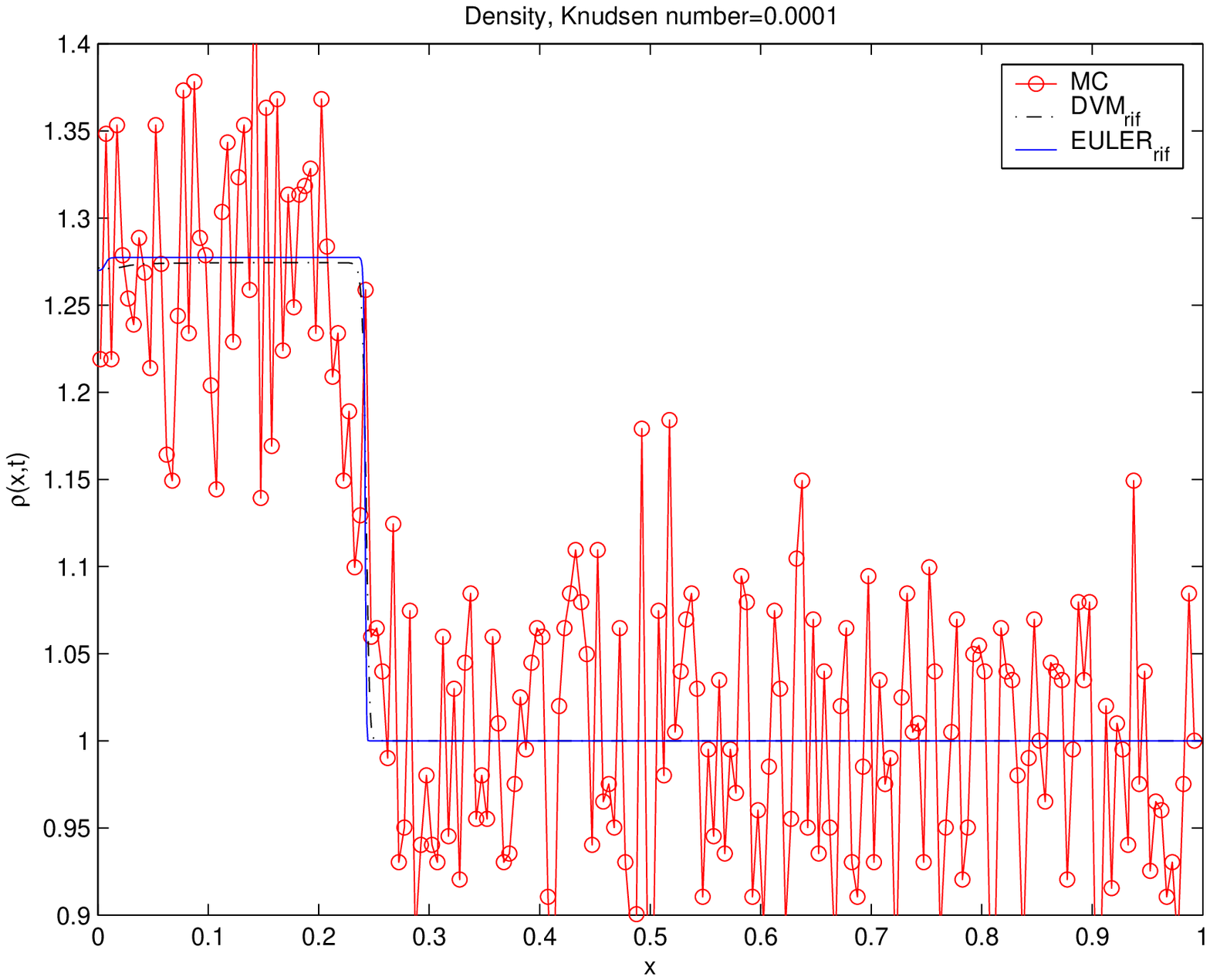}
\includegraphics[scale=0.39]{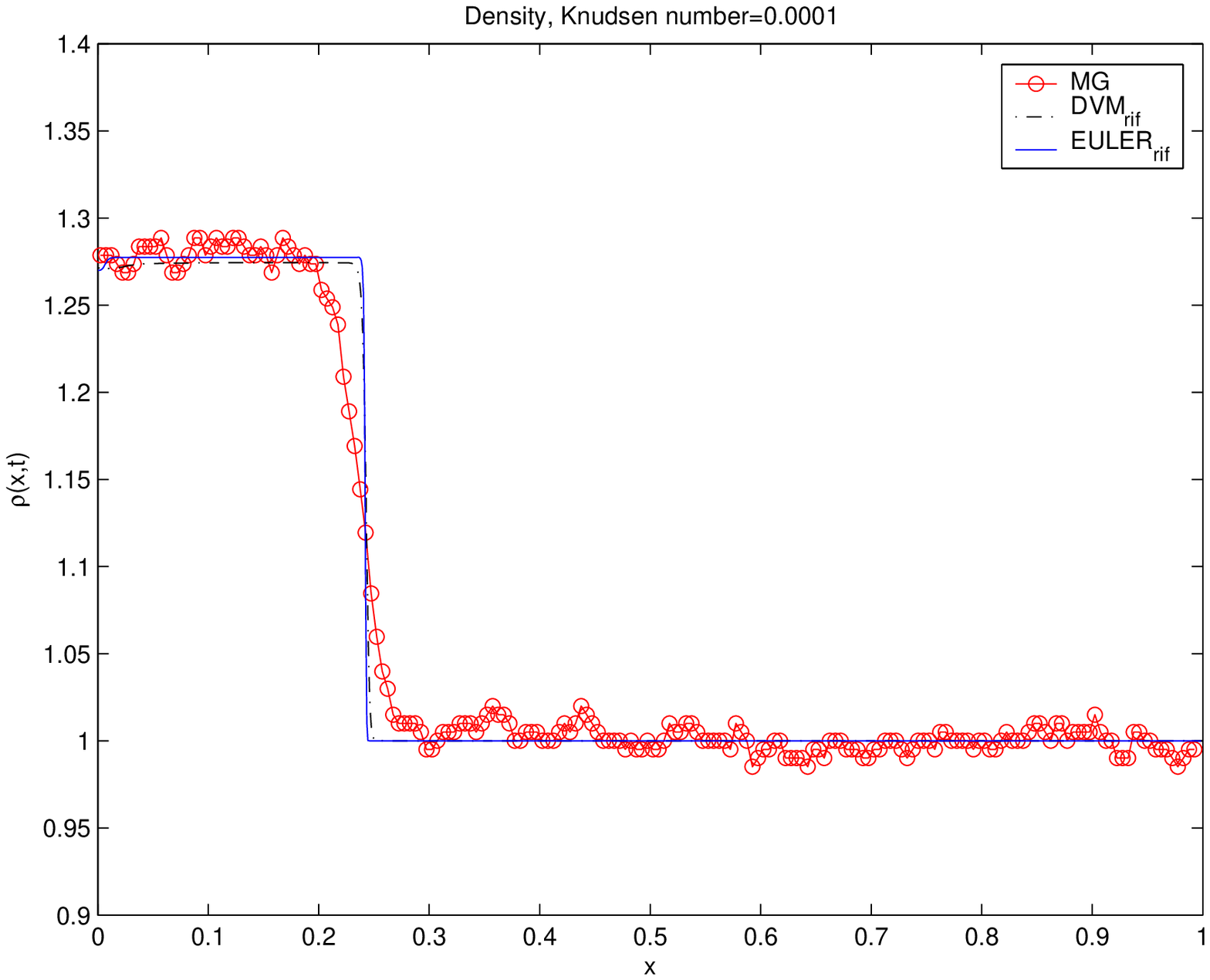}
\includegraphics[scale=0.39]{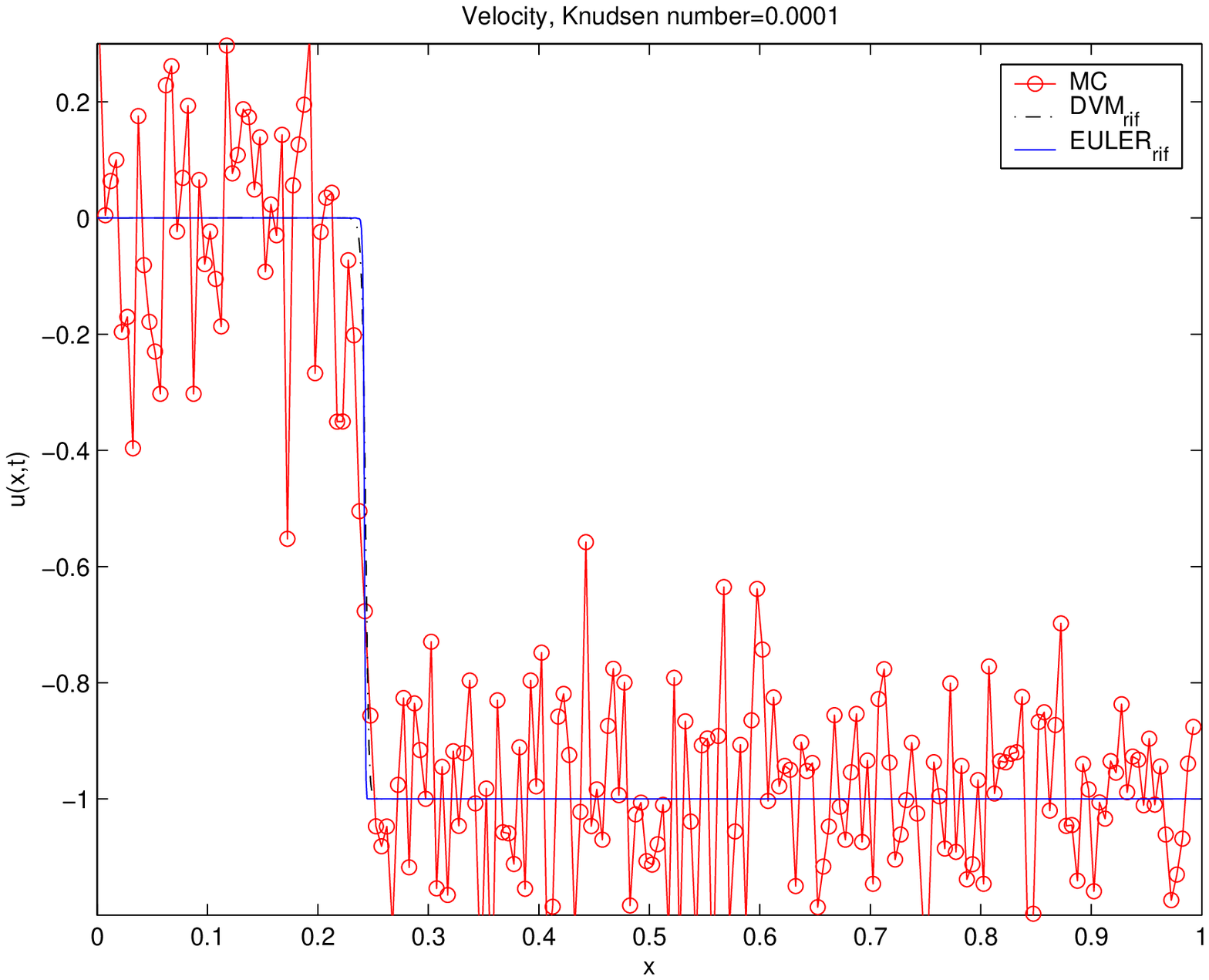}
\includegraphics[scale=0.39]{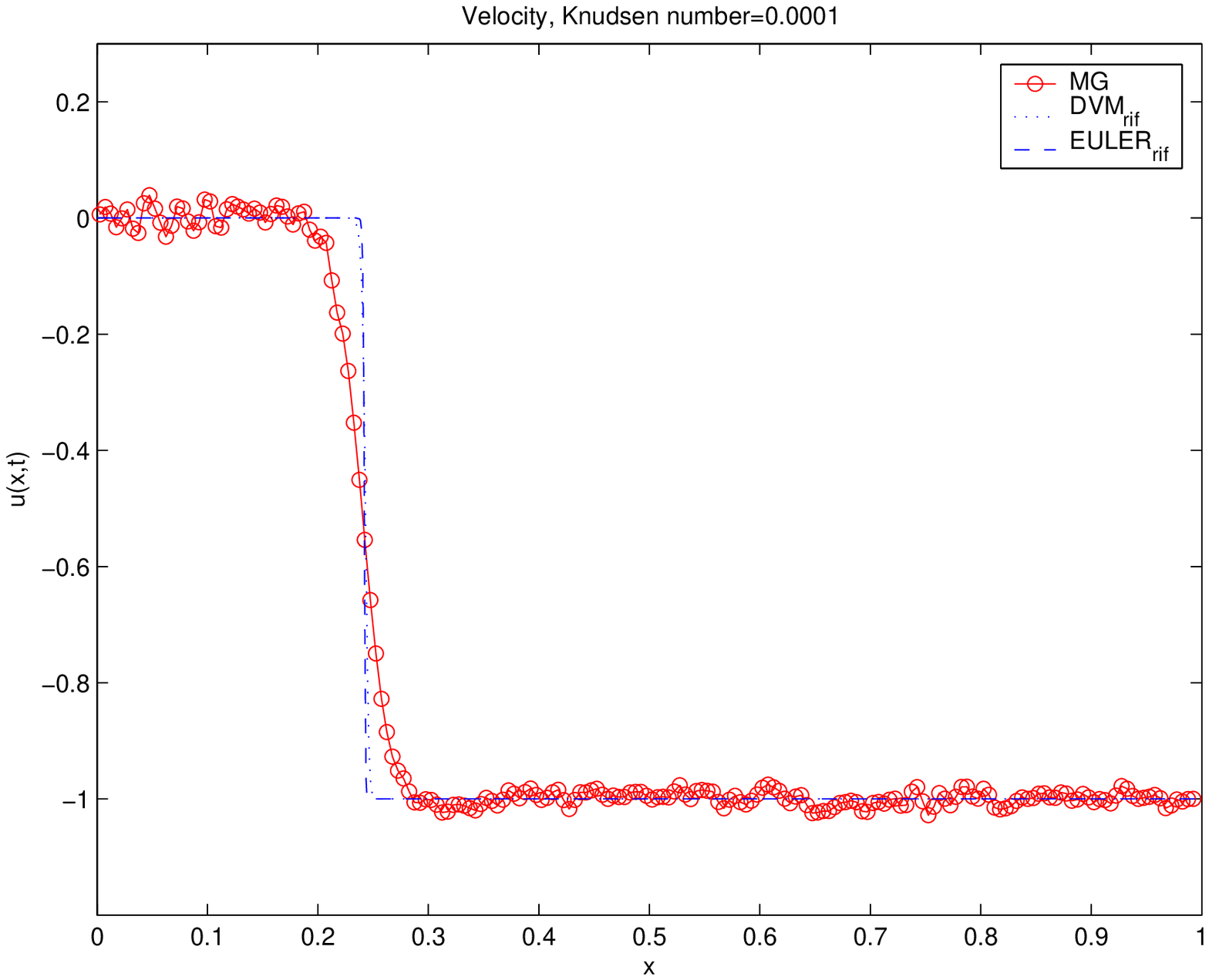}
\includegraphics[scale=0.39]{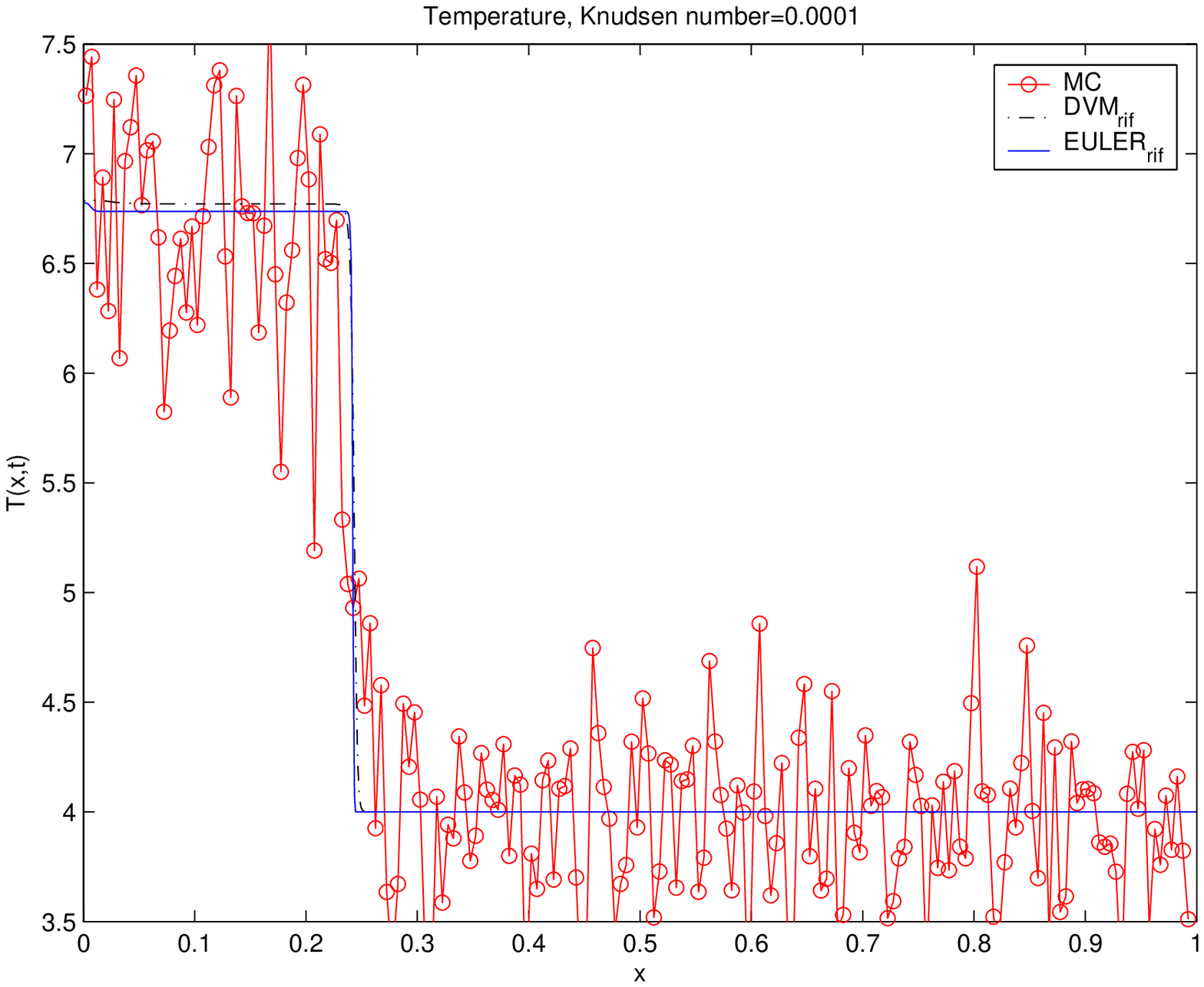}
\includegraphics[scale=0.39]{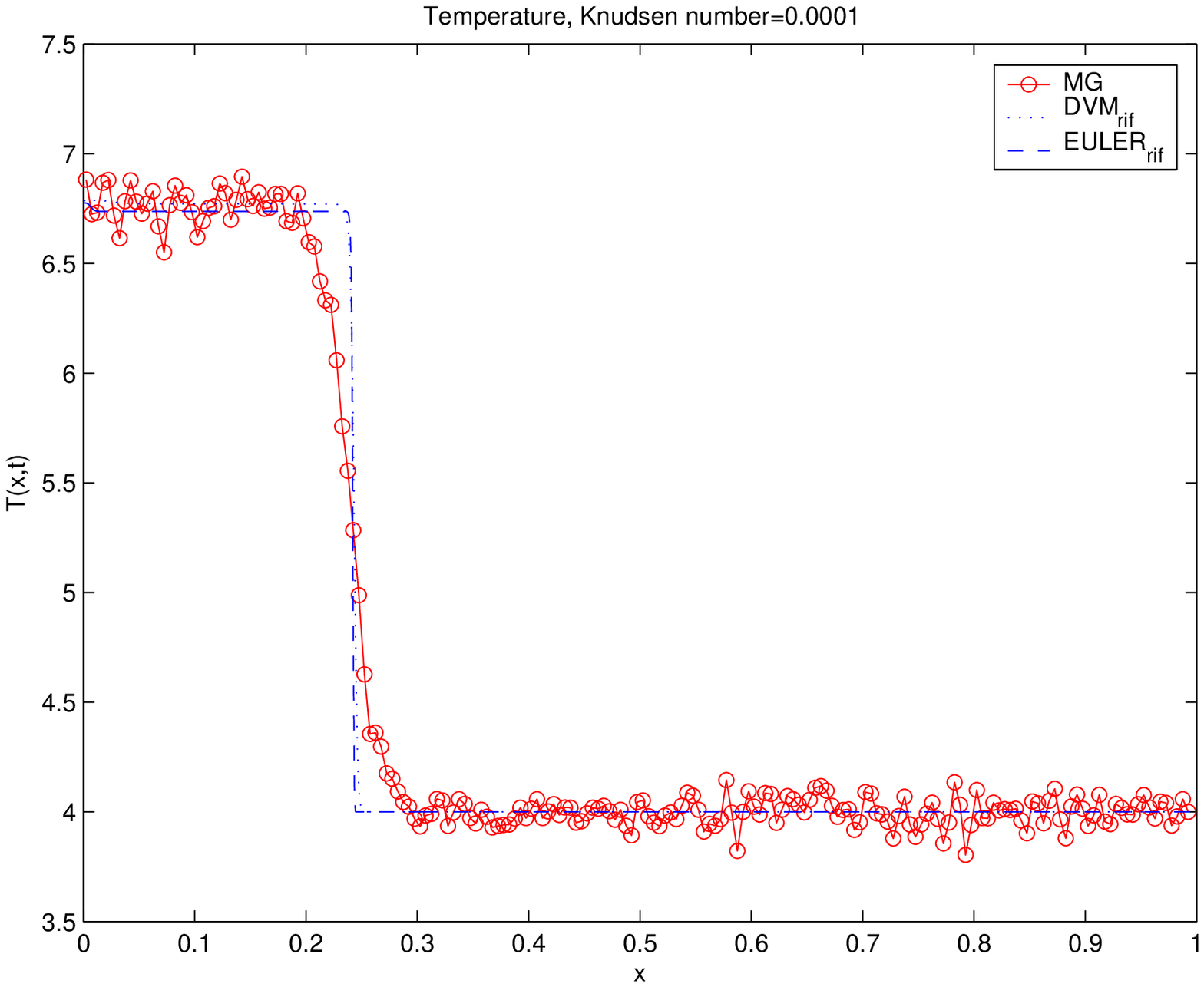}
\caption{Unsteady Shock: Solution at $t=0.065$ for the density
(top), velocity (middle) and temperature (bottom). MC method (left),
Moment Guided method MG (right). Knudsen number
$\varepsilon=10^{-4}$. Reference solution: dash dotted line. Euler
solution: continuous line. Monte Carlo or Moment Guided: circles
plus continuous line.} \label{US3}
\end{center}
\end{figure}

\begin{figure}
\begin{center}
\includegraphics[scale=0.39]{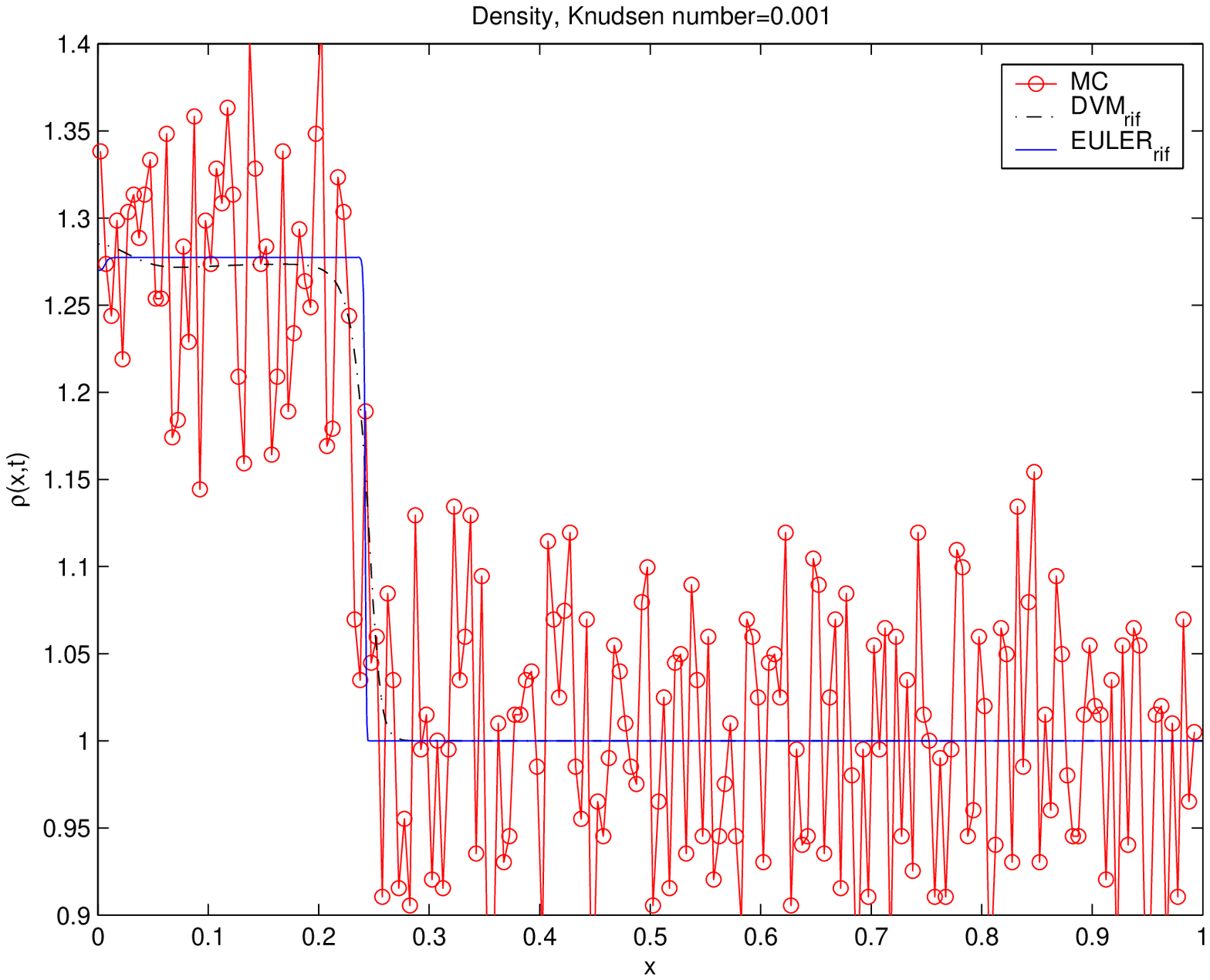}
\includegraphics[scale=0.39]{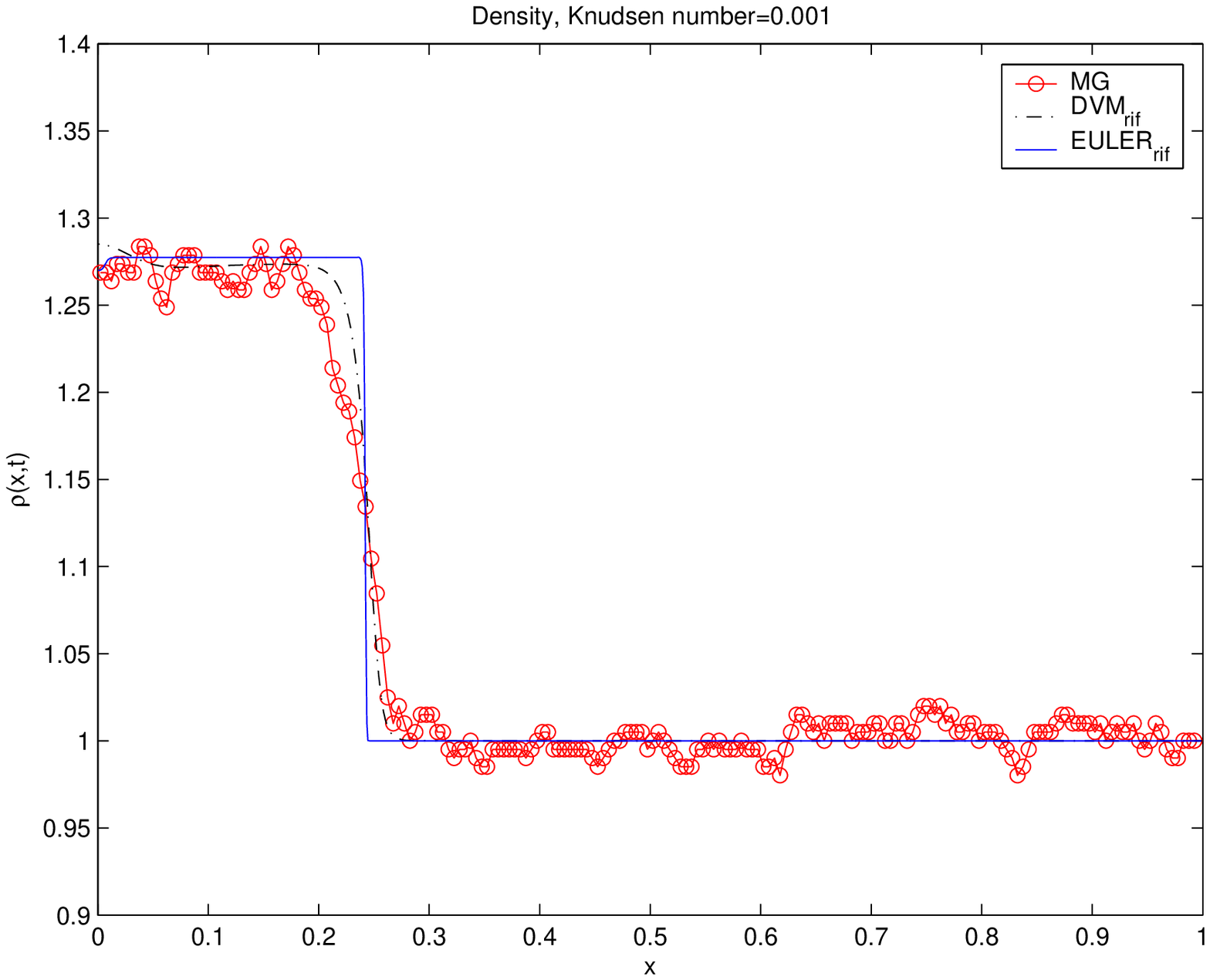}
\includegraphics[scale=0.39]{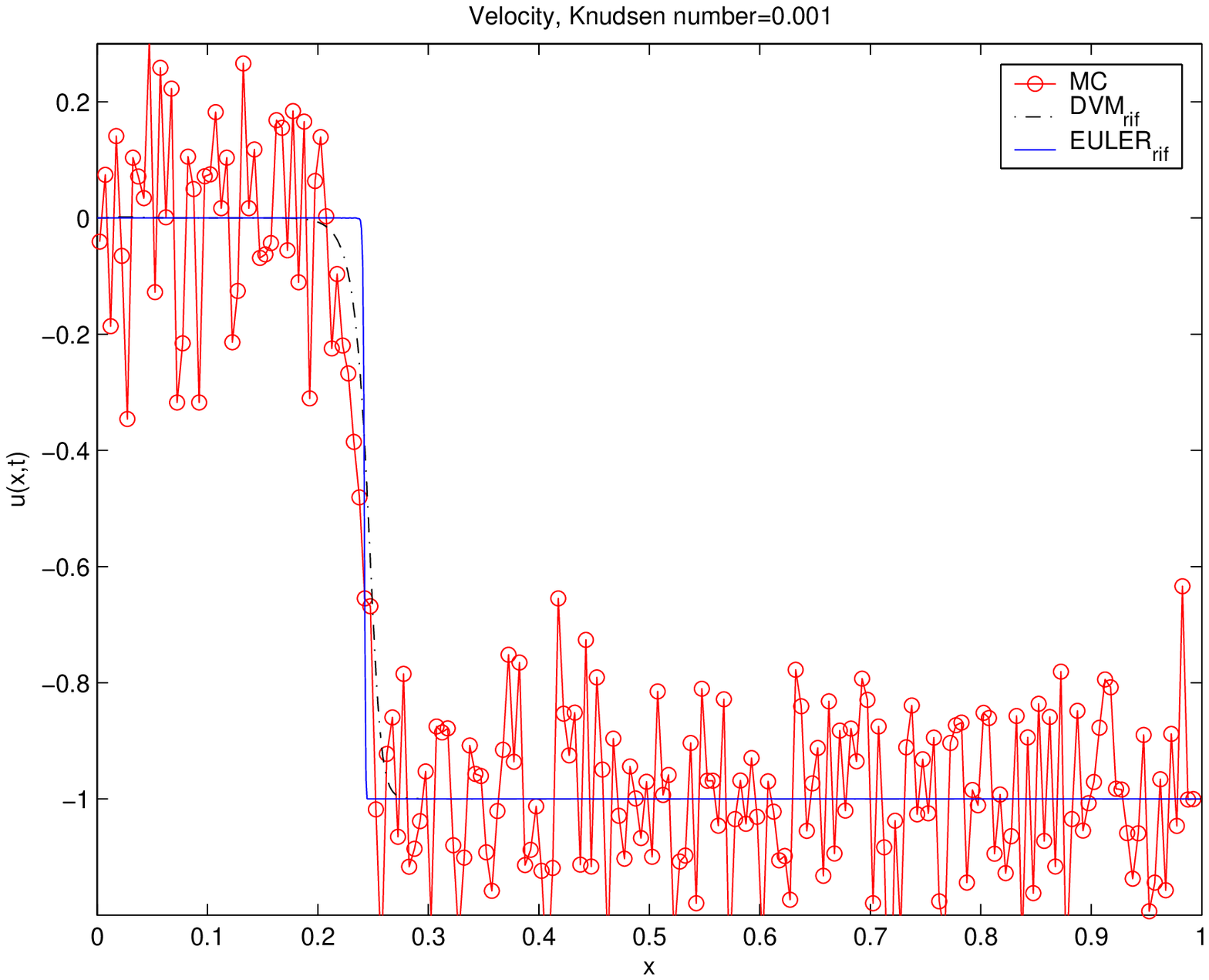}
\includegraphics[scale=0.39]{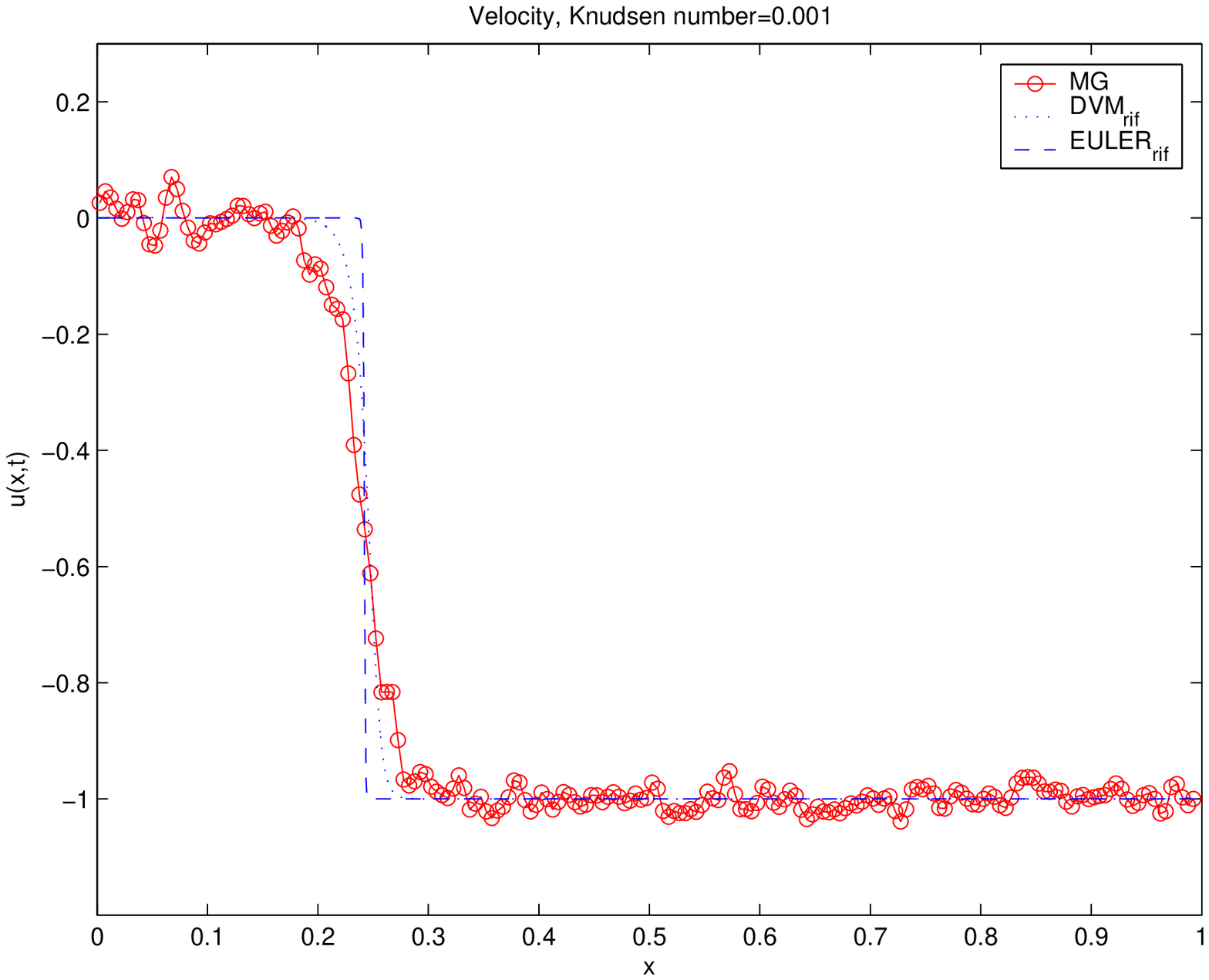}
\includegraphics[scale=0.39]{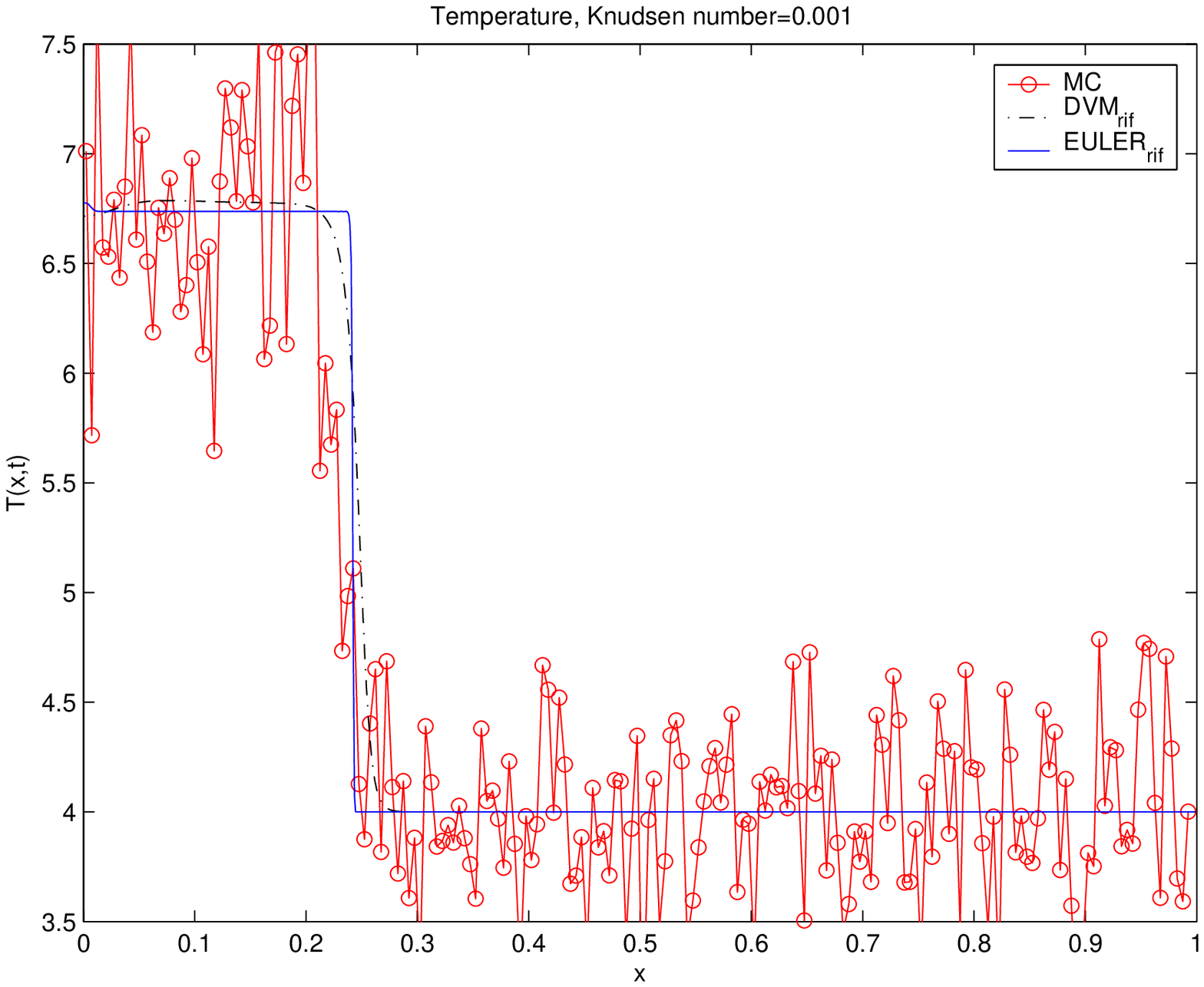}
\includegraphics[scale=0.39]{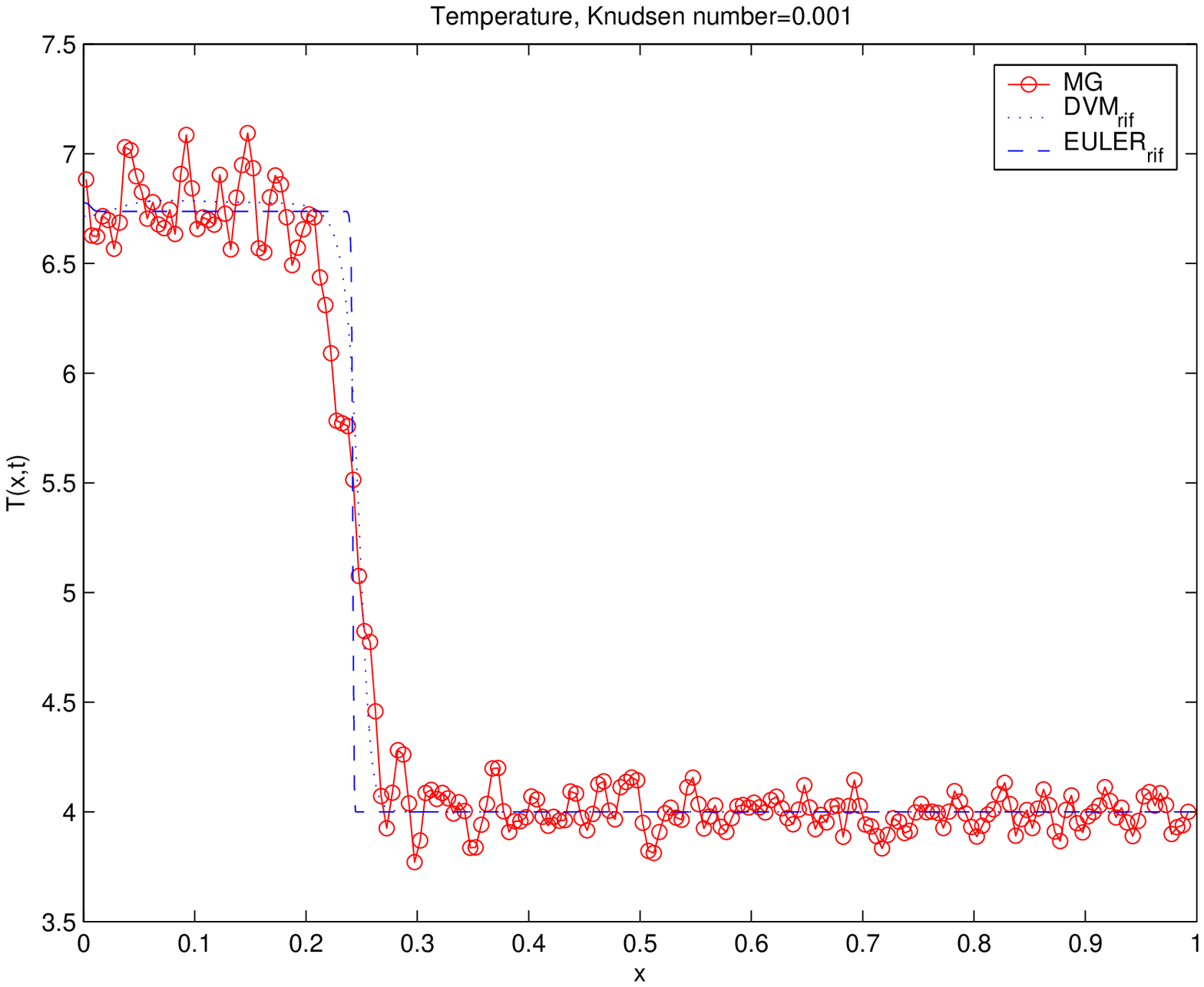}
\caption{Unsteady Shock: Solution at $t=0.065$ for the density
(top), velocity (middle) and temperature (bottom). MC method (left),
Moment Guided method MG (right). Knudsen number
$\varepsilon=10^{-3}$. Reference solution: dash dotted line. Euler
solution: continuous line. Monte Carlo or Moment Guided: circles
plus continuous line.} \label{US2}
\end{center}
\end{figure}

\begin{figure}
\begin{center}
\includegraphics[scale=0.39]{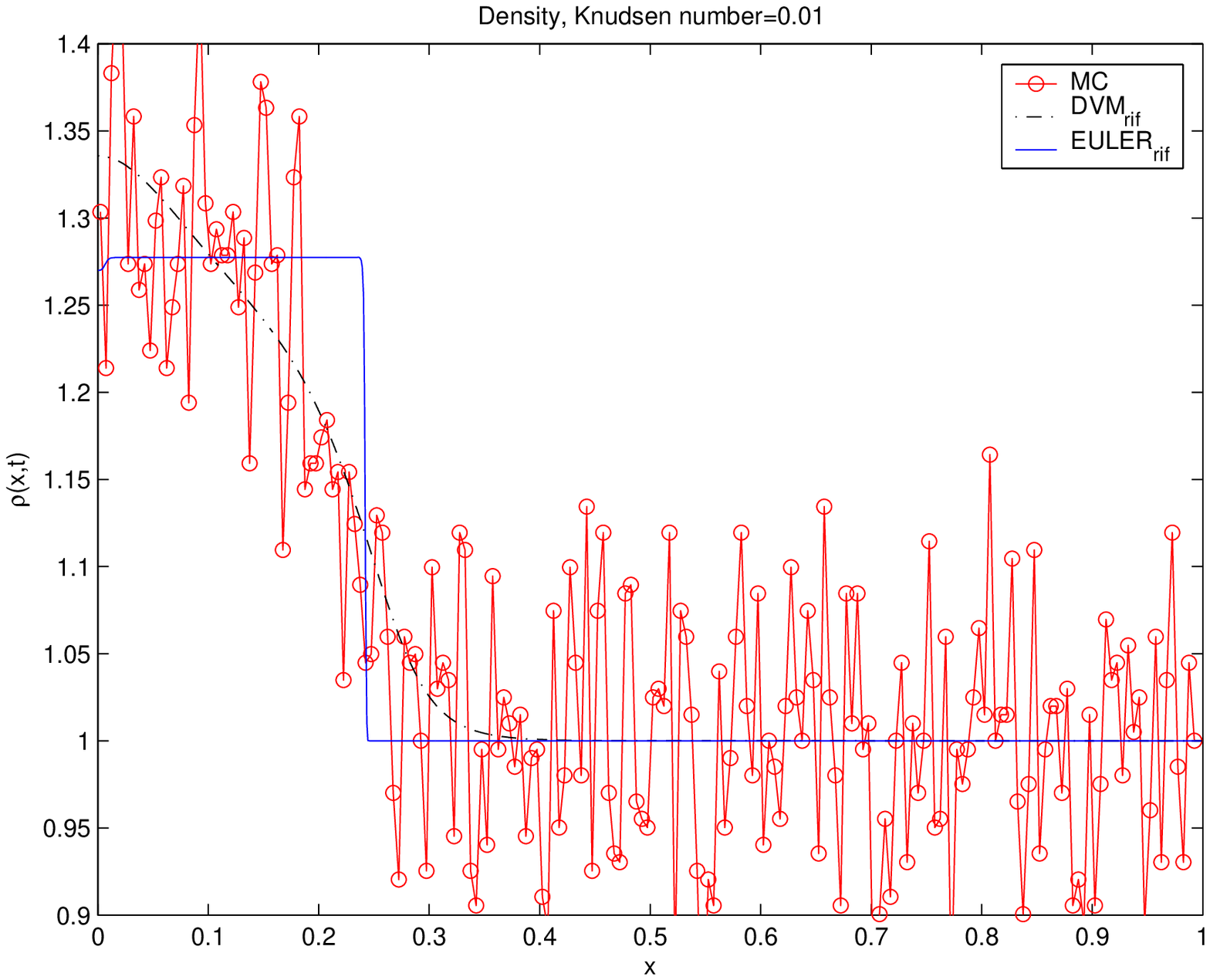}
\includegraphics[scale=0.39]{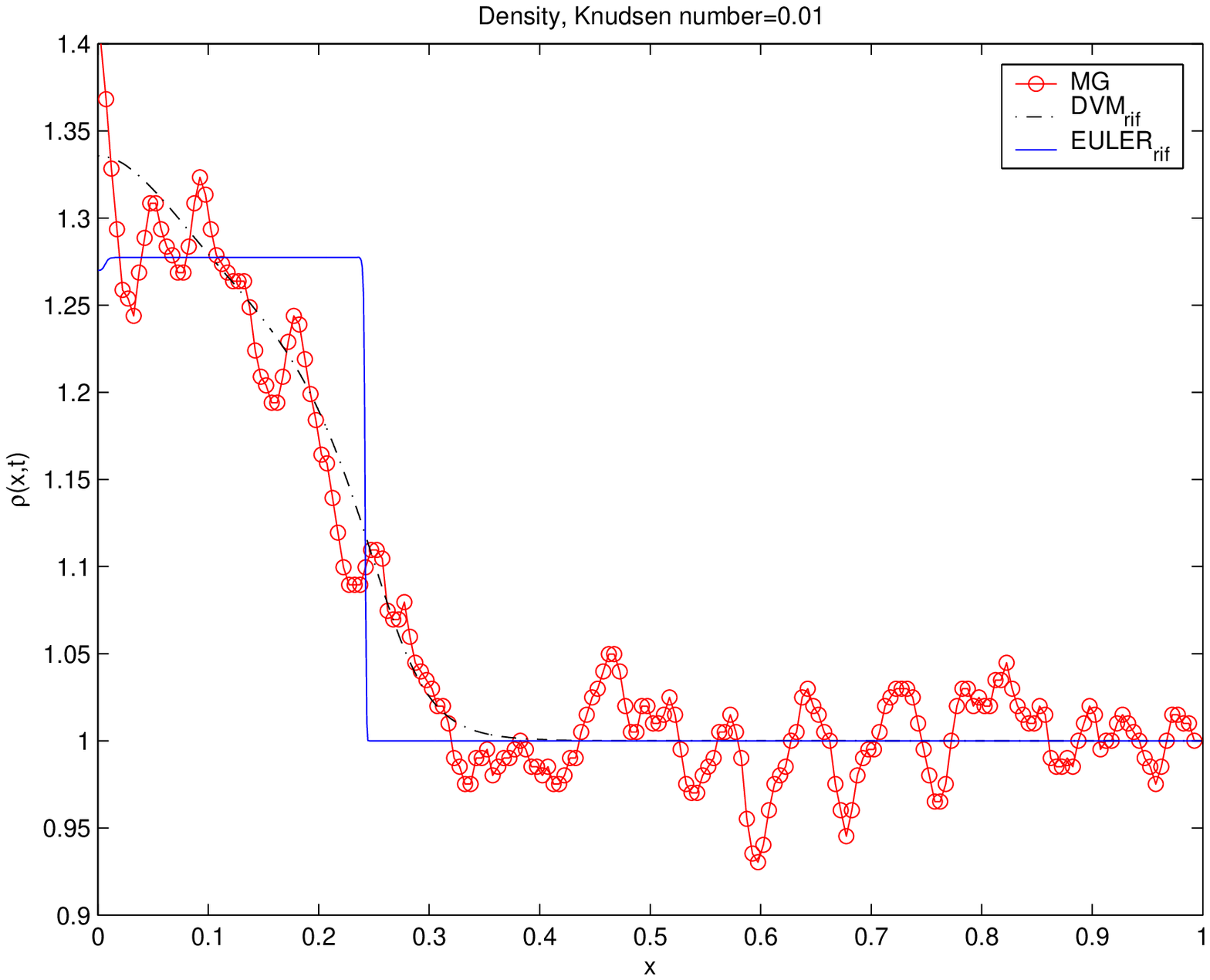}
\includegraphics[scale=0.39]{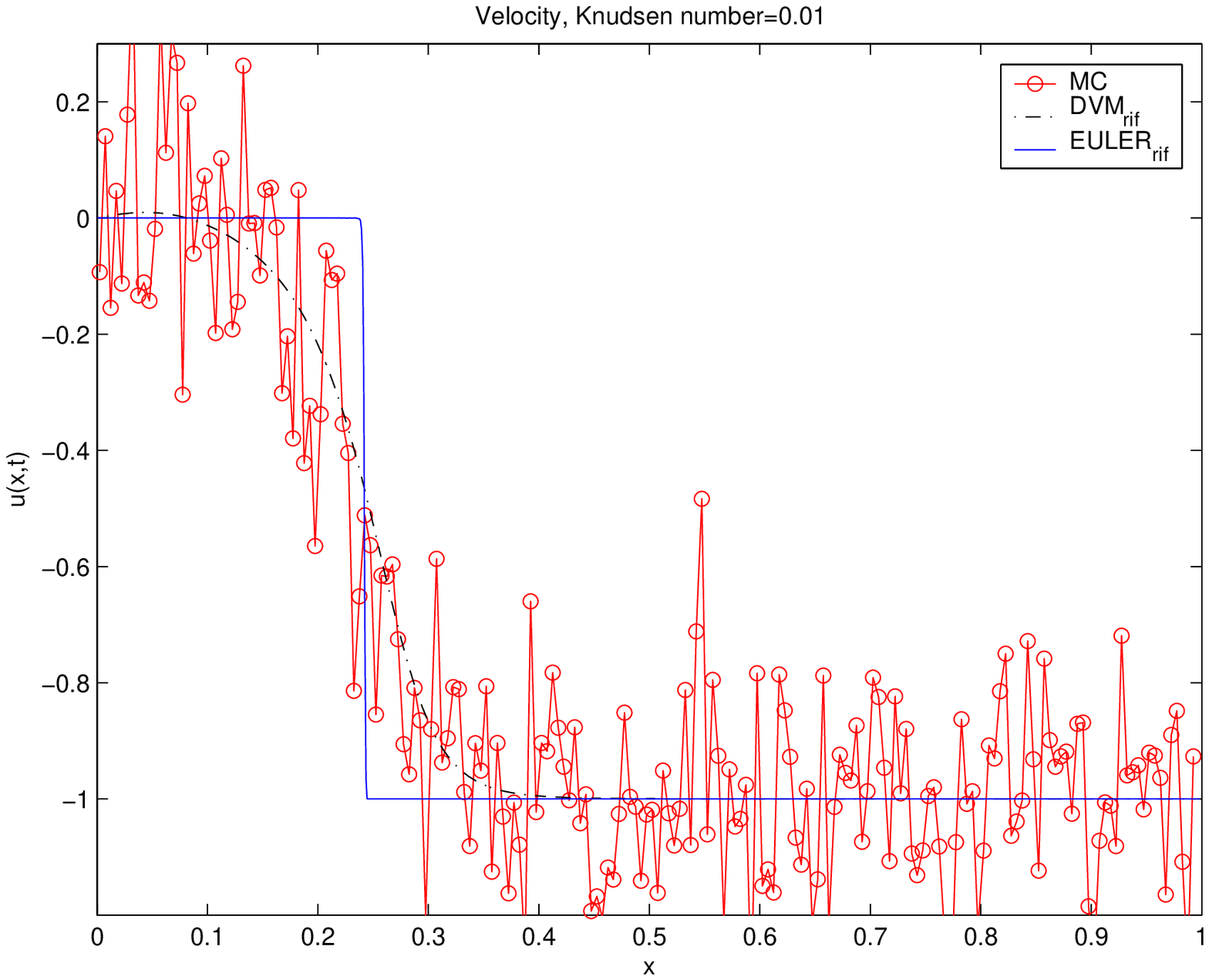}
\includegraphics[scale=0.39]{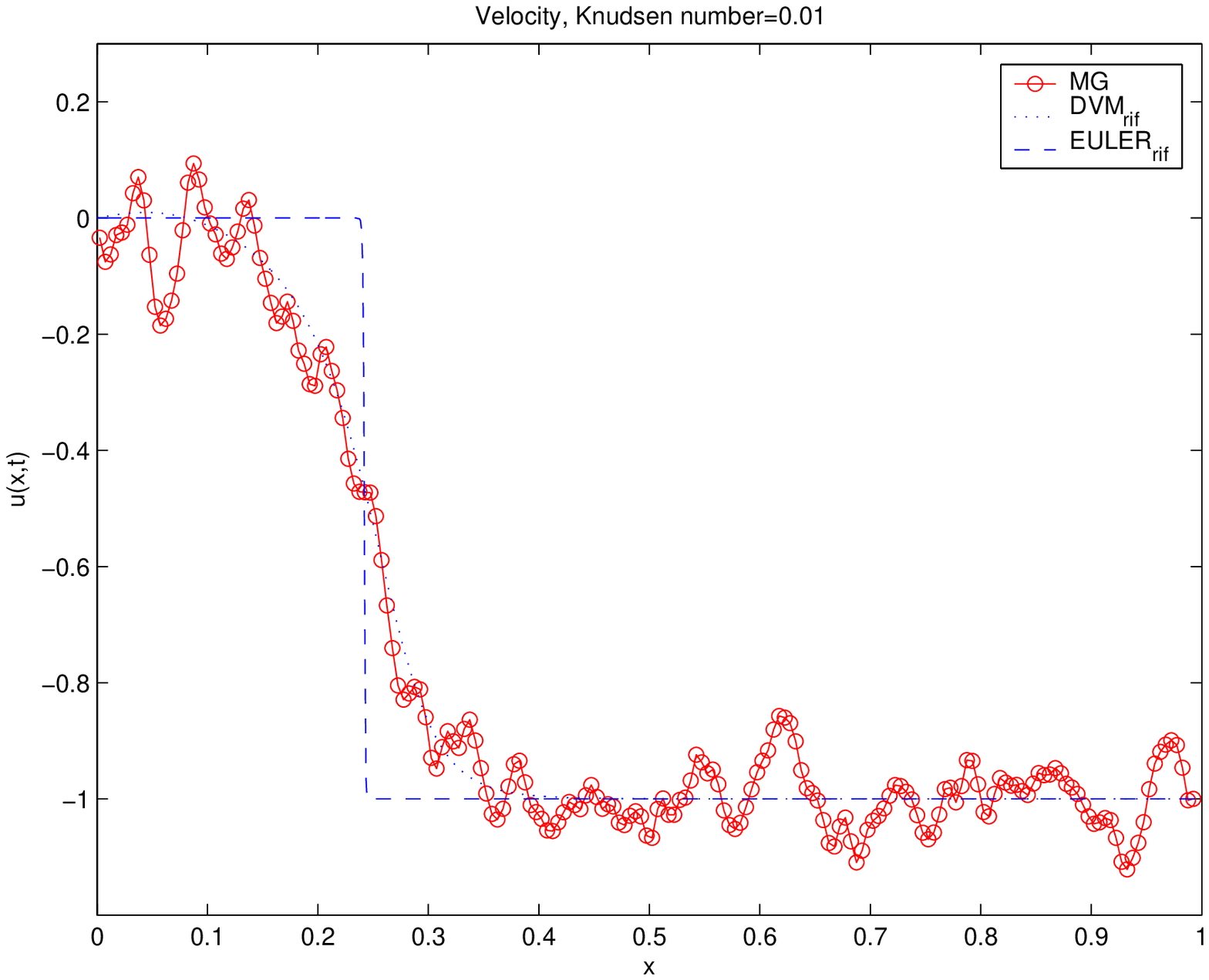}
\includegraphics[scale=0.39]{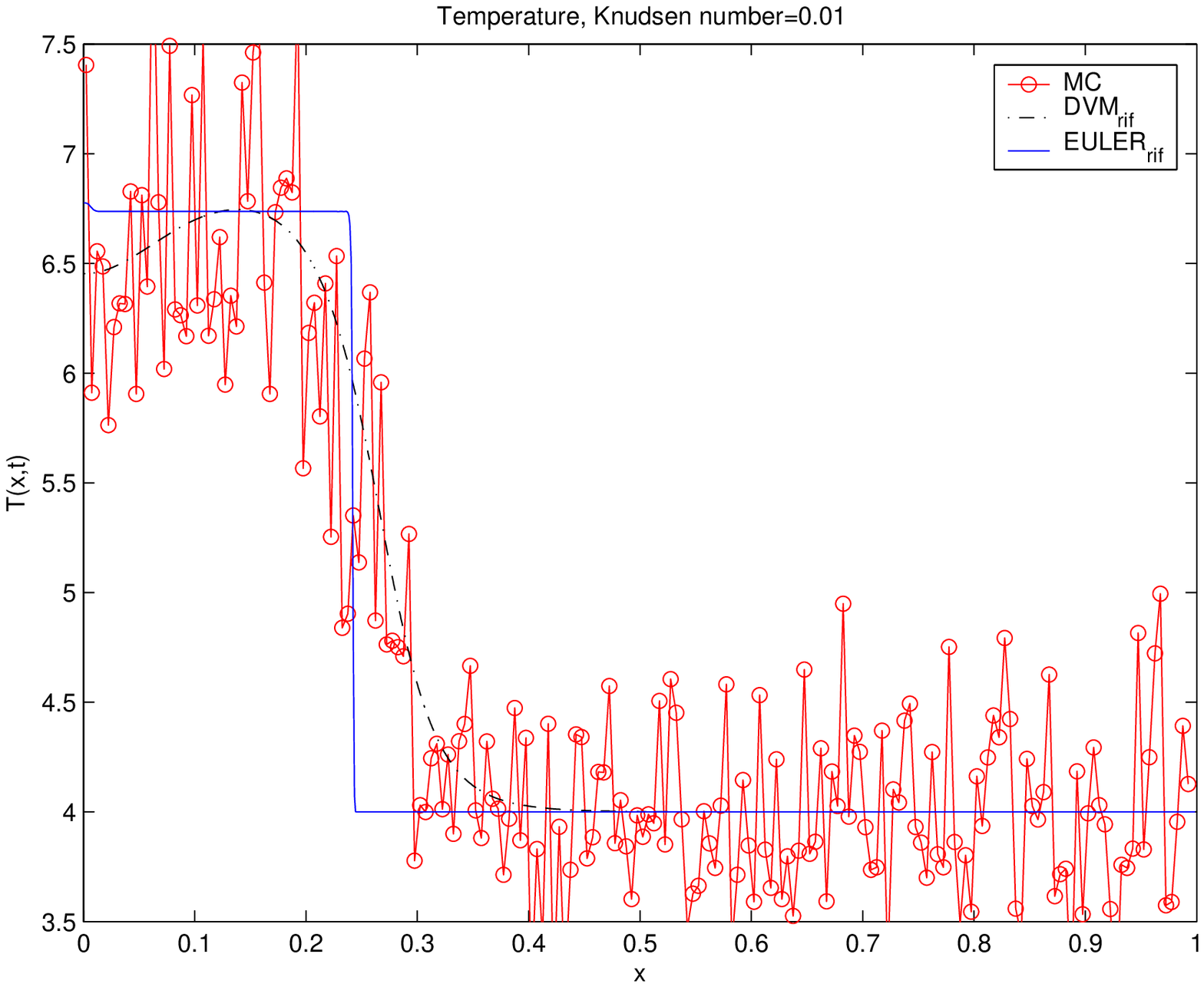}
\includegraphics[scale=0.39]{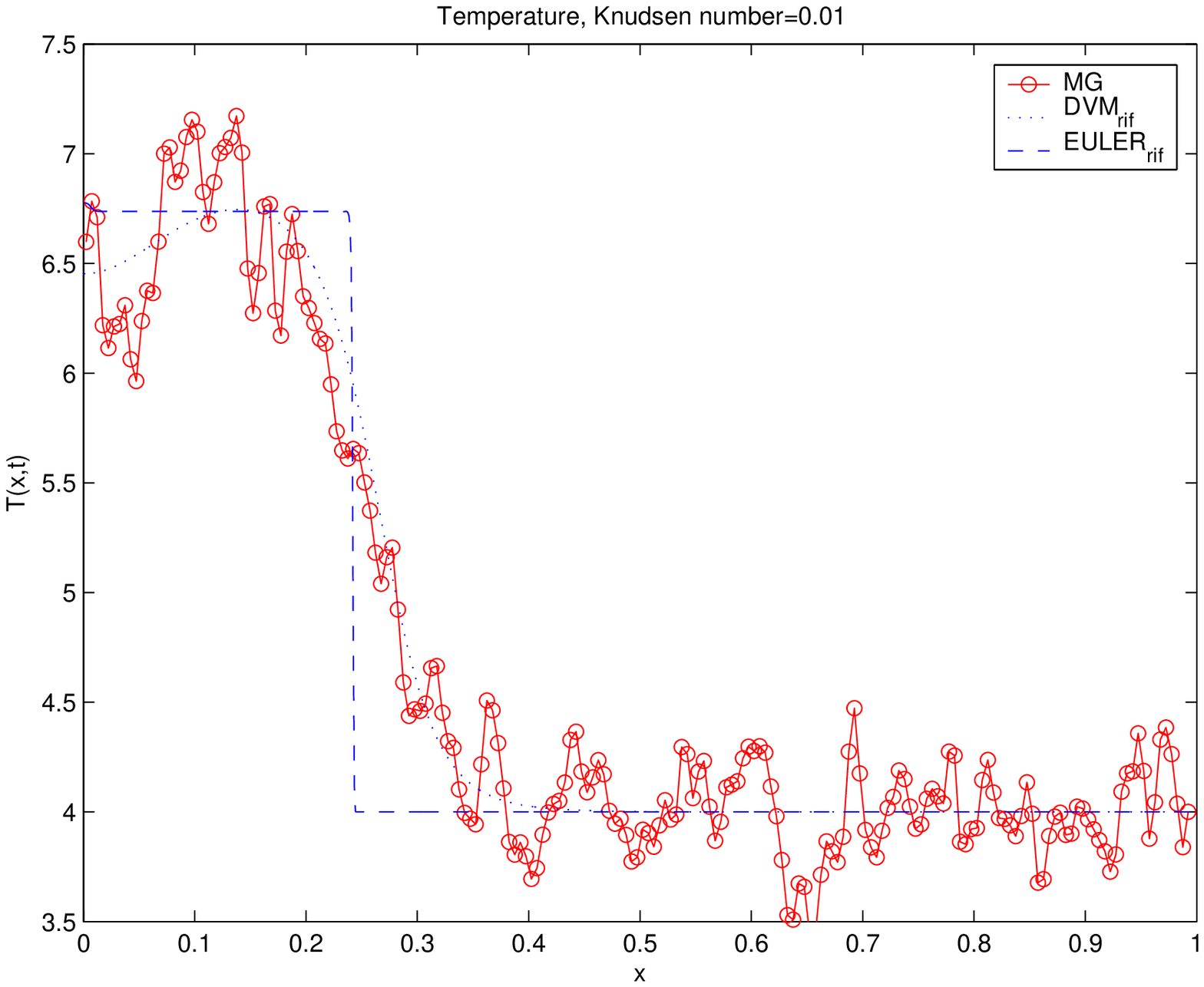}
\caption{Unsteady Shock: Solution at $t=0.065$ for the density
(top), velocity (middle) and temperature (bottom). MC method (left),
Moment Guided method MG (right). Knudsen number
$\varepsilon=10^{-2}$. Reference solution: dash dotted line. Euler
solution: continuous line. Monte Carlo or Moment Guided: circles
plus continuous line.} \label{US1}
\end{center}
\end{figure}

\begin{figure}
\begin{center}
\includegraphics[scale=0.39]{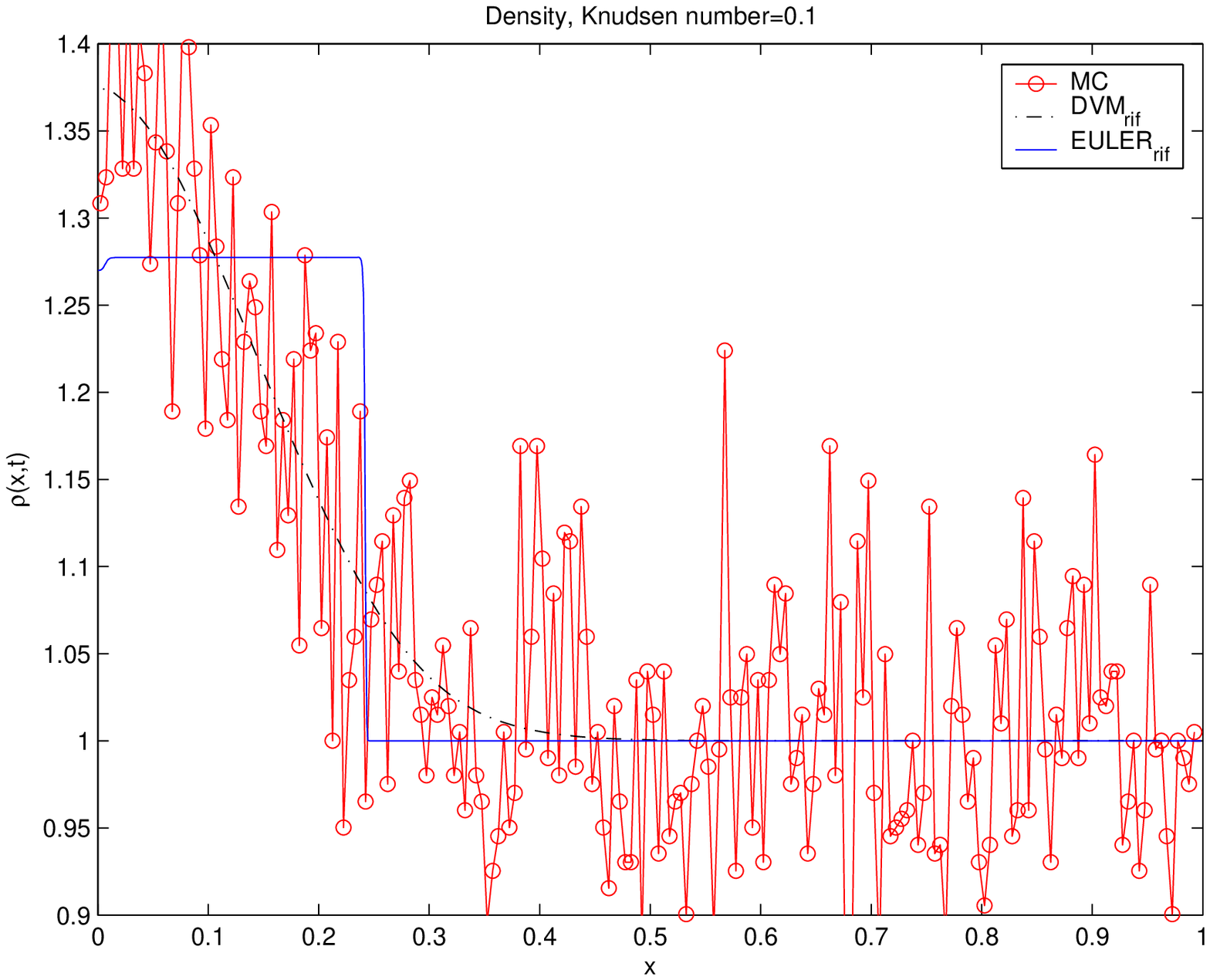}
\includegraphics[scale=0.39]{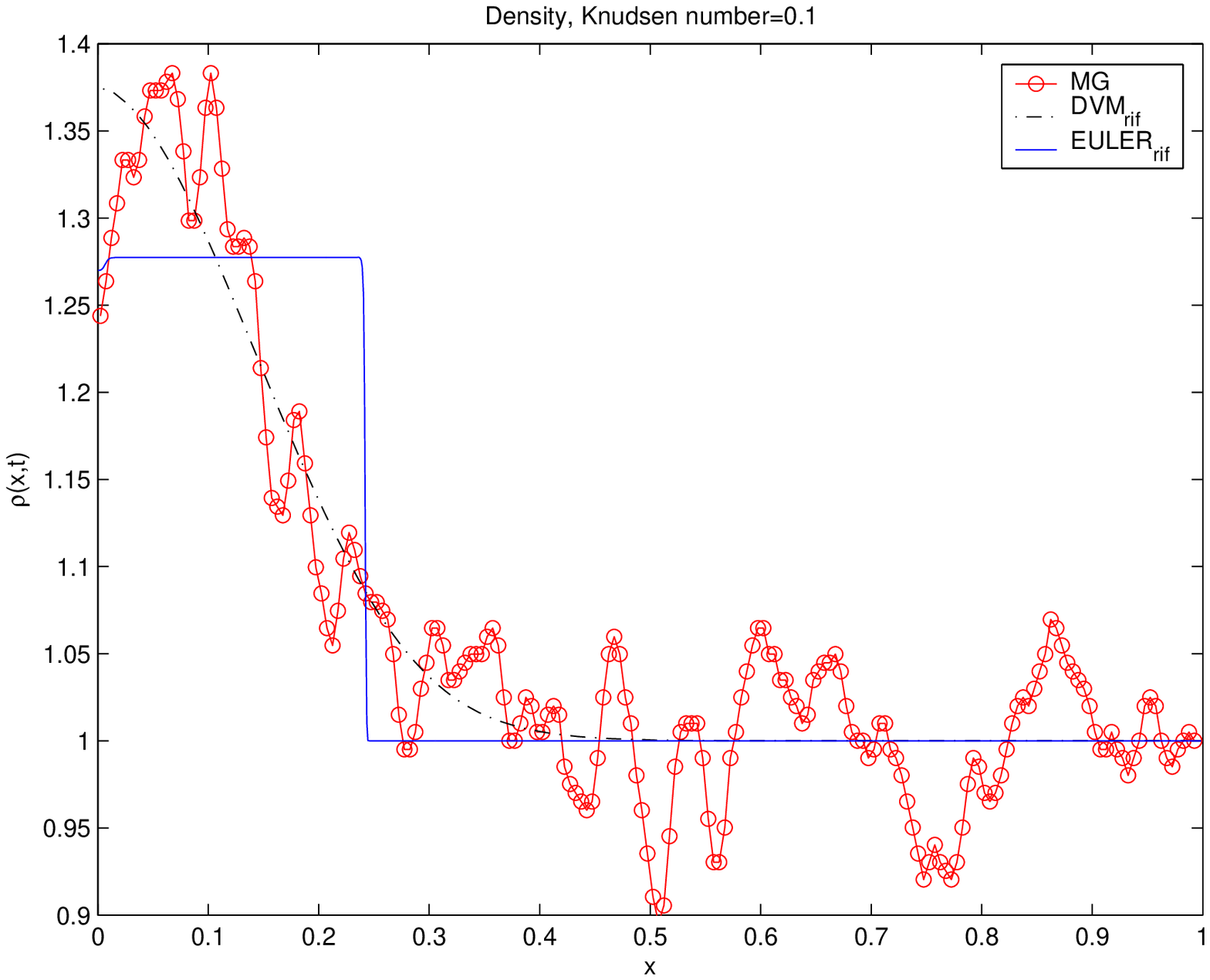}
\includegraphics[scale=0.39]{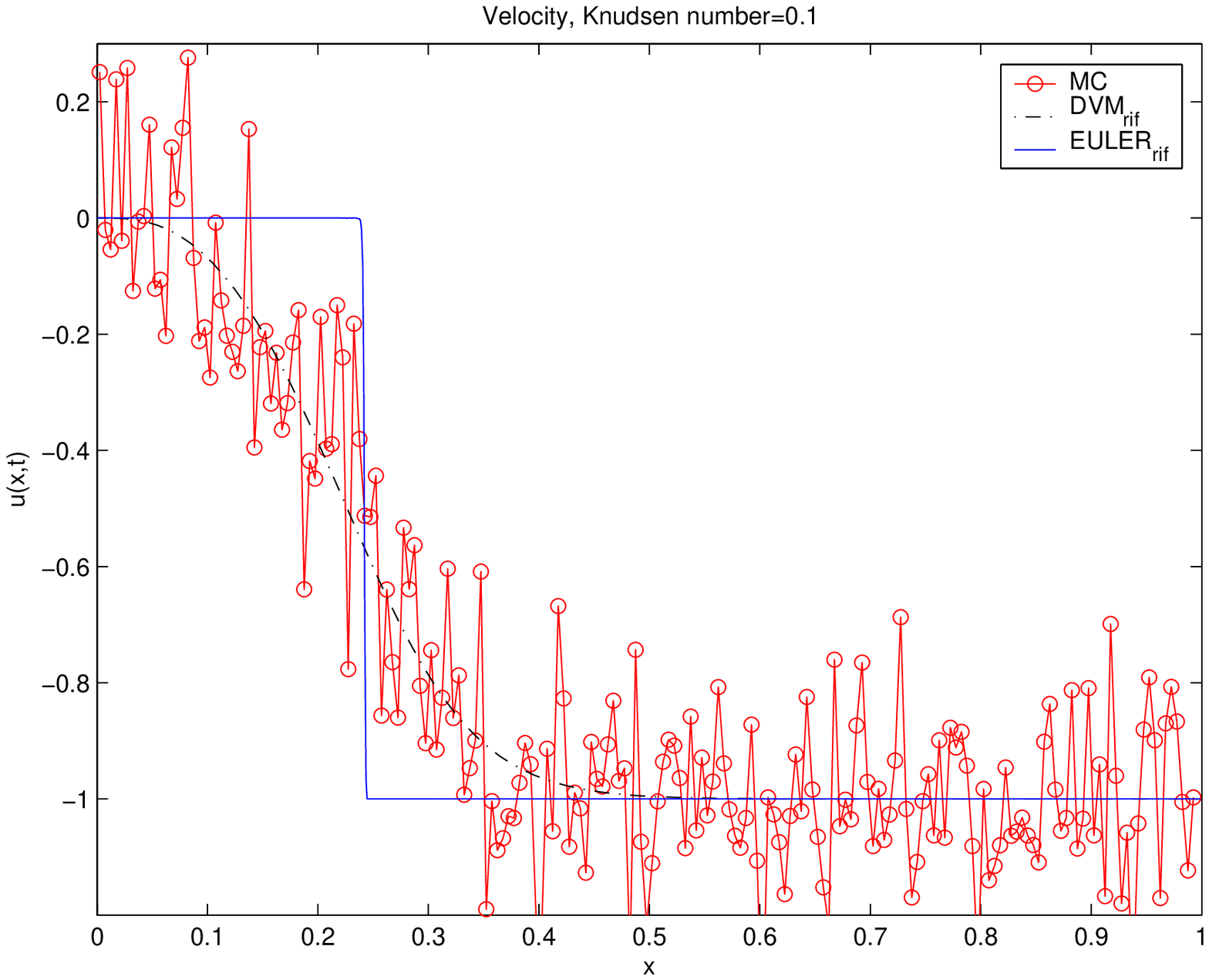}
\includegraphics[scale=0.39]{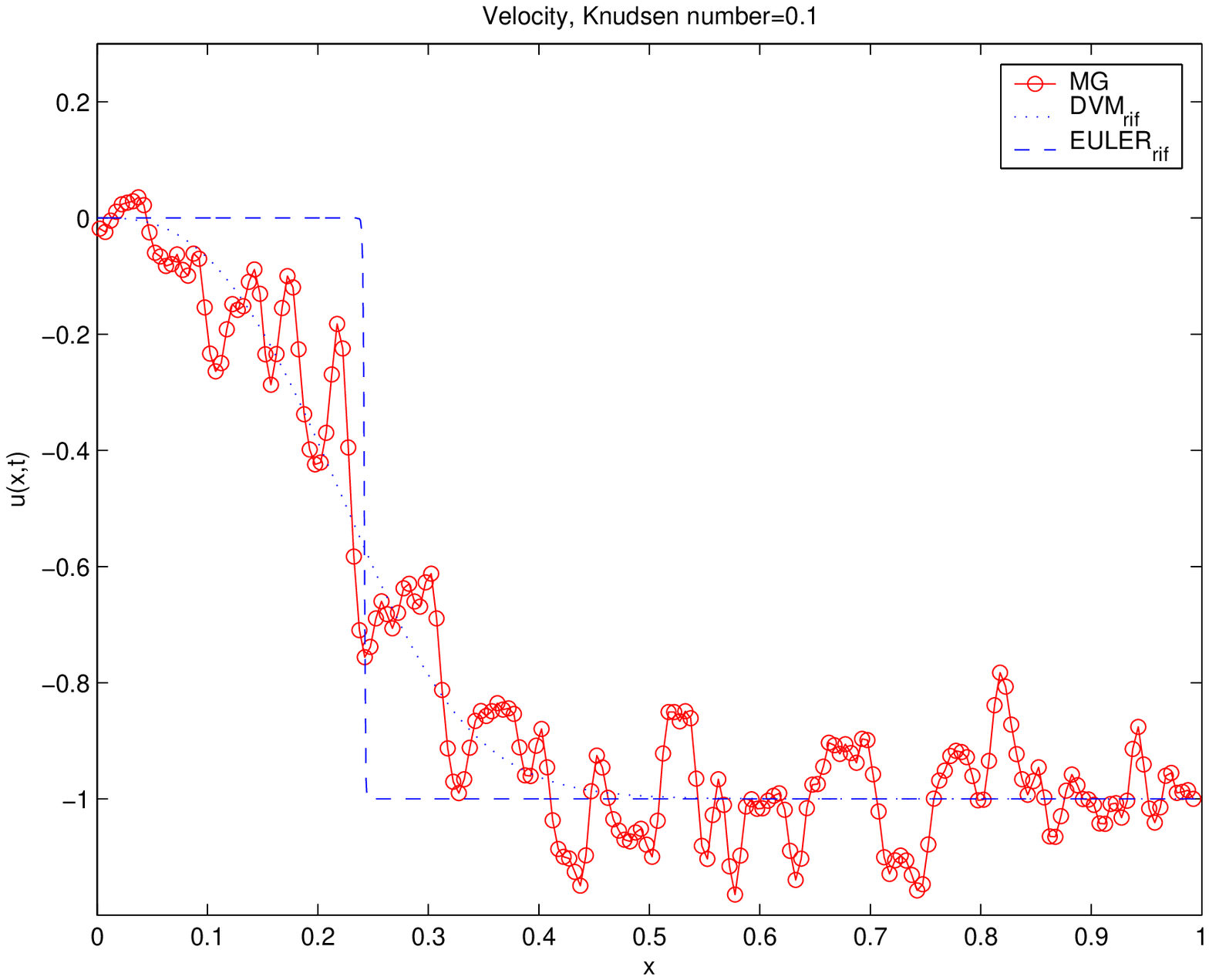}
\includegraphics[scale=0.39]{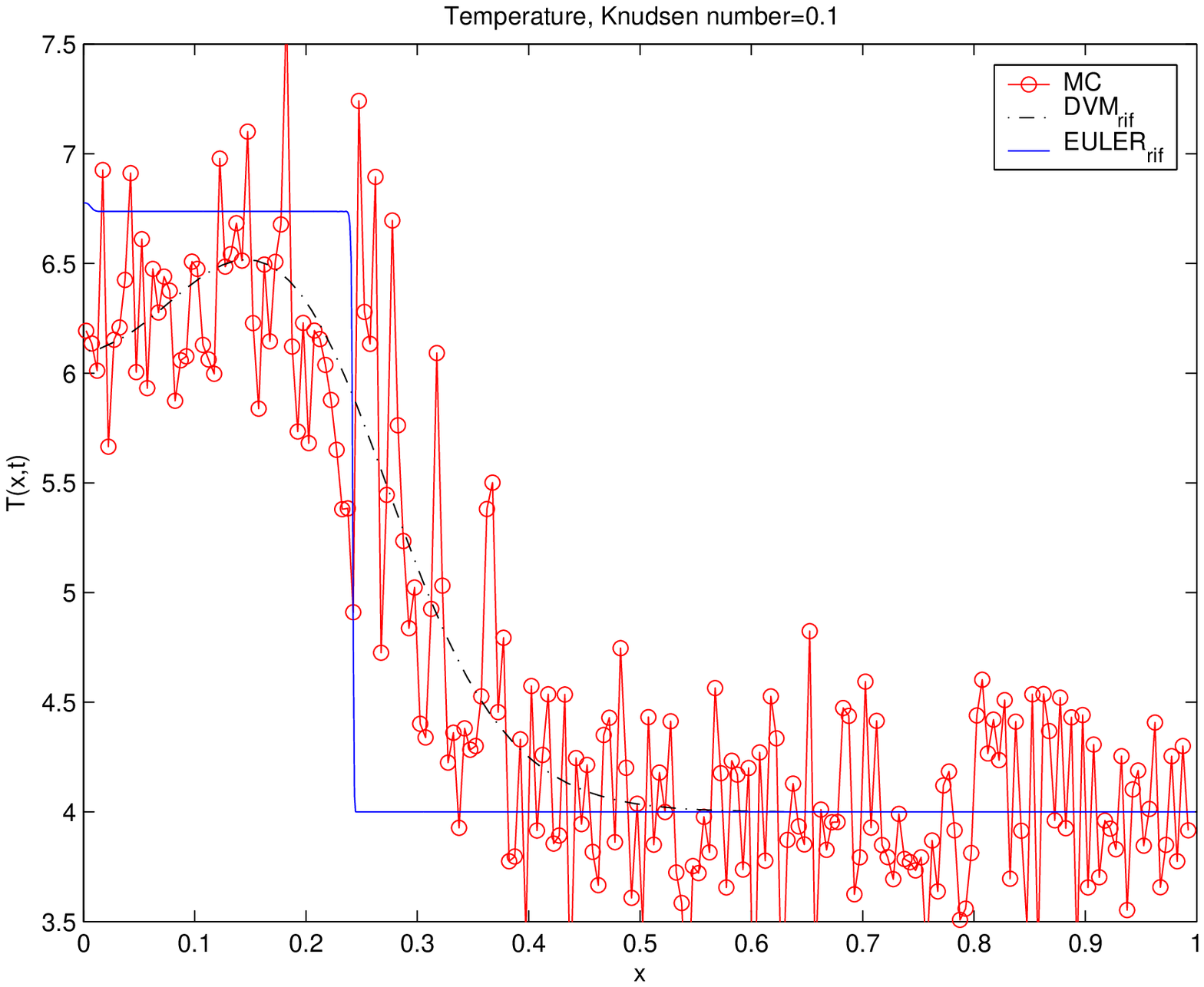}
\includegraphics[scale=0.39]{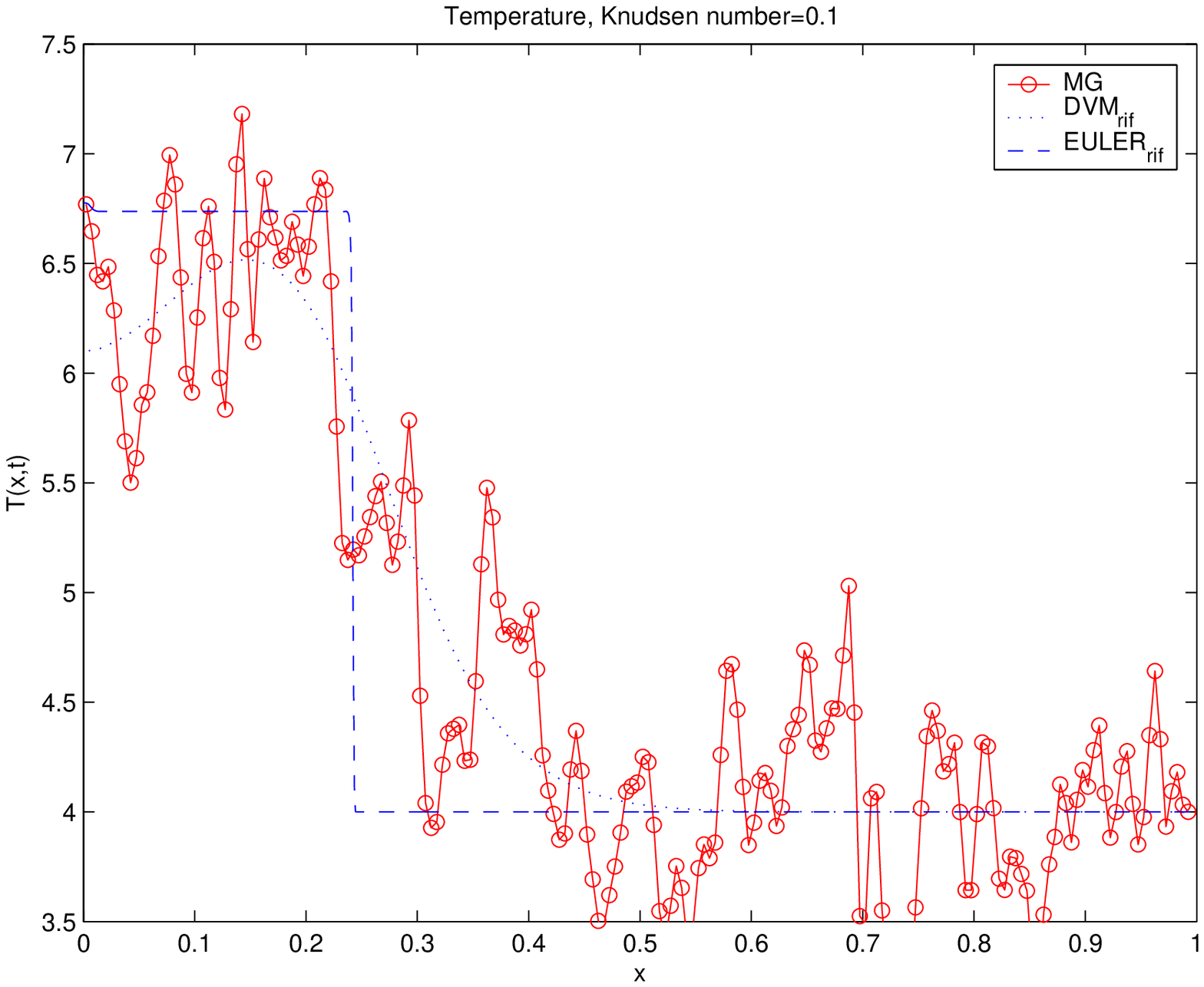}
\caption{Unsteady Shock: Solution at $t=0.065$ for the density
(top), velocity (middle) and temperature (bottom). MC method (left),
Moment Guided method MG (right). Knudsen number
$\varepsilon=10^{-1}$. Reference solution: dash dotted line. Euler
solution: continuous line. Monte Carlo or Moment Guided: circles
plus continuous line.} \label{US0}
\end{center}
\end{figure}

Figures \ref{US3} to \ref{US0} consider the same initial data for
the density, mean velocity and temperature with different initial
Knudsen number values, ranging from $\varepsilon=10^{-4}$ to
$\varepsilon=10^{-1}$. $100$ particles per cell are used and
solutions are averaged over two different realizations. Each Figure
depicts the density, mean velocity and temperature from top to
bottom, with the pure Monte Carlo solver (on the left) and Moment
Guided method (on the right). In addition, we represent solutions of
the compressible Euler equations and as reference solution we used a
discrete velocity model for the BGK equation \cite{Mieussens}. These
Figures show a large reduction of fluctuations especially for small
Knudsen numbers.

\subsection*{Sod shock tube}

Finally we look at the classical Sod Shock Tube test. For this test
case, we consider the possibility of using second order fluid
solvers. We observe that this choice has the effect of increasing
fluctuations far from thermodynamical equilibrium. This is natural
since we miss the smoothing effect of a first order scheme. However
solutions obtained with first order schemes can be unsatisfactory in
some situations because of their large numerical diffusion
especially close to thermodynamical equilibrium. The solution which
is adopted here consists in switching from the first order to the
second order scheme according to the ratio of the thermodynamical
flux with respect to the non equilibrium flux. In practice, in each
cell, the scheme automatically uses a second order MUSCL scheme when
the kinetic term is small and a first order scheme otherwise.

\begin{figure}
\begin{center}
\includegraphics[scale=0.39]{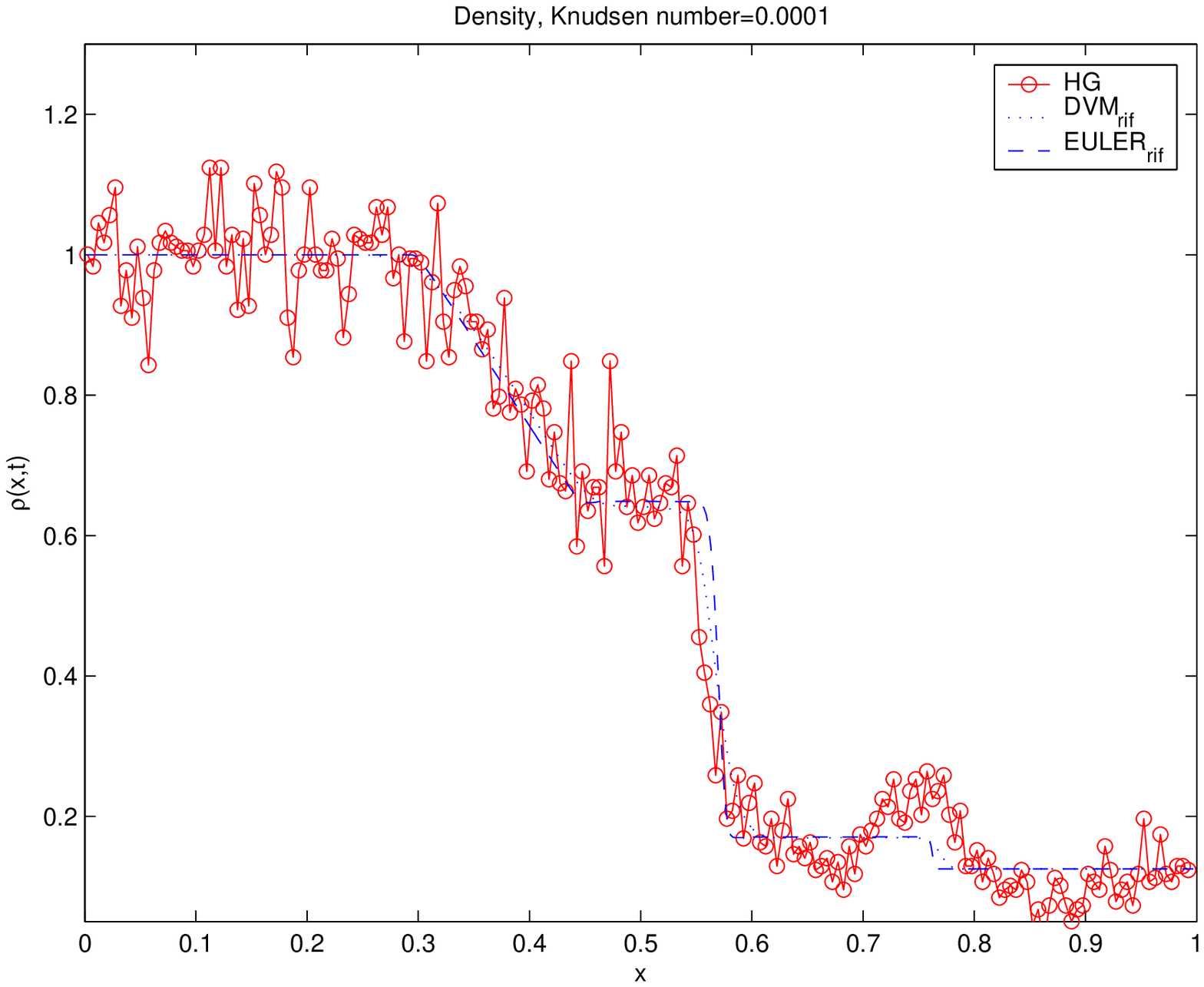}
\includegraphics[scale=0.39]{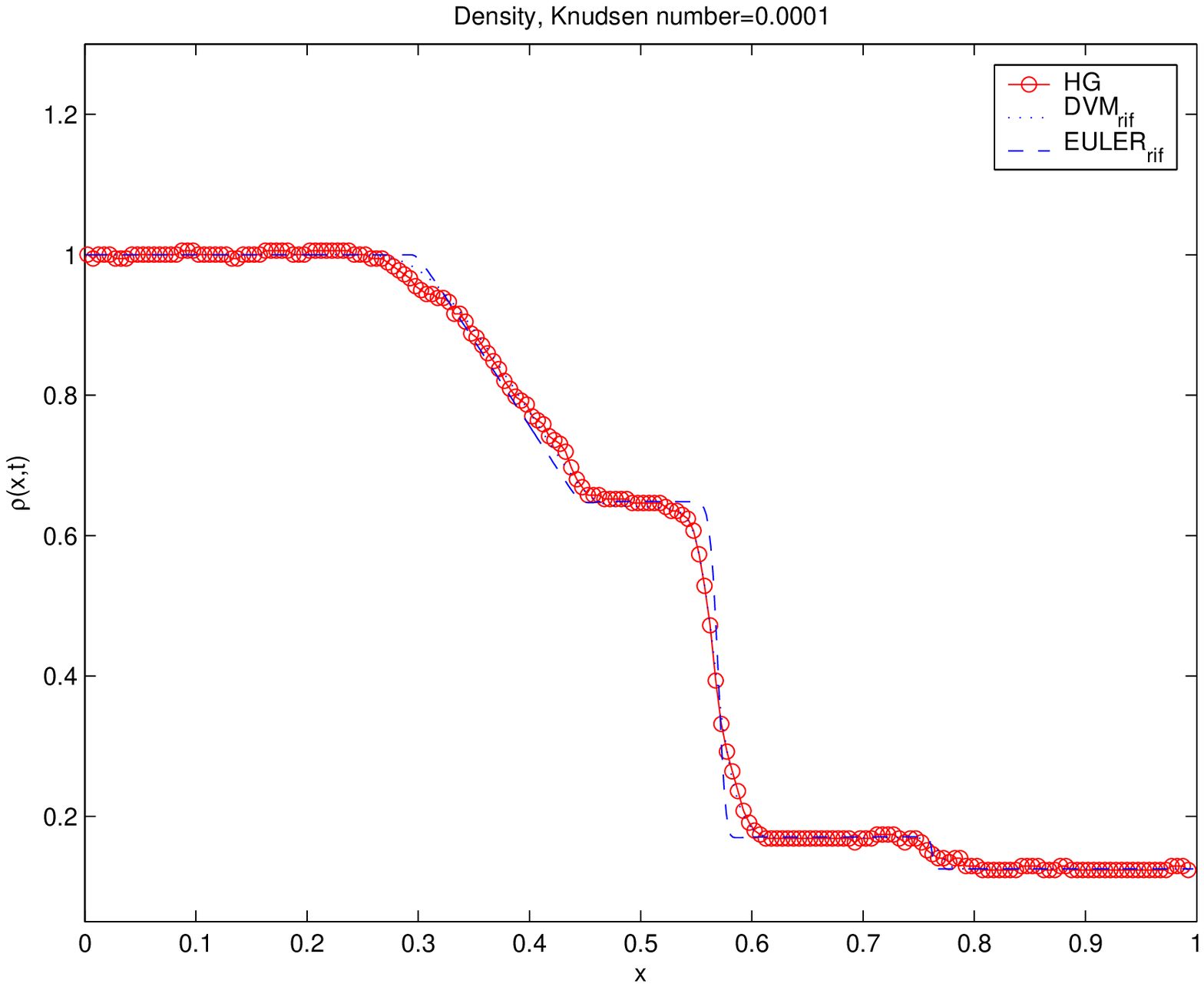}
\includegraphics[scale=0.39]{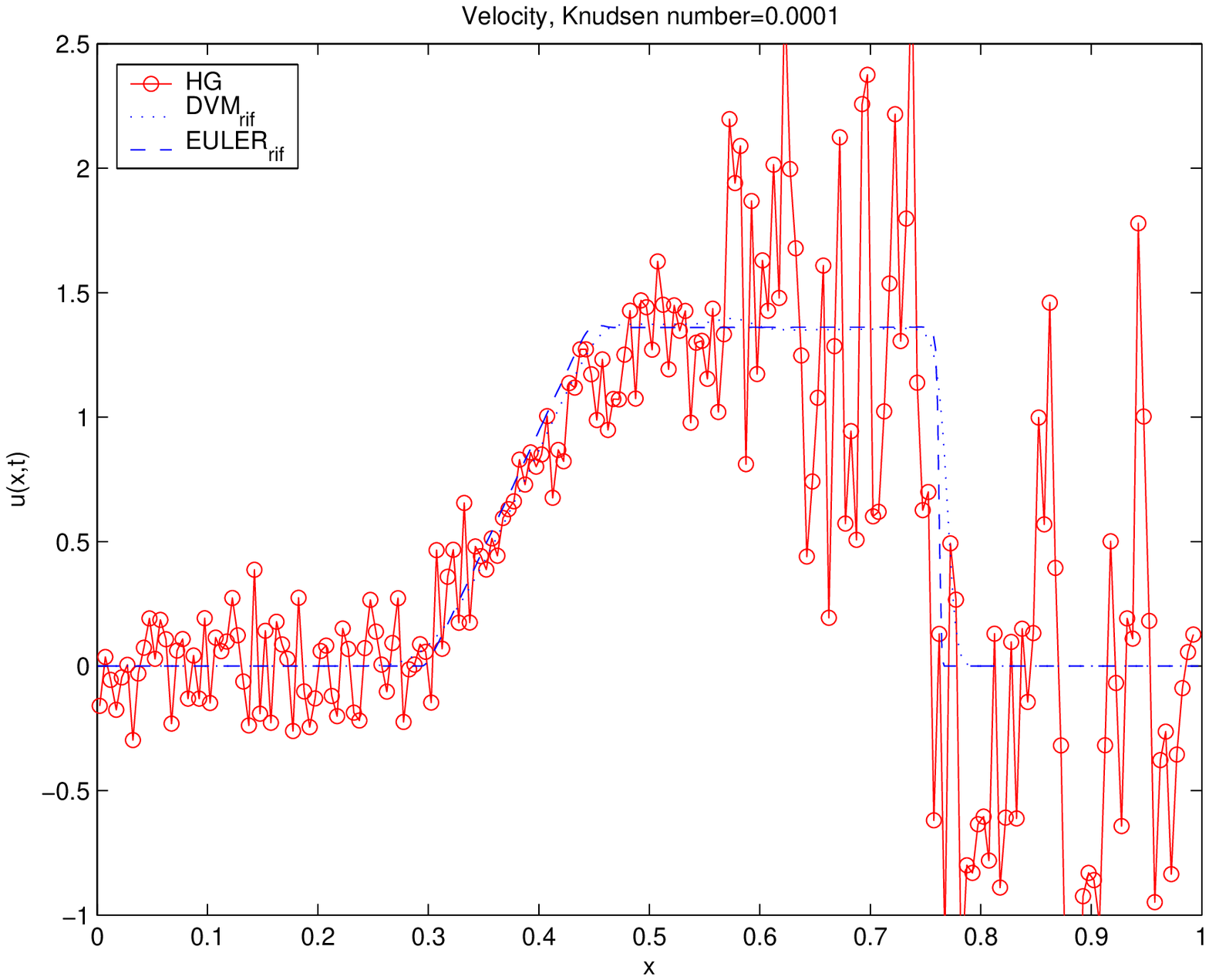}
\includegraphics[scale=0.39]{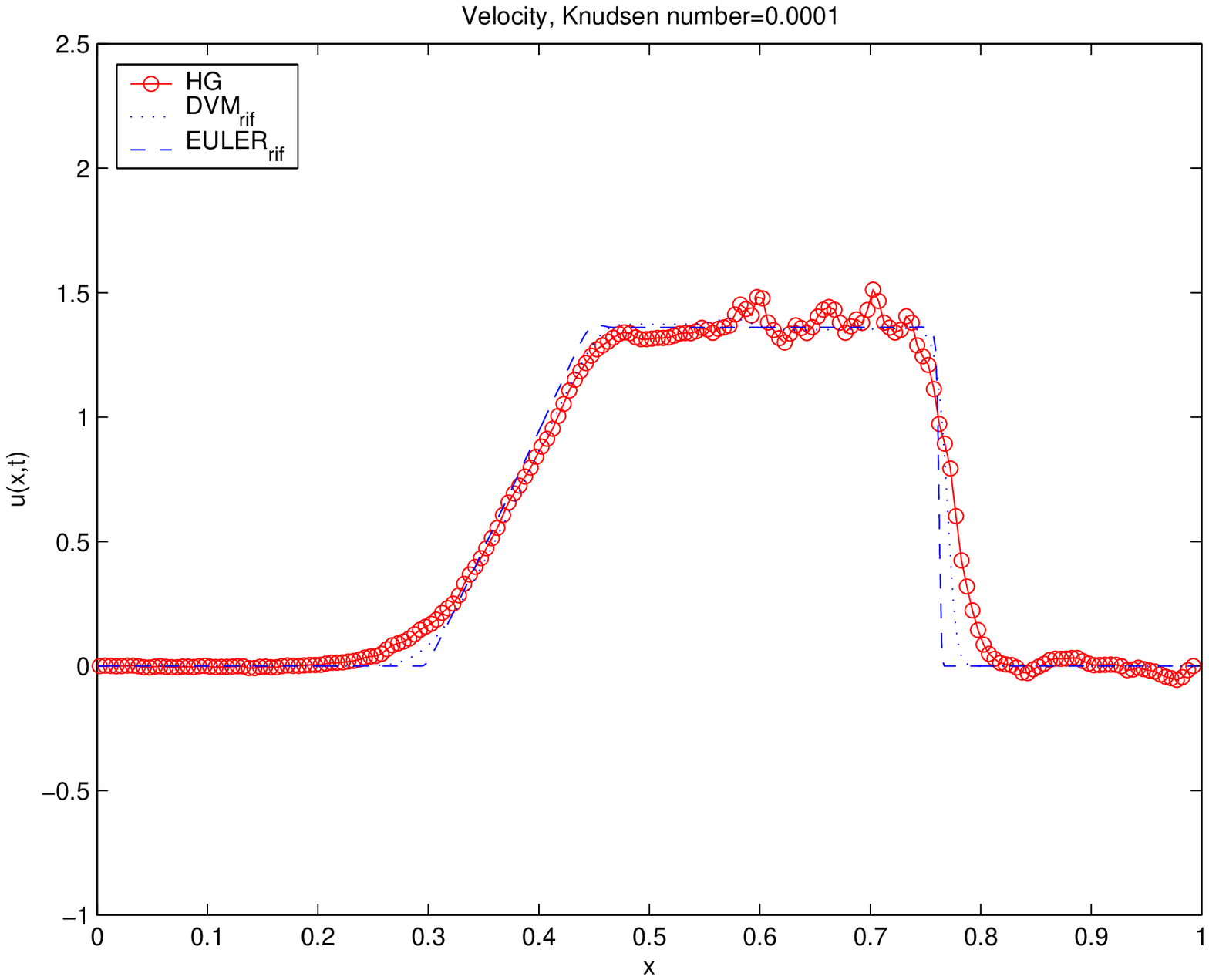}
\includegraphics[scale=0.39]{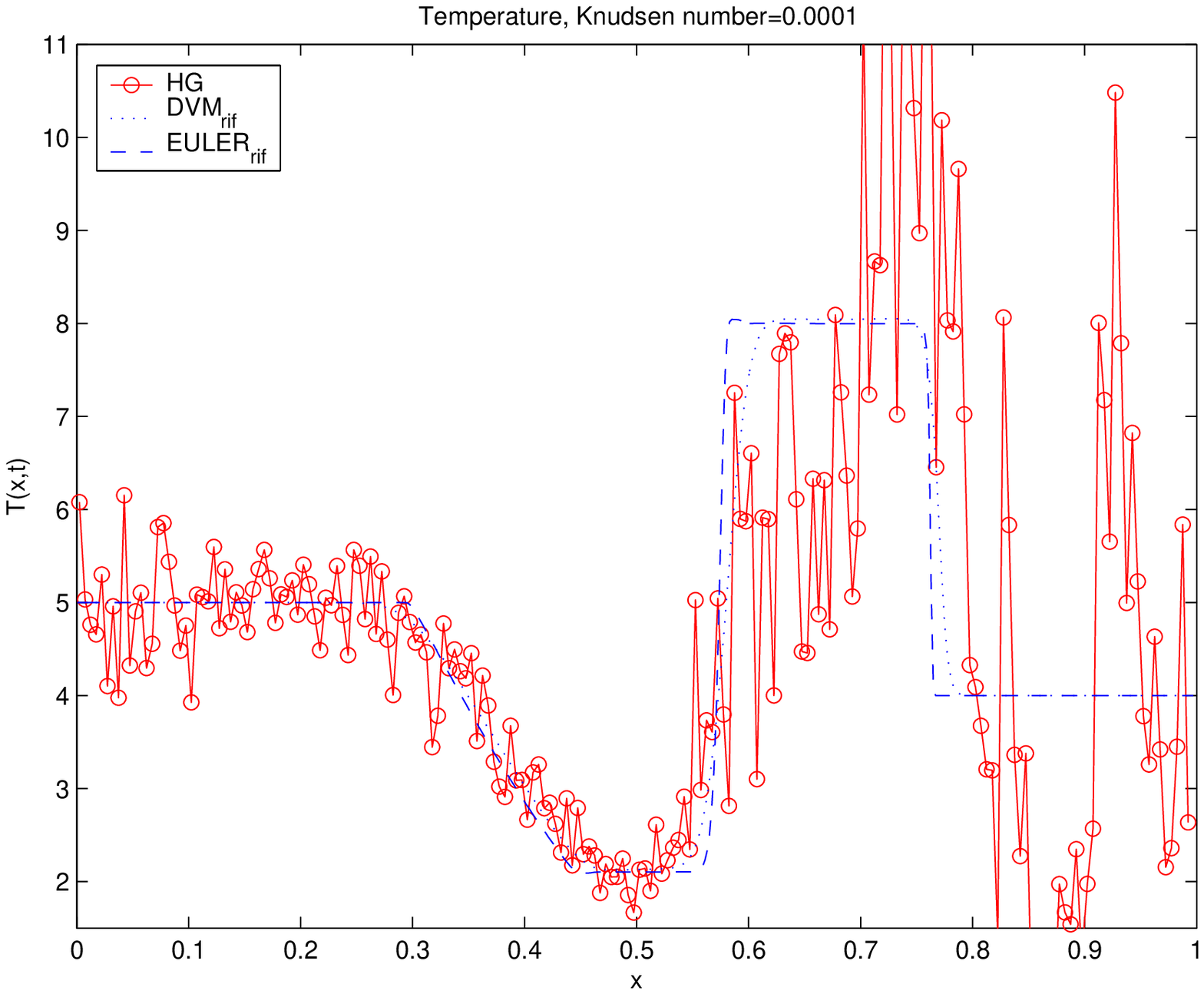}
\includegraphics[scale=0.39]{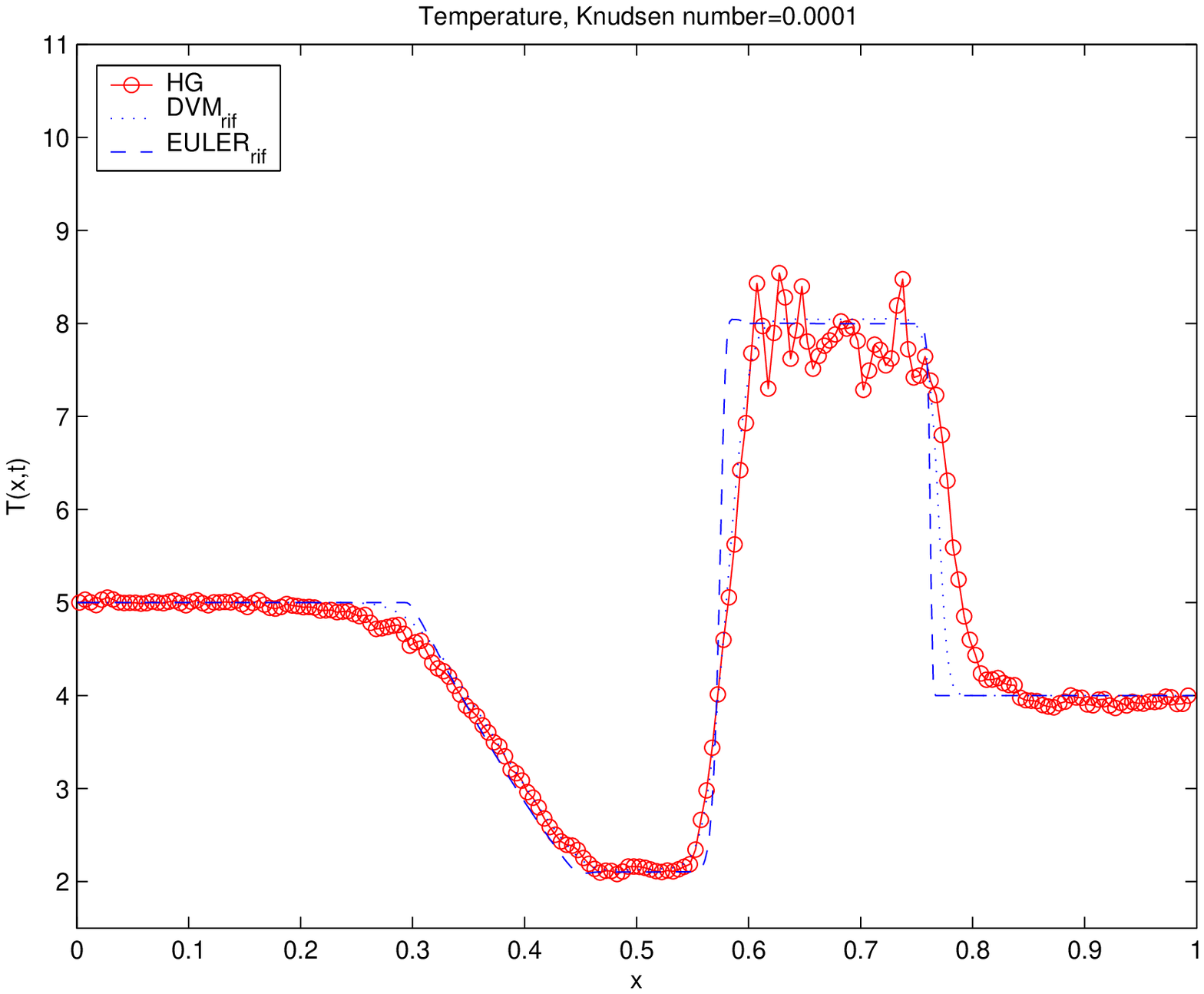}
\caption{Sod Shock Tube Test: Solution at $t=0.05$ for the density
(top), velocity (middle) and temperature (bottom). MC method (left),
Moment Guided MG method (right). Knudsen number
$\varepsilon=10^{-4}$. Reference solution: dash dotted line. Euler
solution: continuous line. Monte Carlo or Moment Guided: circles
plus continuous line.} \label{ST3}
\end{center}
\end{figure}

\begin{figure}
\begin{center}
\includegraphics[scale=0.39]{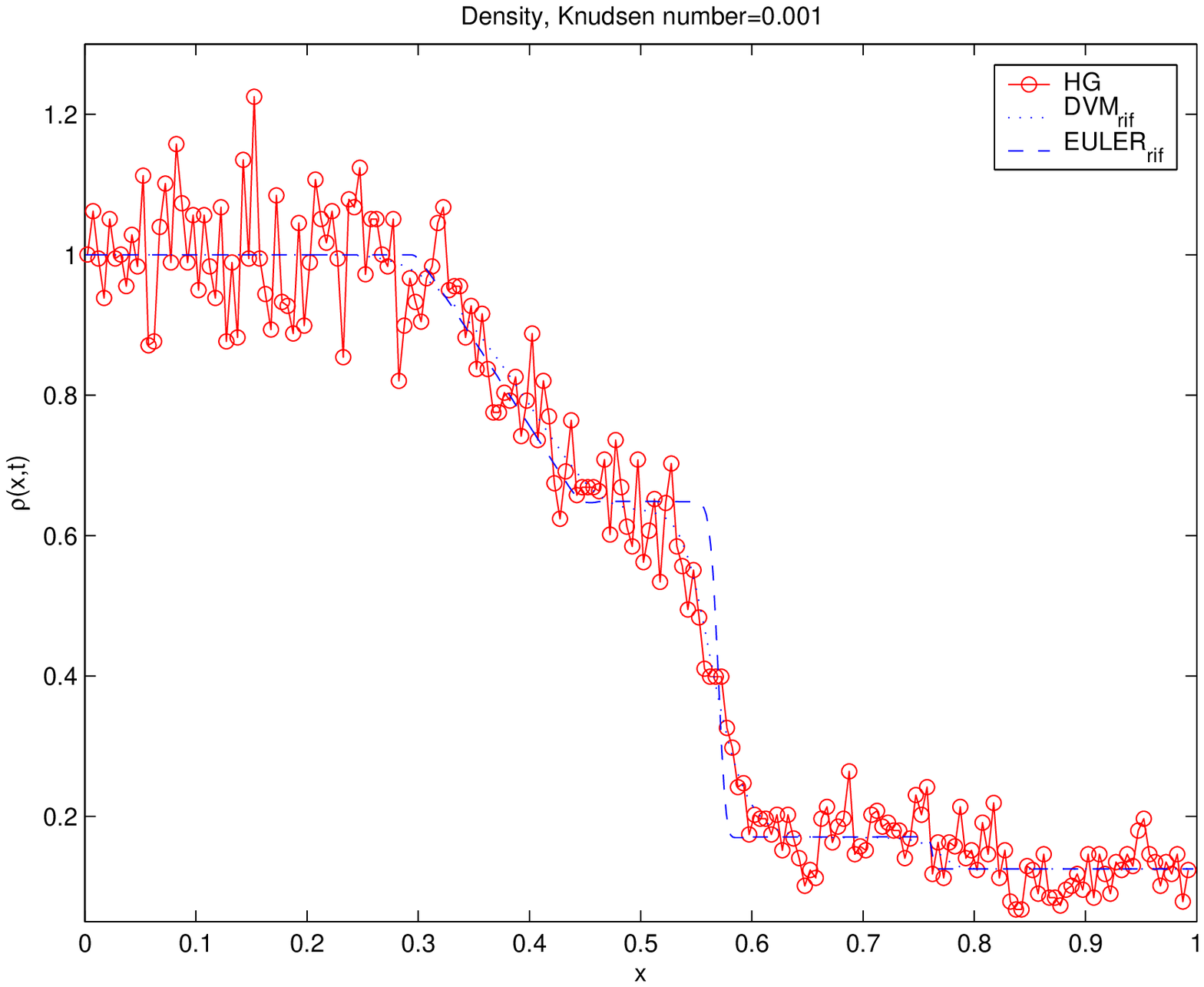}
\includegraphics[scale=0.39]{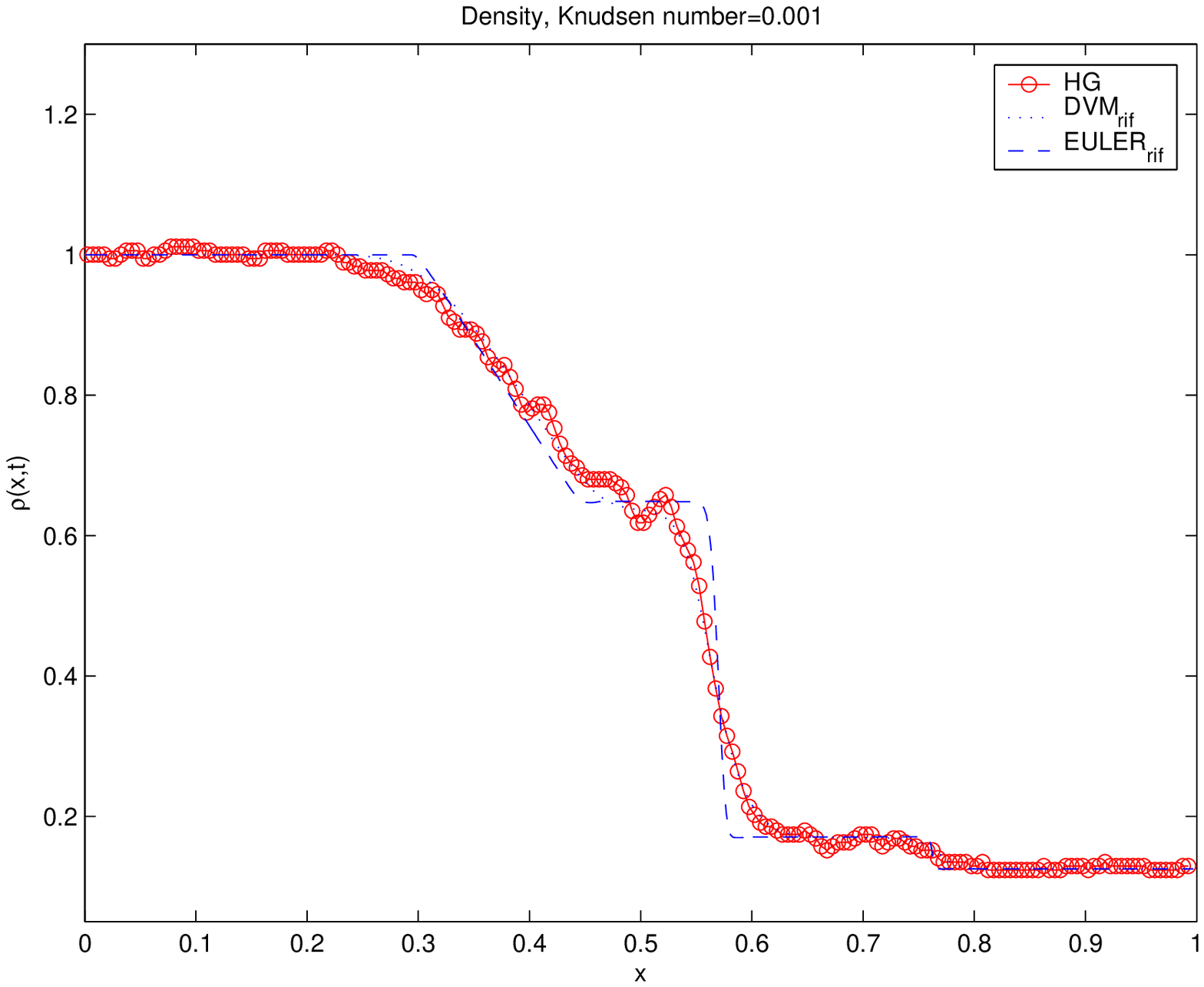}
\includegraphics[scale=0.39]{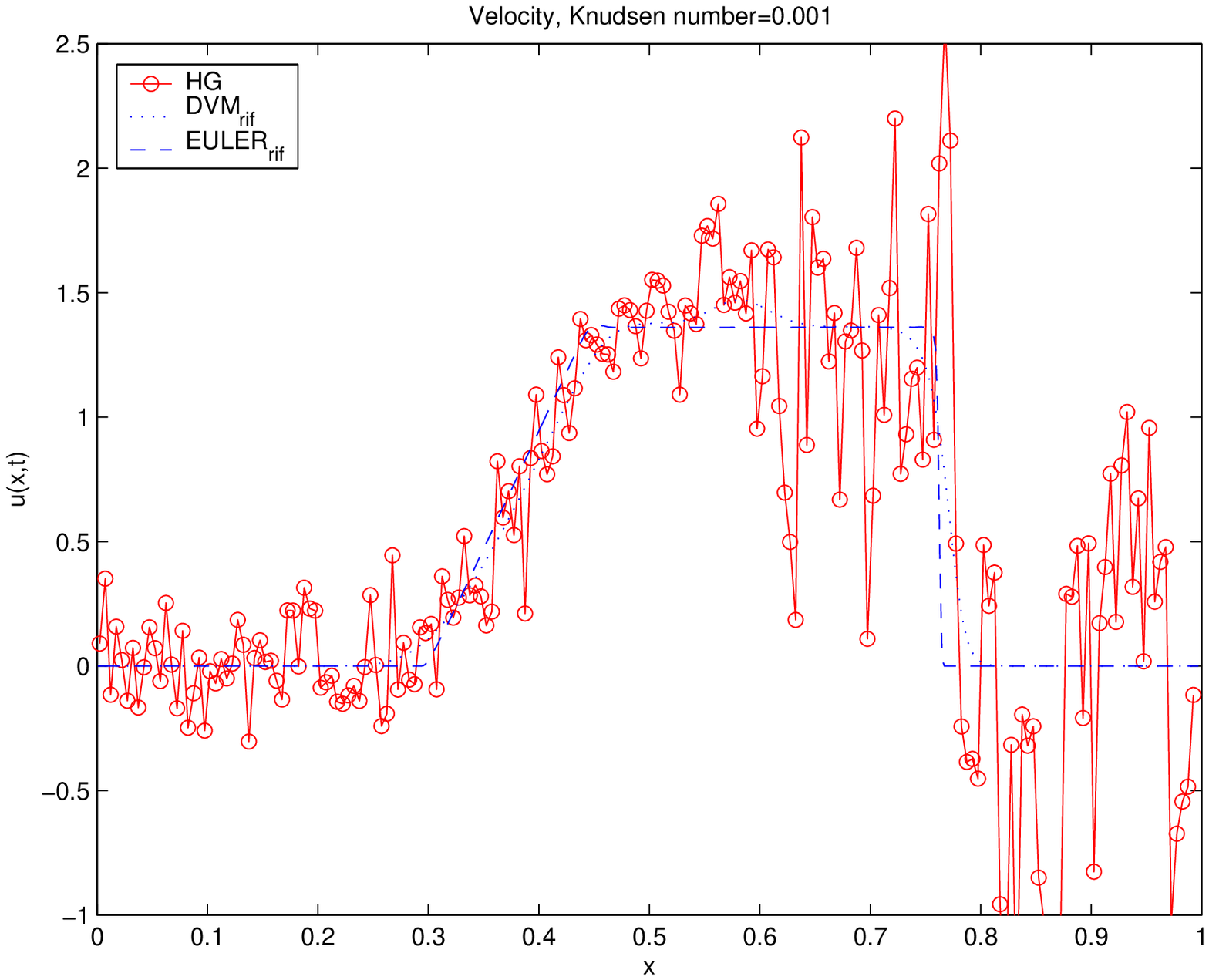}
\includegraphics[scale=0.39]{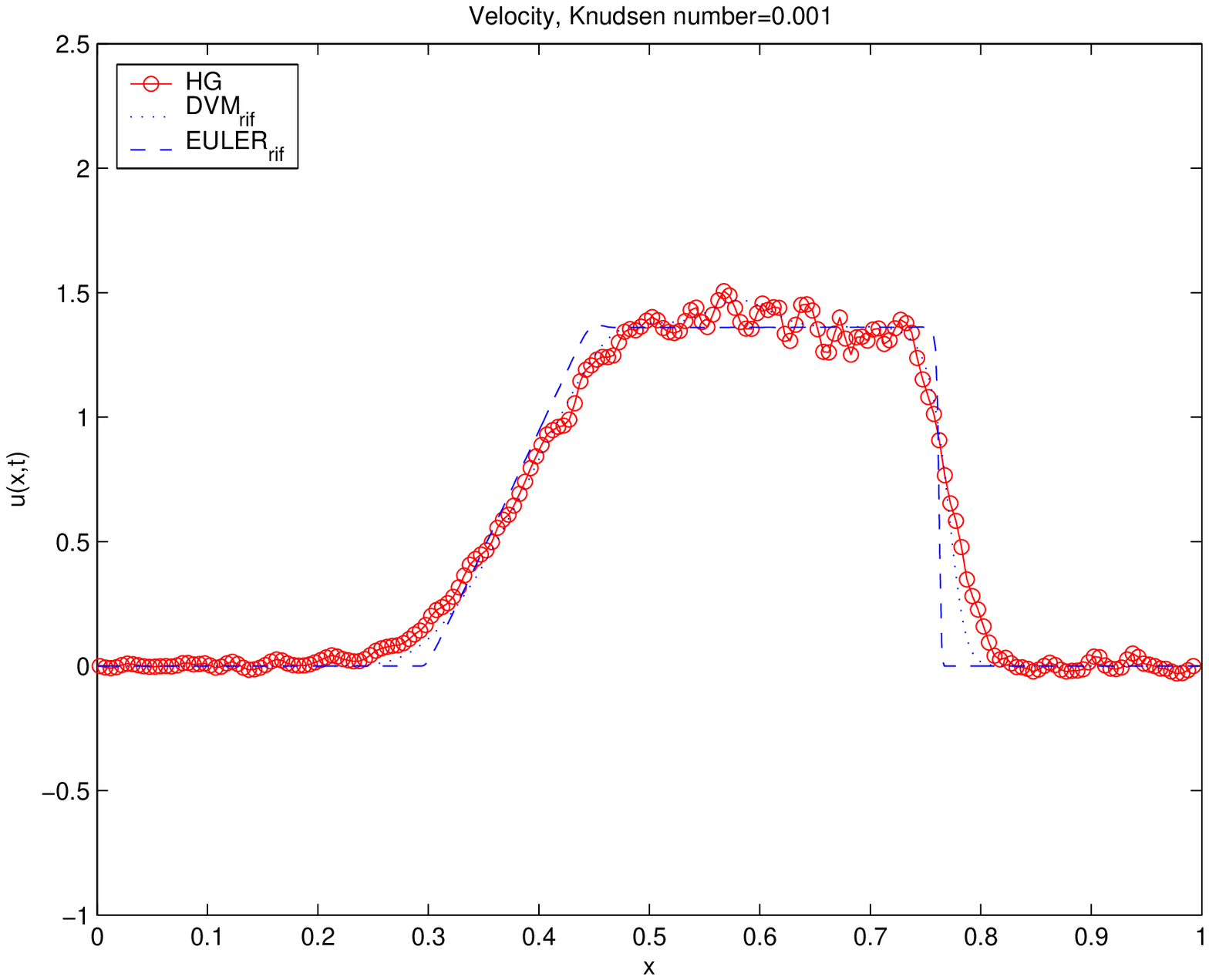}
\includegraphics[scale=0.39]{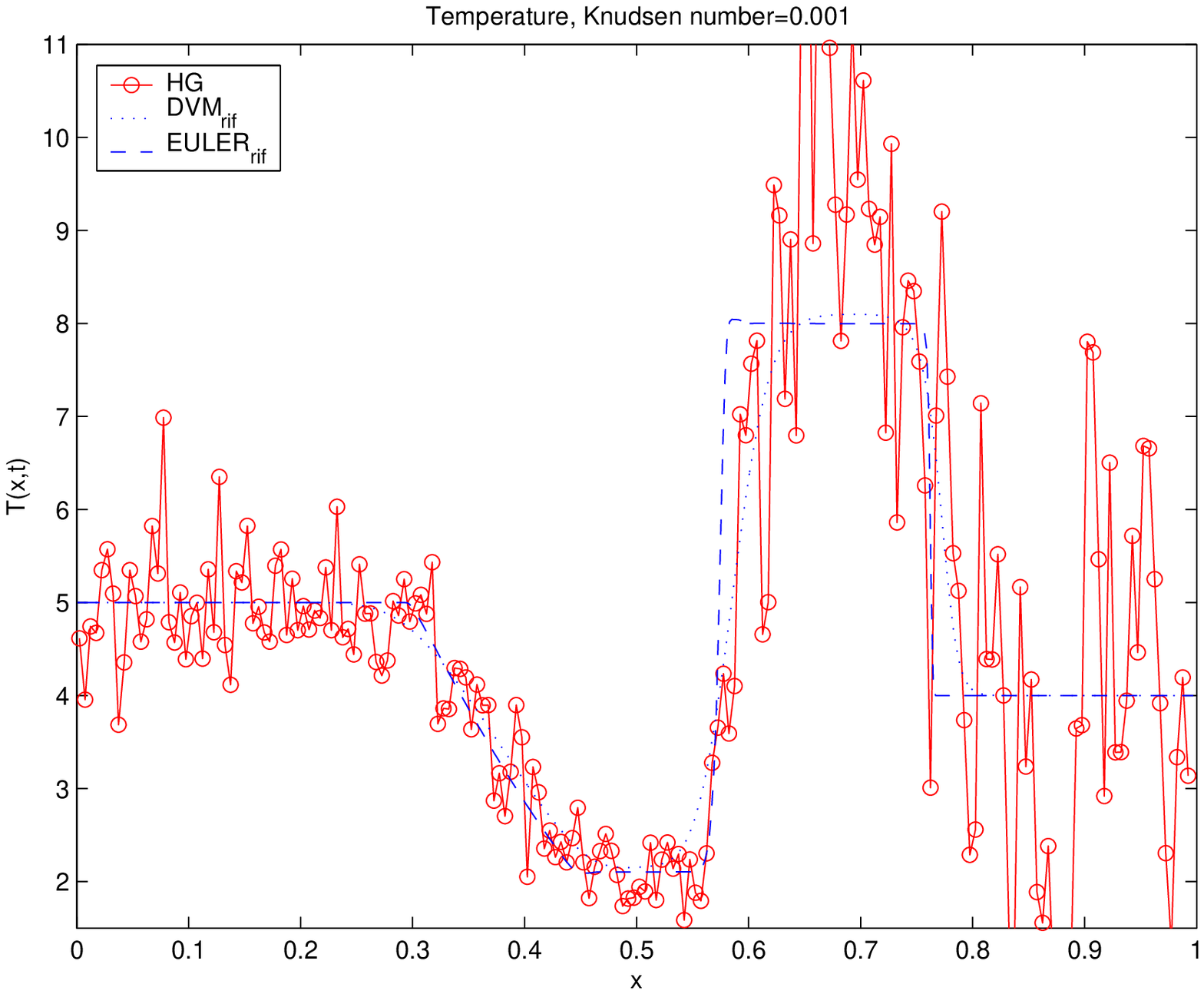}
\includegraphics[scale=0.39]{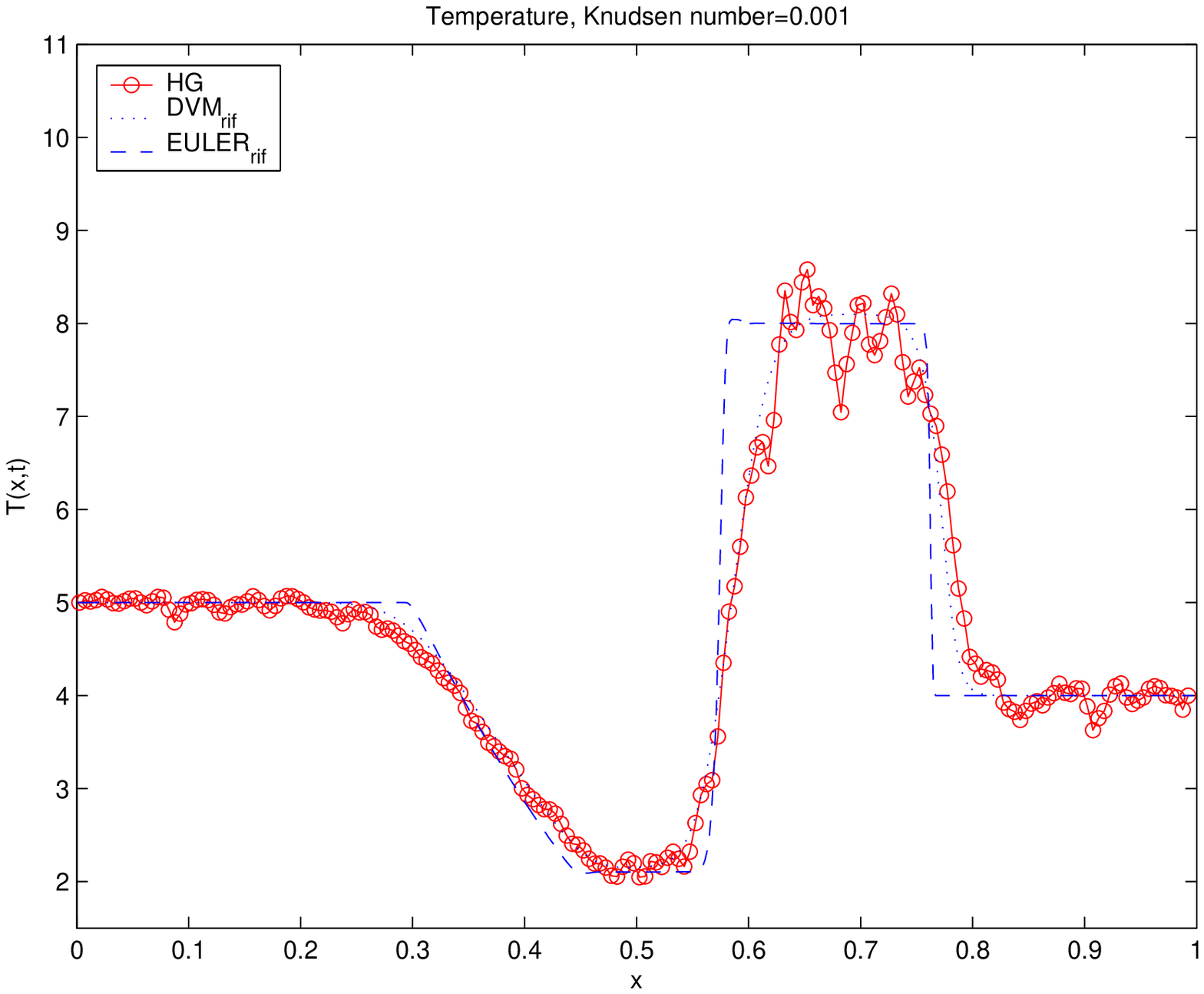}
\caption{Sod Shock Tube Test: Solution at $t=0.05$ for the density
(top), velocity (middle) and temperature (bottom). MC method (left),
Moment Guided MG method (right). Knudsen number
$\varepsilon=10^{-3}$. Reference solution: dash dotted line. Euler
solution: continuous line. Monte Carlo or Moment Guided: circles
plus continuous line.} \label{ST2}
\end{center}
\end{figure}

\begin{figure}
\begin{center}
\includegraphics[scale=0.39]{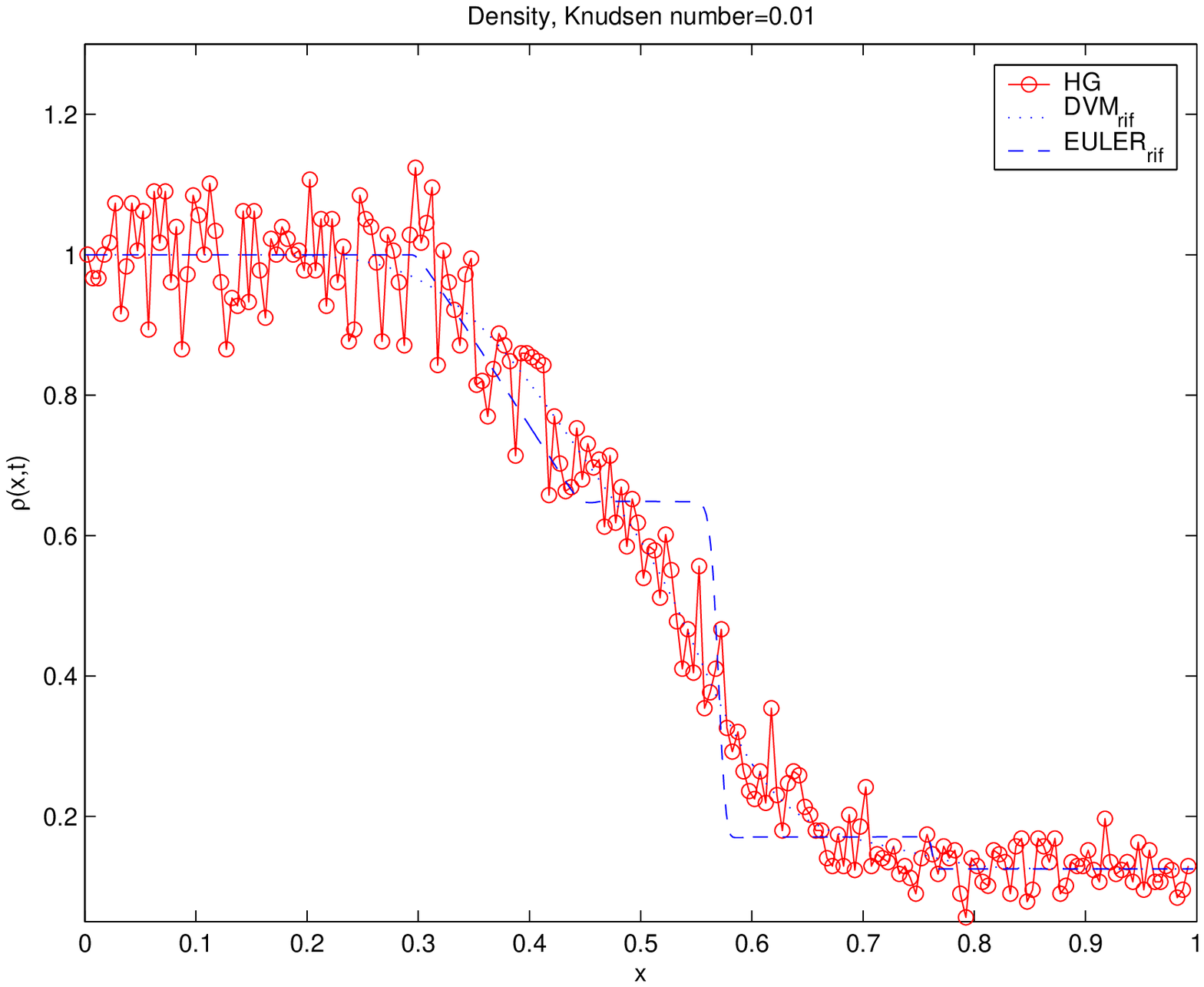}
\includegraphics[scale=0.39]{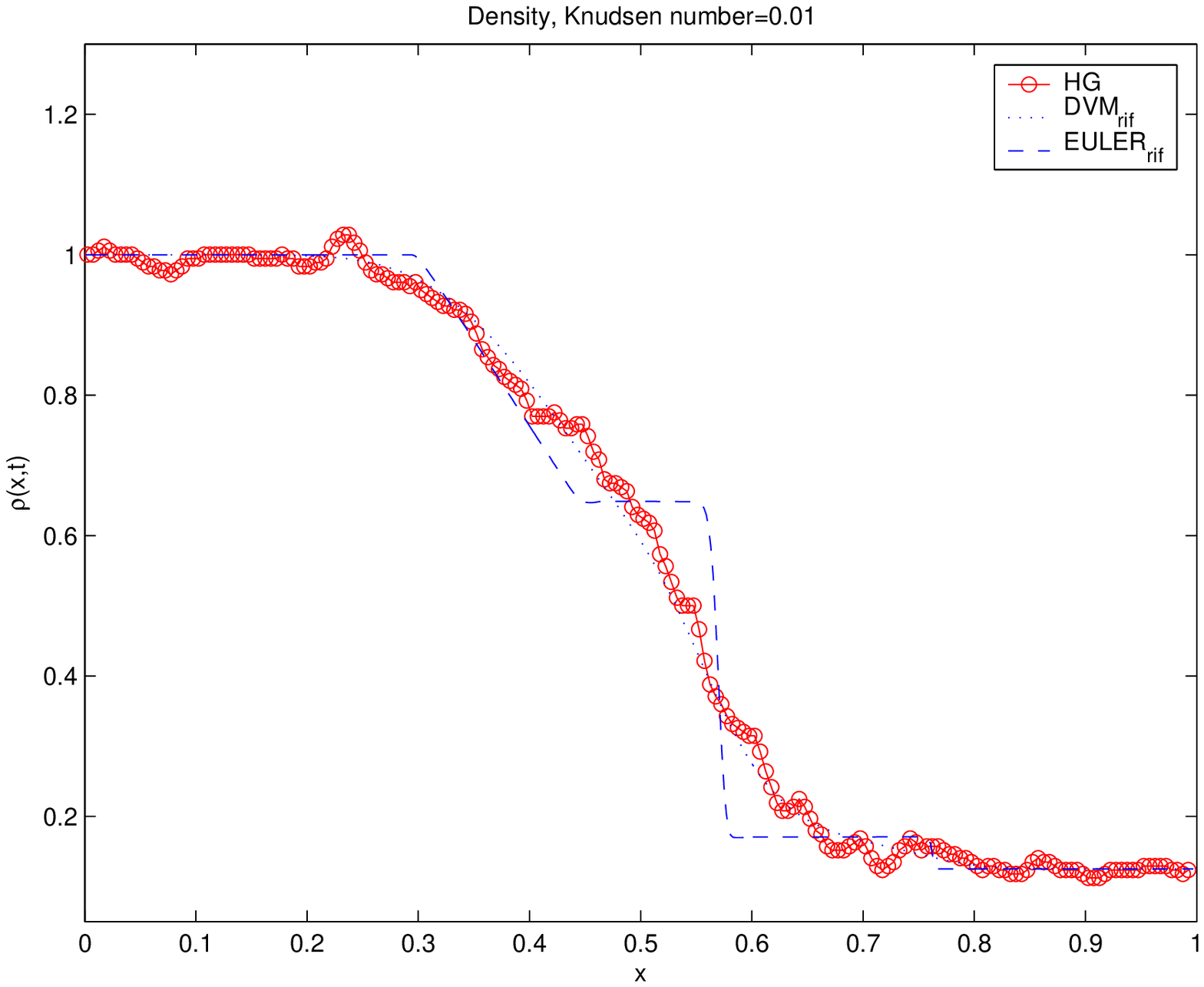}
\includegraphics[scale=0.39]{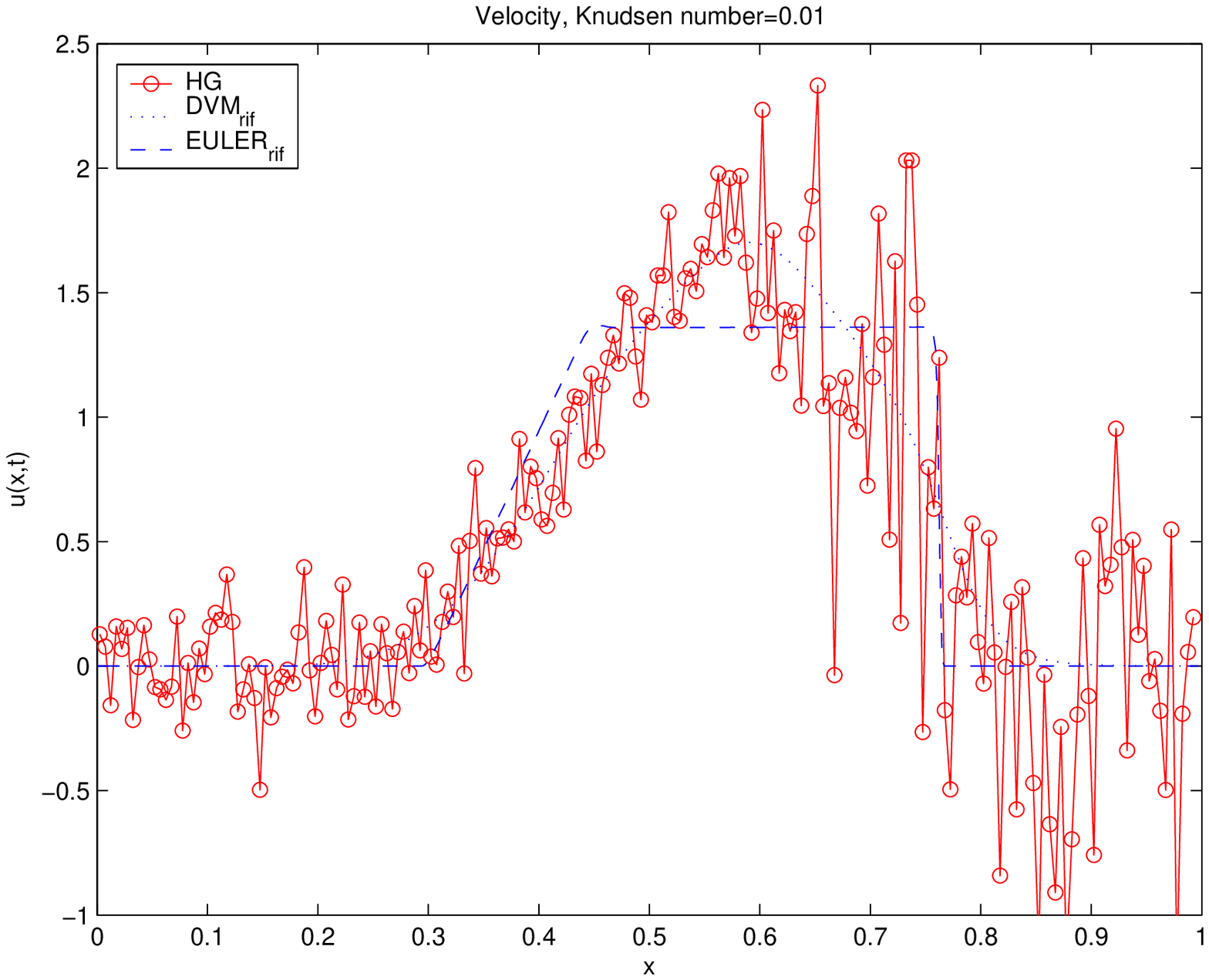}
\includegraphics[scale=0.39]{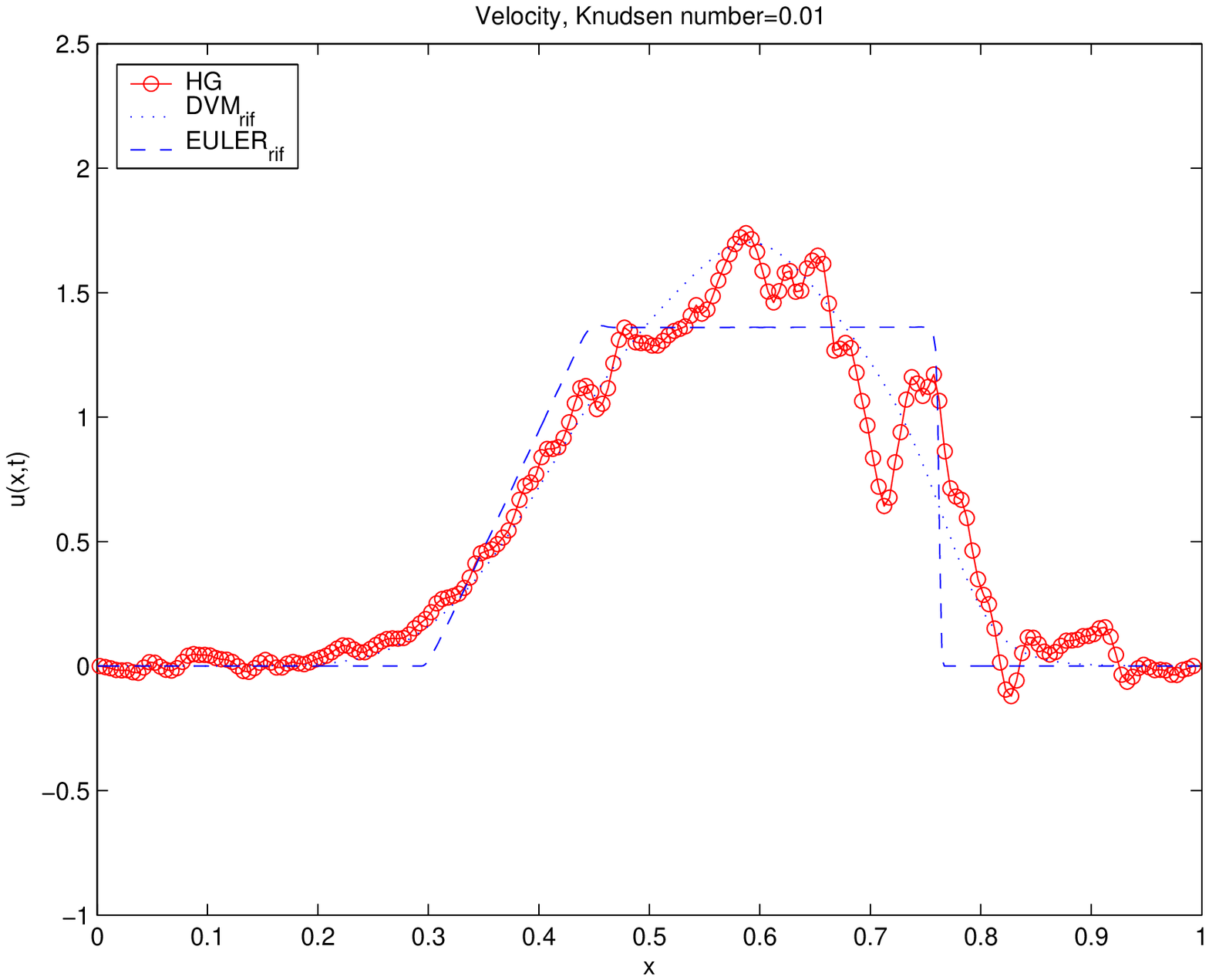}
\includegraphics[scale=0.39]{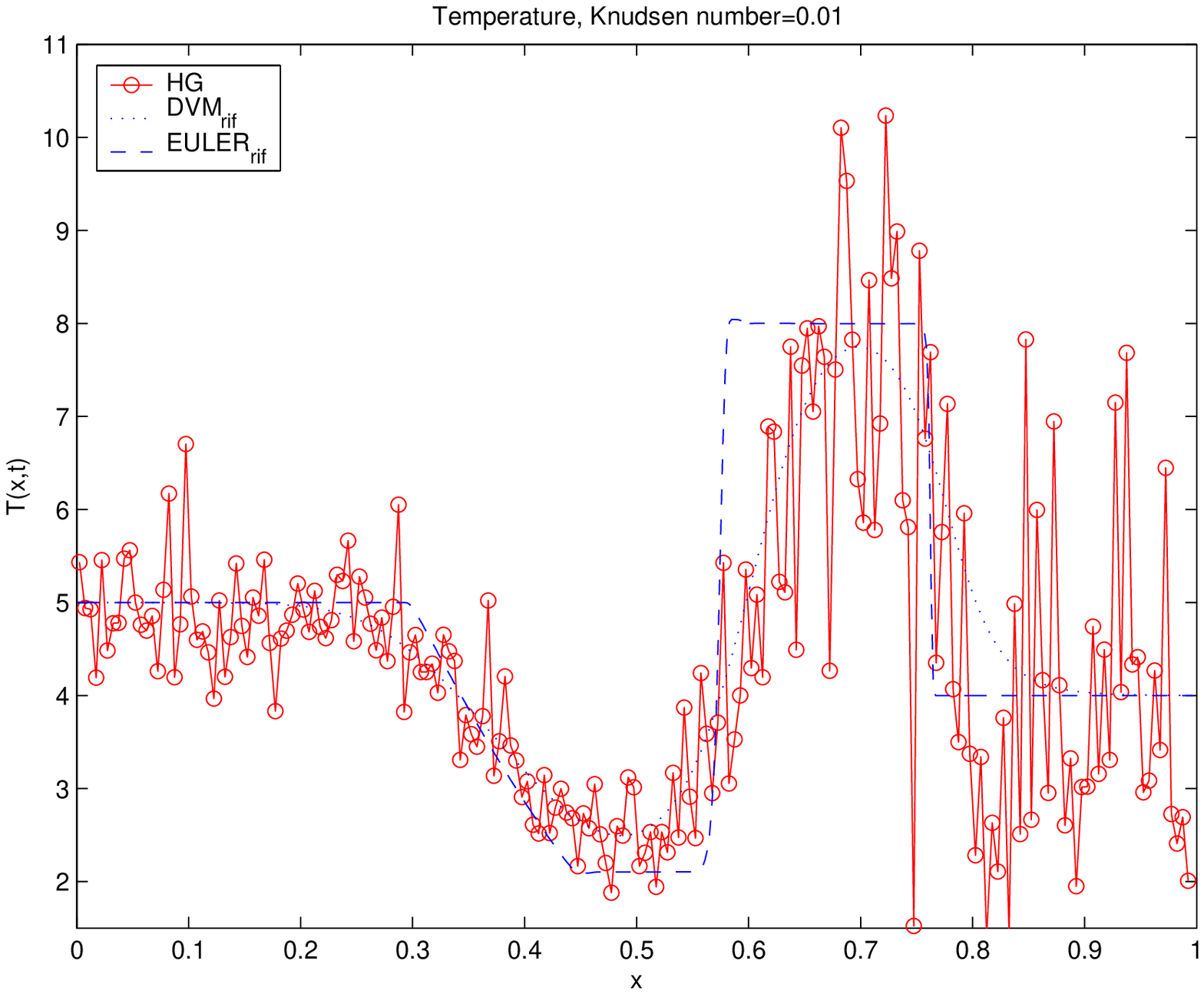}
\includegraphics[scale=0.39]{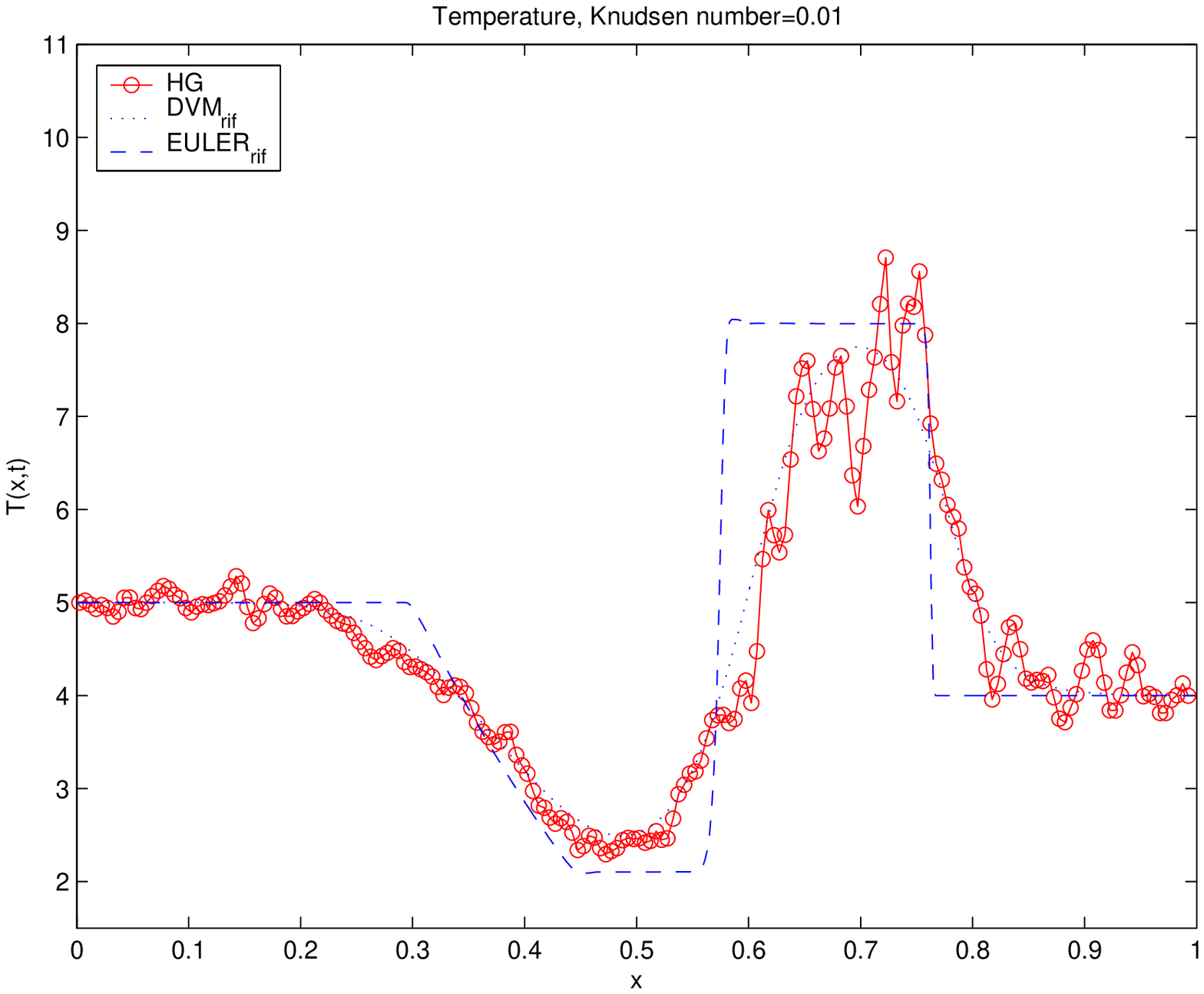}
\caption{Sod Shock Tube Test: Solution at $t=0.05$ for the density
(top), velocity (middle) and temperature (bottom). MC method (left),
Moment Guided MG method (right). Knudsen number
$\varepsilon=10^{-2}$. Reference solution: dash dotted line. Euler
solution: continuous line. Monte Carlo or Moment Guided: circles
plus continuous line.} \label{ST1}
\end{center}
\end{figure}

\begin{figure}
\begin{center}
\includegraphics[scale=0.39]{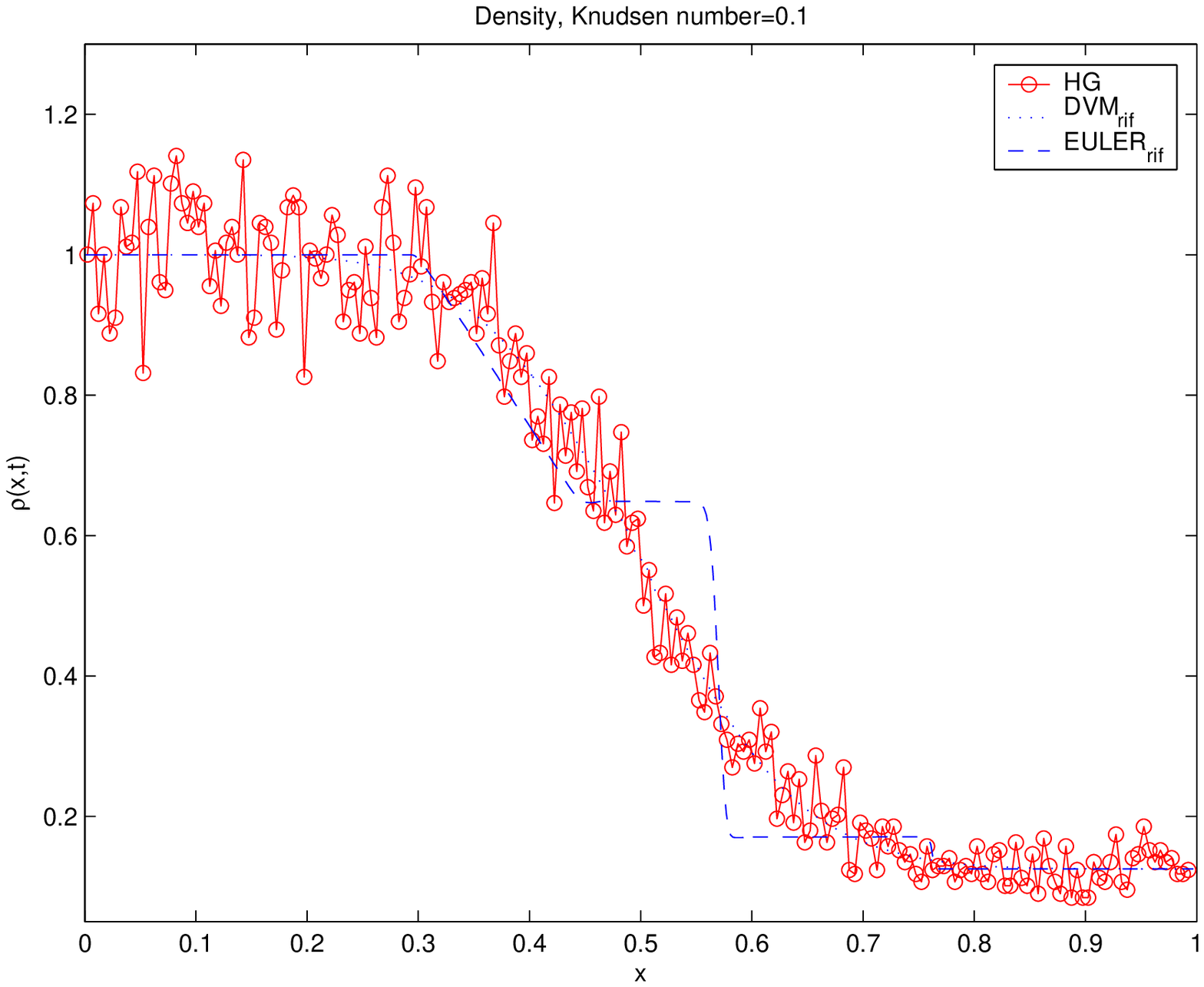}
\includegraphics[scale=0.39]{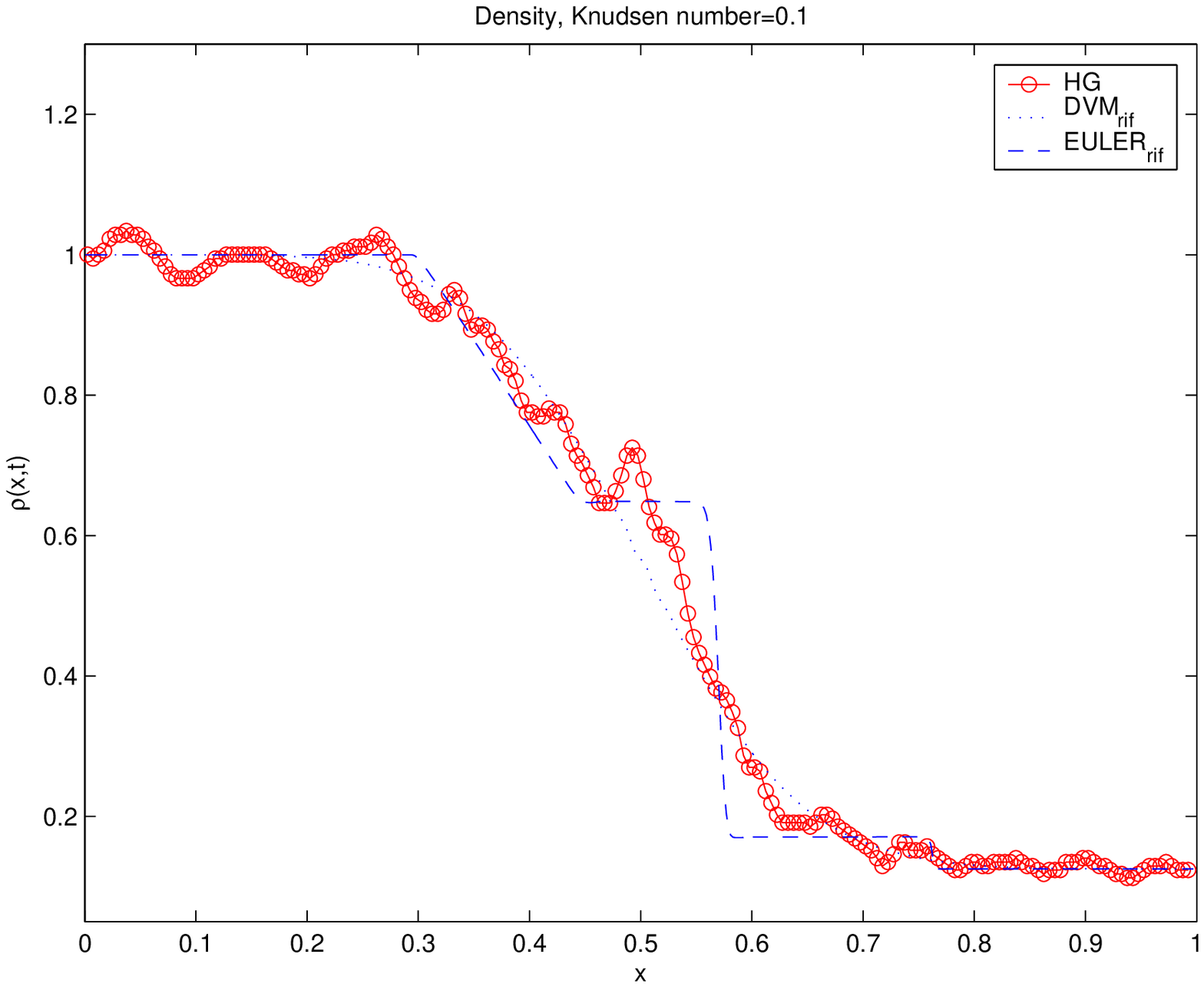}
\includegraphics[scale=0.39]{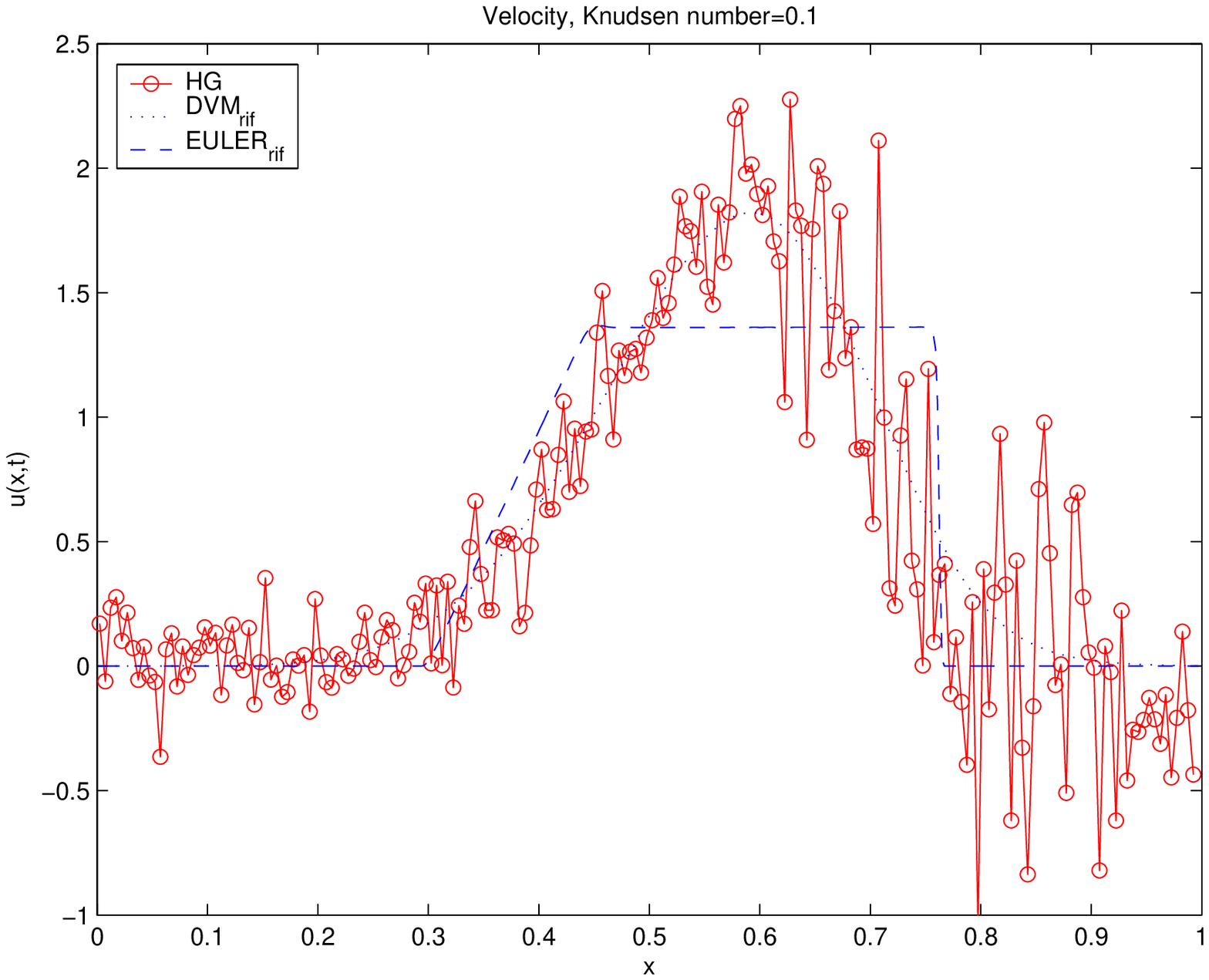}
\includegraphics[scale=0.39]{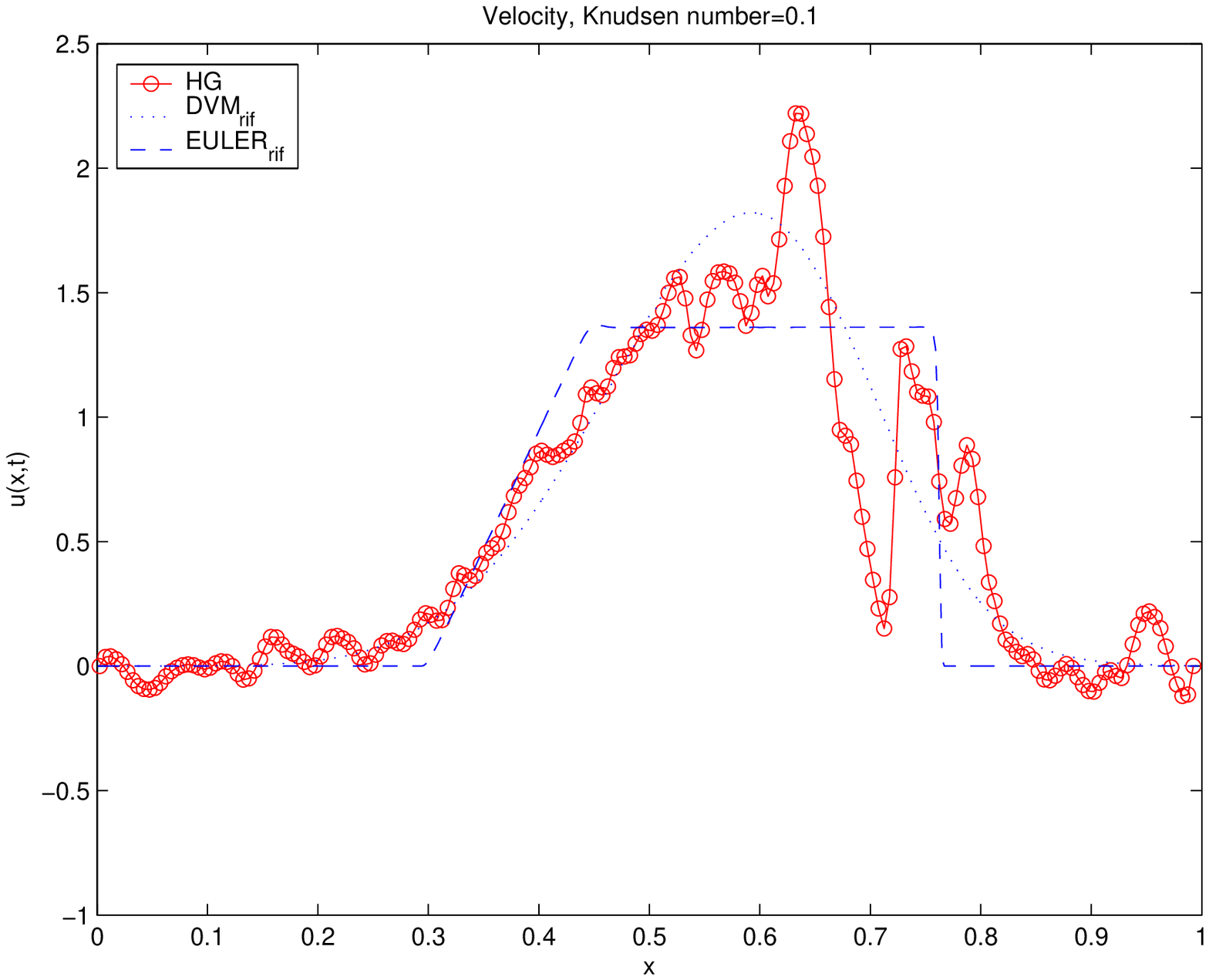}
\includegraphics[scale=0.39]{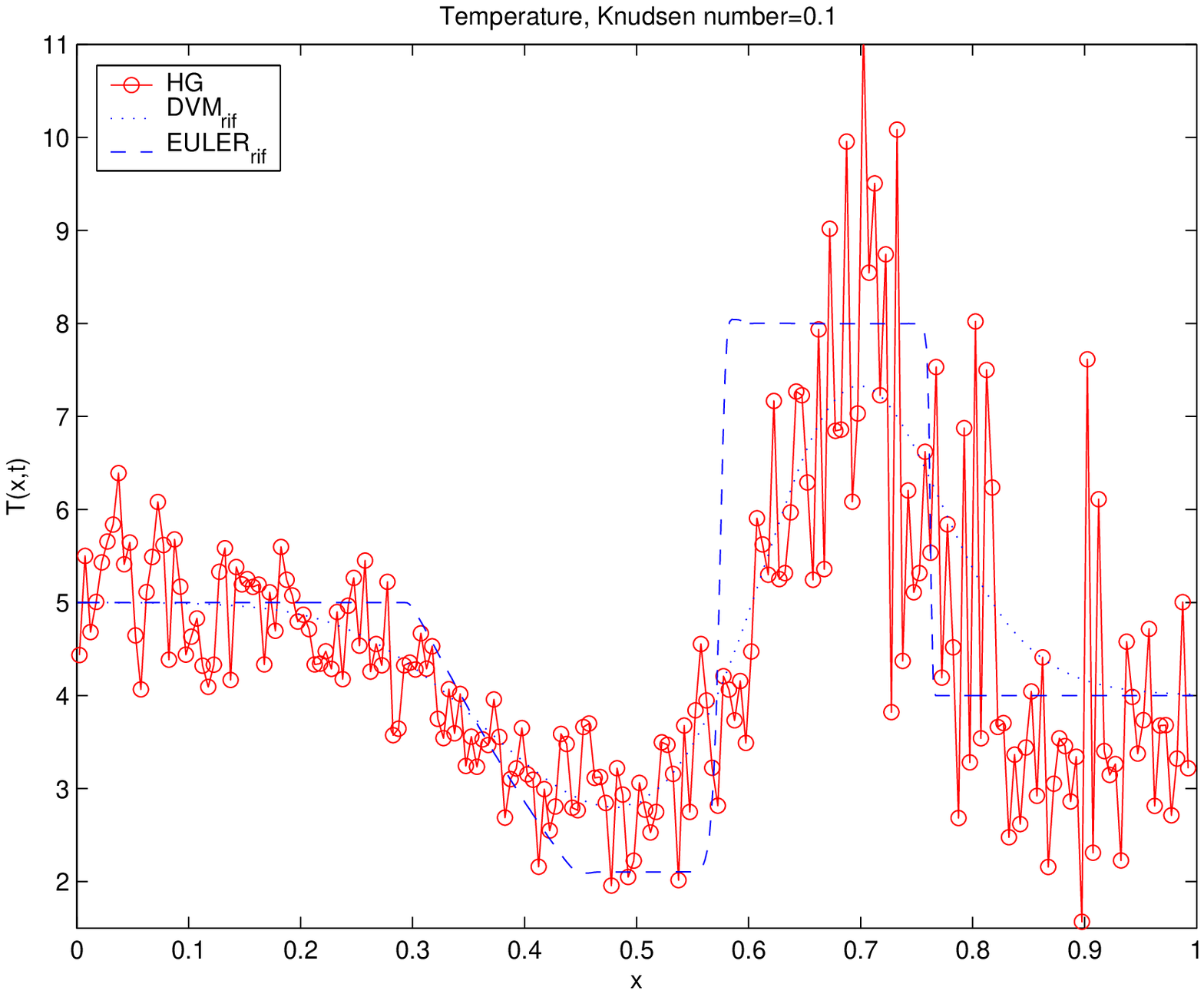}
\includegraphics[scale=0.39]{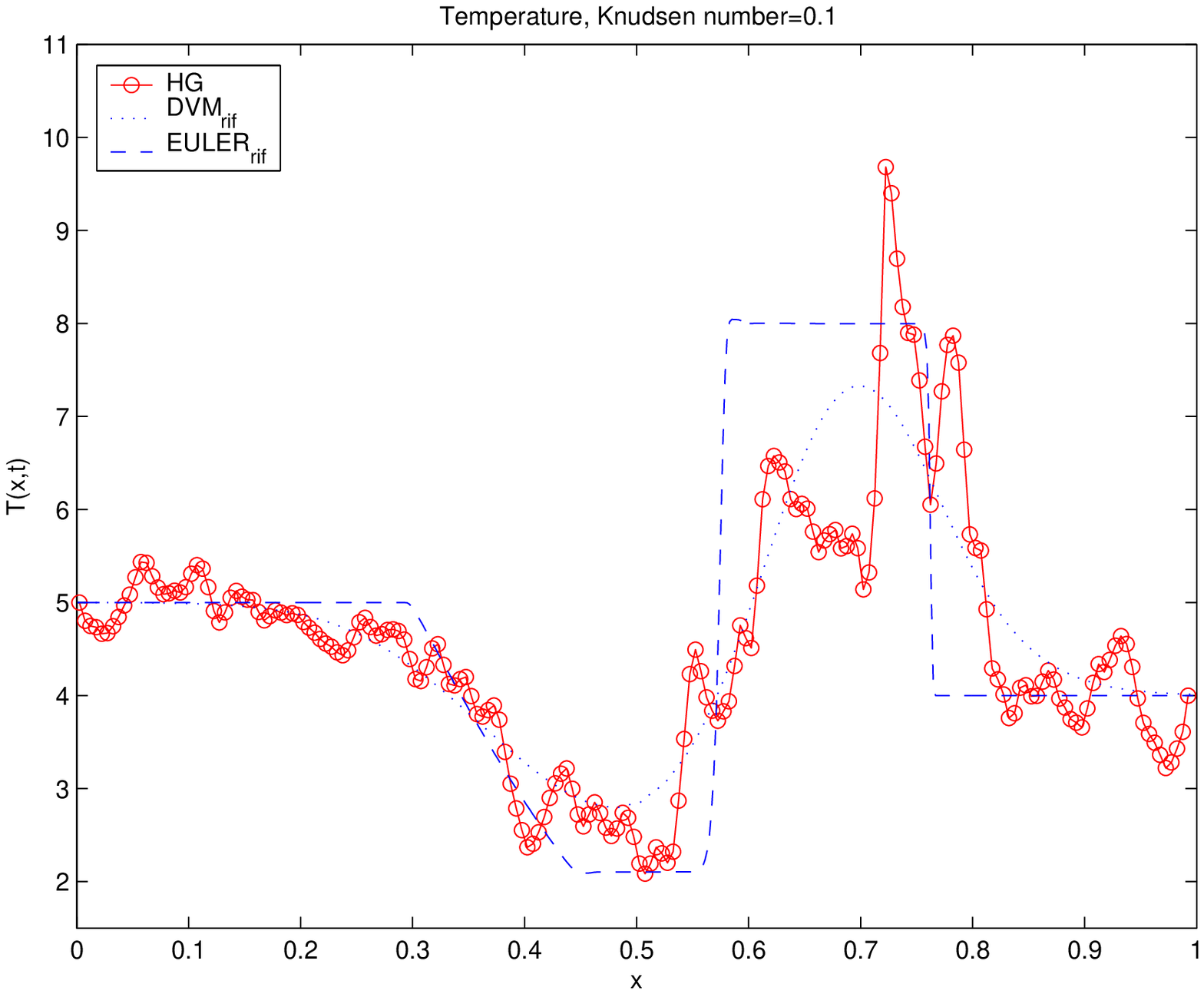}
\caption{Sod Shock Tube Test: Solution at $t=0.05$ for the density
(top), velocity (middle) and temperature (bottom). MC method (left),
Moment Guided MG method (right). Knudsen number
$\varepsilon=10^{-1}$. Reference solution: dash dotted line. Euler
solution: continuous line. Monte Carlo or Moment Guided: circles
plus continuous line.} \label{ST0}
\end{center}
\end{figure}

Figures \ref{ST3} to \ref{ST0} consider the same initial data for
the density, mean velocity and temperature with different initial
Knudsen number values, which range from $\varepsilon=10^{-4}$ to
$\varepsilon=10^{-1}$. $100$ per cell are used and only one
realization is considered. As for the unsteady shock test each
figure depicts the density, mean velocity and temperature from top
to bottom, with the pure Monte Carlo solver (left) and the Moment
Guided method (right). A reference solution obtained through a
discrete velocity scheme \cite{Mieussens} is represented in each
figure as well as the solution of the compressible Euler equations.
The figures shows good results for all ranges of Knudsen numbers in
terms of reduction of fluctuations. The high order solver does not
seem to increase the variance but improves the solution in the fluid
limit.

\section{Conclusions} We have developed a new class of hybrid methods
which aim at reducing the variance in Monte Carlo schemes. The key
idea consists in driving particle positions and velocities in such a
way that moments given by the solution of the kinetic equation
exactly match moments given by the solution of an appropriate set of
moment equations. It is important to point out that the schemes
which can be derived through this technique can be easily
implemented in existing Monte Carlo codes through few modifications:
adding a fluid solver and a routine for the moment matching.

Preliminary numerical results show reductions of fluctuations in all
regimes compared to DSMC. The reduction becomes stronger as we
approach equilibrium. Numerical convergence tests show better
performances of the proposed method, in terms of stochastic error,
compared to pure Monte Carlo schemes. For these problems the moment
guided method seems very promising, leading to solutions which
contain less fluctuations at a computational cost which is
comparable to the cost of a traditional Monte Carlo solver, and the
addition of the cost of a macroscopic solver for the compressible
Euler equations, which is usually computationally less expensive
than the Monte Carlo method.

Currently, we are working on extensions of the present method to the
full Boltzmann equation in the multidimensional case. To this aim we
plan to use both classical Monte Carlo methods like Bird or Nanbu
methods \cite{Nanbu80} and time relaxed Monte Carlo (TRMC)
techniques \cite{PR}. Moreover we plan to explore other possible
algorithms which can possibly further reduce fluctuations, such as
matching higher order moments and/or using higher order closure of
the hierarchy in order to solve a larger set of hydrodynamics
equations, or using hybrid representations of the distribution
function \cite{CPima}. We hope to be able to present other results
supporting this methodology in the near future \cite{dimarco1}.


\end{document}